\documentclass[twocolumn]{aastex701}

\usepackage{threeparttable, tablefootnote}
\usepackage[version=3]{mhchem}
\usepackage{gensymb}
\usepackage{siunitx} 
\usepackage{rotating}  
\usepackage[table]{xcolor}
\usepackage[most]{tcolorbox} 
\usepackage{array}
\usepackage{multirow}
\usepackage{rotating}  
\usepackage{ragged2e}  

\newtcbox{\myboxi}[1][]{nobeforeafter,tcbox raise base,colframe=green!50!black,colback=green!50!black,height=8pt,valign=center,raster valign=center,
  box align=base,sharp corners,top=0pt,bottom=0pt,left=0pt,right=2pt,
  boxrule=0pt,boxsep=2.5pt,before upper=\strut,#1}

\newtcbox{\myboxxi}[1][]{nobeforeafter,tcbox raise base,colframe=black!50!white,colback=black!10!white,height=8pt,valign=center,raster valign=center,
  box align=base,sharp corners,top=0pt,bottom=0pt,left=0pt,right=2pt,
  boxrule=0pt,boxsep=2.5pt,before upper=\strut,#1}

\newtcbox{\myboxxxi}[1][]{nobeforeafter,tcbox raise base,colframe=yellow!85!black,colback=yellow!85!black!,height=8pt,valign=center,raster valign=center,
  box align=base,sharp corners,top=0pt,bottom=0pt,left=0pt,right=2pt,
  boxrule=0pt,boxsep=2.5pt,before upper=\strut,#1}

\usepackage{array}
\newcolumntype{P}[1]{>{\centering\arraybackslash}p{#1}}
\begin{document}

\title{JWST-MIRI's multi-dimensional view of mass loss in the irradiated disks of NGC 1977}

\correspondingauthor{Alice S. Booth} 
\email{alice.booth@cfa.harvard.edu}

\author[0000-0003-2014-2121]{Alice S. Booth} 
\altaffiliation{Clay Postdoctoral Fellow}
\affiliation{Center for Astrophysics \textbar\, Harvard \& Smithsonian, 60 Garden St., Cambridge, MA 02138, USA}
\email{alice.booth@cfa.harvard.edu}

\author[0000-0002-2028-8860]{Qijia Zhou} 
\affiliation{Center for Astrophysics \textbar\, Harvard \& Smithsonian, 60 Garden St., Cambridge, MA 02138, USA}
\email{qijia.zhou@cfa.harvard.edu}

\author[0000-0001-6072-9344]{Jinyoung Serena Kim} 
\affiliation{Steward Observatory, University of Arizona, 933 N. Cherry Ave, Tucson, AZ 85721-0065, USA}
\email{serena00@arizona.edu}

\author[0000-0001-8798-1347]{Karin Öberg}
\affiliation{Center for Astrophysics \textbar\, Harvard \& Smithsonian, 60 Garden St., Cambridge, MA 02138, USA}
\email{koberg@cfa.harvard.edu }

\author[0000-0002-0150-0125]{Jenny Calahan}
\affiliation{Center for Astrophysics \textbar\, Harvard \& Smithsonian, 60 Garden St., Cambridge, MA 02138, USA}
\email{jenny.calahan@cfa.harvard.edu}

\author[0000-0001-7552-1562]{Klaus Pontoppidan}
\affiliation{Jet Propulsion Laboratory, California Institute of Technology, 4800 Oak Grove Drive, Pasadena, CA 91109, USA}
\affiliation{Division of Geological and Planetary Sciences, California Institute of Technology, MC 150-21, 1200 E California Boulevard, Pasadena,
CA 91125, USA}
\email{klaus.m.pontoppidan@jpl.nasa.gov}

\author[0000-0002-9593-7618]{Thomas J. Haworth}
\affiliation{Astronomy Unit, Department of Physics and Astronomy, Queen Mary University of London, Mile End Road, London E1 4NS, UK}
\email{apw688@qmul.ac.uk}

\author[0000-0001-7962-1683]{Ilaria Pascucci}
\affiliation{Lunar and Planetary Laboratory, The University of Arizona, Tucson, AZ 85721, USA}
\email{pascucci@arizona.edu}

\author[0000-0002-0150-0125]{Ryan Boyden}
\altaffiliation{Jansky Fellow of the National Radio Astronomy Observatory}
\affiliation{National Radio Astronomy Observatory, Charlottesville, VA 22903, USA}
\affiliation{Department of Astronomy, University of Virginia, Charlottesville, VA 22904, USA}
\email{rboyden@virginia.edu}

\author[0000-0002-0554-1151]{Mayank Narang}
\affiliation{Jet Propulsion Laboratory, California Institute of Technology, 4800 Oak Grove Drive, Pasadena, CA 91109, USA}
\email{mayankn1154@gmail.com}

\author[0000-0001-8284-4343]{Karina Maucó}
\affiliation{European Southern Observatory, Karl-Schwarzschild-Strasse 2, 85748 Garching bei München, Germany}
\email{kmaucoco@eso.org}

\author[0000-0001-8407-4020]{Aditya M. Arabhavi}
\affiliation{Kapteyn Astronomical Institute, Rijksuniversiteit Groningen, Postbus 800, 9700AV Groningen, The Netherlands}
\email{arabhavi@astro.rug.nl}

\author[0000-0002-4276-3730]{Nicholas Ballering}
\affiliation{Space Science Institute, Boulder, CO 80301, USA}
\email{nballering@spacescience.org}

\begin{abstract}
The evolution of protoplanetary disks, and consequently the outcomes of planet formation, are thought to be significantly altered in regions containing massive stars. Extreme cases in the Orion Nebula Cluster (ONC) demonstrate the impact of external irradiation  (FUV$\gtrsim10^{4}~G_{0}$) on disk evolution, but intermediate environments remain less observationally constrained. We present JWST/MIRI Medium Resolution Spectroscopy (MRS) observations of seven proplyds in NGC~1977 exposed to an external FUV field of $10^{3}-10^{5}~G_{0}$ from the B1V star 42~Orionis (42~Ori). We characterize emission from molecular (\ce{H2}) and atomic (e.g., [Ne II], [Ar II], HI) species, and in some cases, MIRI reveals extended emission tracing the proplyd ionization front and wind. The closest disk to 42~Ori, KCFF\#1, is undergoing extreme mass loss, traced by a 1000’s-of-au-long dusty tail, and lacks clear \ce{H_2} or HI emission, indicating an advanced stage of dispersal. The remaining six disks exhibit two-temperature components of \ce{H2} emission (500–700~K and 1000–1500~K), likely tracing the disk molecular layer and a photoevaporative wind, alongside HI lines which are used to estimate mass accretion rates. When comparing KCFF\#2 and \#6, which have similar host stars, KCFF\#2 (closer to 42~Ori) is dominated by externally driven mass loss, with extended molecular and atomic emission, whereas KCFF\#6 only shows extended \ce{H2} emission, with roughly equal contributions from accretion and external mass loss. While the sample is small, this work demonstrates how JWST/MIRI can assess environmental impacts on disk evolution, with NGC~1977 bridging strongly irradiated disks in the ONC and the more local population.

\end{abstract}

\keywords{}

\section{Introduction} 

The diverse outcomes of planet formation are expected to reflect their formation environments, which include not only the disks of gas, dust, and ice from which planetary systems assemble from \citep{2023ARA&A..61..287O}, but also external factors such the proximity to O- and B-type stars \citep{2022EPJP..137.1132W, 2025OJAp....8E..54A}. The strong ultraviolet (UV) radiation fields produced by these massive stars heat and ionize nearby disks, processes predicted to alter their physical and chemical evolution and significantly accelerate their dispersal \citep[e.g.,][]{2007MNRAS.376.1350C, 2013ApJ...766L..23W, 2018MNRAS.481..452H, 2020MNRAS.492.1279S, 2021MNRAS.503.4172H, 2023MNRAS.526.4315H, 2025MNRAS.537..598K,2025ApJ...991...94C}. Quantifying the impacts of these effects observationally is essential for connecting disk properties to the observed diversity of mature exoplanetary systems.

The most fundamental impact of external irradiation is the induction of winds and the resulting mass loss from the disk \citep{2022EPJP..137.1132W}. These photoevaporative winds are distinct from internal MHD or star-driven thermal winds \citep{2023ASPC..534..567P} and are a compounding factor to consider in disk evolution. The former are dominated by far-ultraviolet (FUV) radiation and can dramatically shorten the disk lifetime and, in turn, set both the mass reservoir available for and the timescale of planet formation \citep{2022MNRAS.515.4287W, 2023MNRAS.522.1939Q,2024A&A...689A.338H,2025ApJ...979..120H}. 
In theory, mass-loss rates scale with the strength of the impinging FUV radiation (typically expressed in units of the interstellar value $G_{0} \equiv 1.6\times10^{-3}~\mathrm{erg~s^{-1}~cm^{-2}}$; \citealt{Habing1968}) and can reach $10^{-6}~M_{\odot}~\mathrm{yr^{-1}}$ in the most extreme environments where $F_{\mathrm{FUV}} > 10^{4}~G_{0}$ \citep{2018MNRAS.481..452H, 2023MNRAS.526.4315H,2024A&A...687A..93A}. Observationally, there is now ample evidence that external photoevaporation of protoplanetary disks is actively occurring in the Orion Nebula. First identified by VLA and HST imaging \citep{1987ApJ...321..516C, 1987ApJ...314..535G, 1993ApJ...410..696O} and termed \textit{proplyds}, these systems clearly show that disk evolution in Orion is strongly shaped by environment, with radiation from nearby massive stars affecting disk lifetimes, masses and sizes as well as their thermal and chemical structures \citep[e.g.,][]{2010ApJ...725..430M, 2014ApJ...784...82M, 2012ASInC...4...35M, 2018ApJ...860...77E,2020ApJ...894...74B,2023ApJ...947....7B,2023A&A...679A..82M,2023ApJ...954..127B, 2023A&A...673L...2V,2024A&A...687A..93A}.

The James Webb Space Telescope (JWST) offers a new view of protoplanetary disks, with the MIRI Medium Resolution Spectroscopy instrument (MIRI/MRS) with integral field unit (IFU) providing both a broad spectral coverage (4.9 to 27.9 $\mu$m) and enabling spatially resolved (0\farcs1-0\farcs3 pixel scales) atomic and molecular emission-line mapping \citep{2015PASP..127..595W, 2015PASP..127..646W,2023PASP..135d8003W,2023A&A...675A.111A}. In young Class~0/I systems, these IFU-data reveal sources of mass accretion, mass loss, and shock chemistry \citep[see][]{2024ApJ...962L..16N,2025A&A...699A.361V,2025ApJ...983..110T} and similarly, extended emissions have also been detected in several (nearby, $<$200~pc) more evolved Class~II disks with \citet{2024ApJ...965L..13A}, \citet{2025ApJ...980..148S}, \citet{2025arXiv250802576K} and \cite{2026arXiv260507016N} reporting \ce{H2} winds and collimated jets traced by forbidden lines (e.g., [Ne~II], [Ar~II], [Fe~II]). While \citet{2024ApJ...965L..13A} attribute the \ce{H_2} mass loss in Tau~042021 to an MHD wind, \citet{2025ApJ...980..148S} favor an internally driven photoevaporative wind for SY~Cha, with mass-loss rates of $\approx \mathrm{10^{-10}-10^{-9}~M_{\odot}~yr^{-1}}$ inferred for both disks. Additionally, \citet{2024AJ....167..127B} report the first detection of a spatially resolved disk wind traced in [Ne~II] likely tracing the inside-out dispersal of the T Cha disk where with model comparisons, \citet{2024AJ....167..223S} infer a mass loss rate of up to $\approx10^{-8}~M_{\odot}~yr^{-1}$. While significant, these current mass loss rates of these Class II systems are likely not sufficient to dominate the dispersal of these disks on $<$~1 Myr time scales indicating there is still ample time for the formation of planetary systems.

\begin{figure*}[t!]
    \centering
    \includegraphics[width=0.975\hsize]{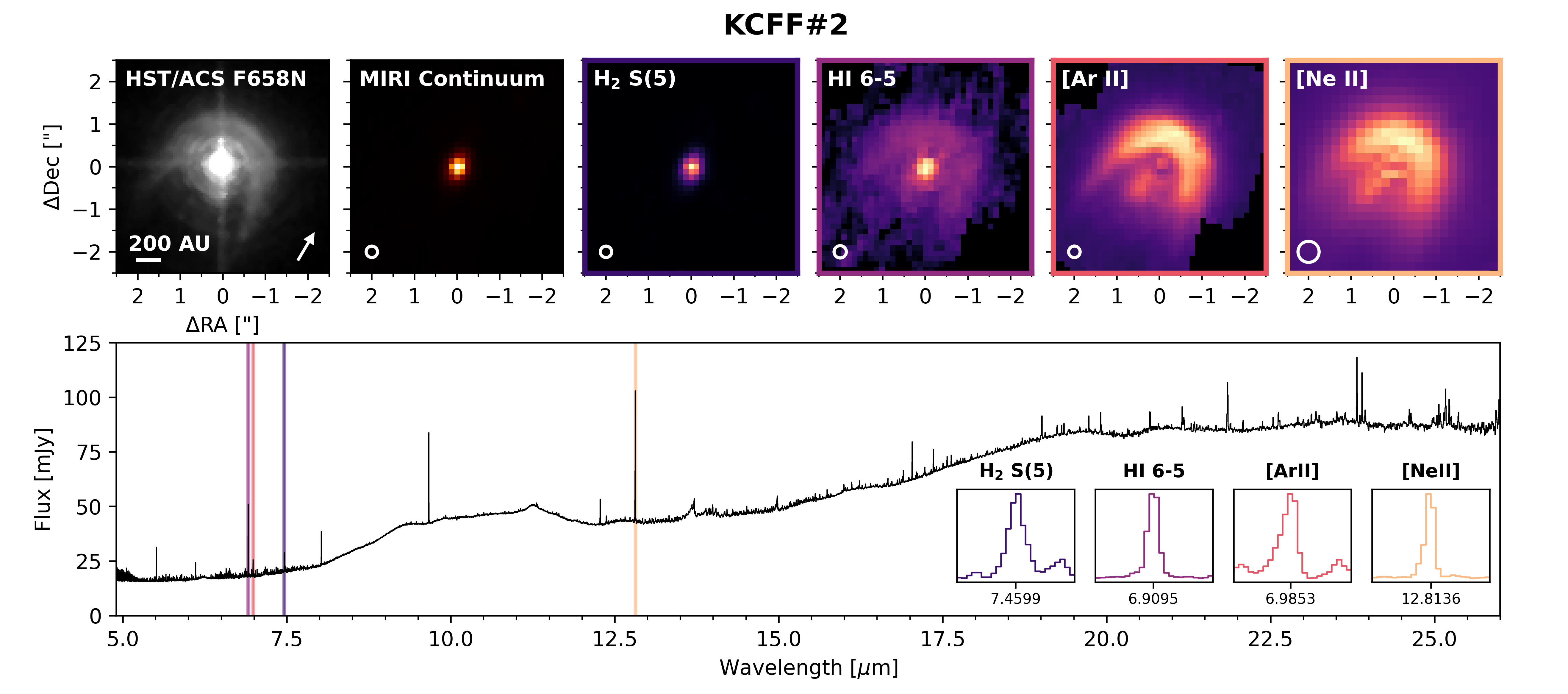}
    \caption{Overview of data of KCFF\#2 where the top panels show the HST data from \citet{2016ApJ...826L..15K} alongside the JWST/MIRI continuum and the integrated intensity maps of a selection of lines after continuum subtraction. The arrow in the bottom right of the HST map shows the direction towards the ionising source 42~Ori. The circle in the bottom-left corner of each JWST/MIRI panel has a diameter of $1.22 \times \lambda/D$. The bottom panel shows the full fringe corrected spectrum extracted from a central disk aperture with inset pannels highlighting the main atomic and molecular lines analysed in this work where the other molecular features are studied in \citet{Zhou2026}.}
    \label{fig:disk2_summary}
\end{figure*}

Looking to disks further afield, recent results from the PDRs4All Early Release Science program targeting the Orion Bar have enabled the characterization of two disks, d203–506 and d203–504, irradiated by the massive stars in the Trapezium Cluster (\citealt{2023Natur.621...56B} and \citealt{2025NatAs...9.1326S}; also see \citealt{2023MNRAS.525.4129H}). These studies demonstrate that FUV fields on the order of $10^{4} G_{0}$ indeed impact disks in ways now observable with JWST. In particular, \citet{2024Sci...383..988B} and \citet{2025NatAs...9.1326S} report mass-loss rates due to external photoevaporation on the order of $10^{-7}$–$10^{-6}~M_{\odot}~\mathrm{yr^{-1}}$ that are derived from \ce{H_2} emission lines. Given that these disks likely had initial masses up to a few tens of Jupiter masses, such mass-loss rates, if sustained, imply disk lifetimes $\ll 1$~Myr, indicating that they are either very young or have only recently been exposed to this level of external irradiation \citep[e.g.; as explored by][]{2025MNRAS.tmp.1907C}, possibly due to prior shielding while cocooned within the Orion Bar \citep{2022MNRAS.512.3788Q,2023MNRAS.520.5331W}.

NGC~1977, a nebula at the northern tip of Orion A molecular cloud complex, provides an excellent testbed for quantifying the impact of external irradiation on planet formation.
\citet{2016ApJ...826L..15K} reported the discovery of seven proplyds using \textit{Spitzer}/IRAC and \textit{HST}/ACS data where disks lie at projected distances of 0.04–0.27~pc from the B1V star 42~Orionis (42~Ori) and exhibit characteristic photoevaporative wind tails 
making this the lowest FUV  radiation environment where proplyds have been discovered. The presence of a single dominant FUV source, 42~Ori, together with a lower, yet more typical, background UV field of $\approx~10^{3}-10^{5}~G_{0}$ \citep{2008ApJ...675.1361F,2025A&A...695A..74A}
makes NGC~1977 a more controlled environment for studying externally irradiated disks under moderate radiation conditions. In this paper, we present JWST–MIRI observations of seven proplyds in NGC~1977, focusing on the atomic emission lines and the molecular \ce{H_2} emission, while the continuum and molecular emissions from the central disks is characterized in \citet{Zhou2026}. Our goals are to understand how these emission tracers relate to the proplyd structures observed at other wavelengths and to constrain the influence of the nearby star 42~Ori on the evolution and planet-forming potential of these systems.

\section{Observations and Analysis} 

The JWST-MIRI data presented here are reported in full detail in \citet{Zhou2026}. In summary, these observations were obtained during JWST Cycle 3 under program ID GO 5269 (P.I. C. Muñoz Romero; co-P.I. A. S. Booth) in February 2025. We targeted the seven known proplyds associated with the B type star 42 Ori, first discovered by \citet{2016ApJ...826L..15K}, using single pointings that cover the full wavelength range of MIRI/MRS including simultaneous imaging with the F500W, F700W and F1000W filters which covered some of the surrounding environment and serendipitously some of our target disks. 

The general properties of the sample are listed in Table~\ref{tab:table1} which include the estimated host star masses \citep{2022MNRAS.512.2594H}, projected separations from 42~Ori \citep{2016ApJ...826L..15K} and an estimate of the FUV they are exposed to. 
We estimated each disk's FUV background from 42 Ori's FUV luminosity and projected distance, following \cite{2025A&A...695A..74A}. Stellar parameters were derived from $T_{eff} = 25400$ \citep{Hohle_42Ori_2010} using MIST isochrones at 1~Myr \citep{Dotter_MIST_2016, Choi_MIST_2016}, and FUV luminosity was calculated by integrating a \cite{Castelli_ATLAS9_2004} atmosphere model over 912-2400~\AA. These fluxes are upper limits, since projected distances underestimate true separations; \cite{2025A&A...695A..74A} provide more precise estimates using the cluster's 3D density distribution, but we use our upper limits here since not all sources are in their catalog.

The data were calibrated using the JWST Disk Infrared Spectral Chemistry Survey (JDISCS) MIRI-MRS pipeline version 9.0 which generally employs the standard JWST Science Calibration Pipeline (version 1.17.1, using CRDS \url{jwst_1322.pmap)} for Stages 0-2b \citep{2022zndo...7041998B, 2024ApJ...963..158P}. We use the images cubes generated here any extract spectra from custom sized areas due to the presence of extended emission where these are noted within the text and captions on a case by case basis. \citet{Zhou2026} 
focus on the molecular emission from the central disks and therefore use the more optimal high-contrast spectra and in particular that recovered from the asteroid fringe-correction as detailed in \citet{2024ApJ...963..158P} but, this only is possible for the center of the field. An example of these spectra for KCFF\#2 is shown in Figure~\ref{fig:disk2_summary} with the full sample shown in \citet{Zhou2026}. 

\begin{figure*}[t!]
    \centering
    \includegraphics[width=0.92\hsize]{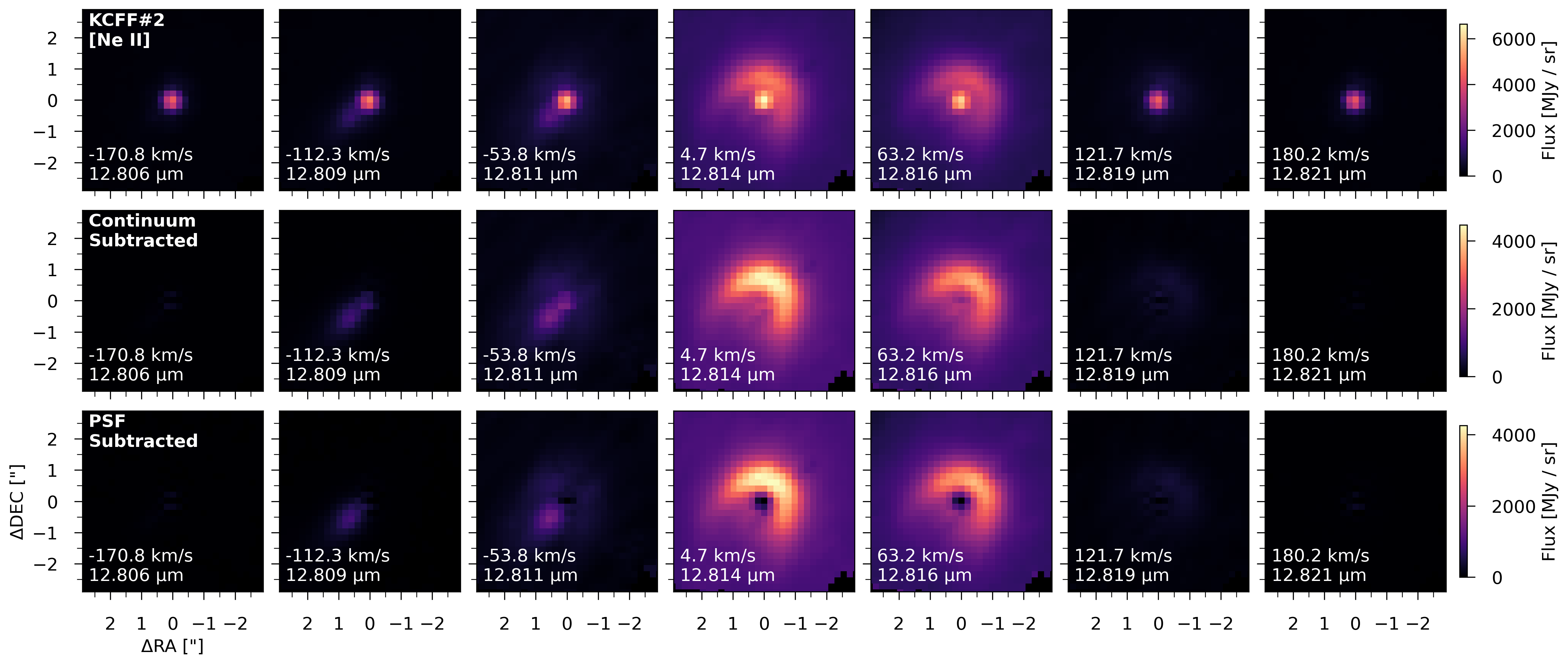}
    \caption{The detection of [Ne II] in KCFF\#2, where the top row shows the native channel maps, the middle row shows the channel maps after a local continuum subtraction, and the bottom row shows the results after a pseudo-PSF subtraction using an average continuum slice scaled to the peak of the line at the image center. Text in the bottom-left corner of each panel indicates the wavelength of the image slice and the corresponding velocity relative to the rest wavelength of the line where, the heliocentric velocity of the source is $\approx$31~km/s.}
    \label{fig:fig2}
\end{figure*}

To visualize and analyze these data, we manipulate the cubes using a range of techniques following the methods outlined in \citet{2008ApJ...676..472B, 2024ApJ...965L..13A, 2025arXiv250802576K}. For each line, we perform a local two-dimensional continuum subtraction in the image plane by averaging the emission in line-free channels on either side of the spectral feature of interest. In addition, we apply a pseudo-point spread function subtraction using the continuum emission. Here, the local average continuum image, which is an unresolved point source for KCFF\#2-7, is scaled to match the maximum intensity of the line peak. Then, this scaled continuum image is subtracted from the data to search for evidence of spatially extended line emission beyond that of the unresolved continuum. Figure~\ref{fig:fig2} illustrates the channel maps and both subtraction methods for the detection of [Ne II] in KCFF\#2 in the sample (see Table~\ref{tab:table1}). We also generate moment maps, including integrated intensity maps, by collapsing the channels over those containing line emission. 

We extract spectra from the calibrated image cubes using circular apertures centered on the continuum peak (or the line peak in the case of weak continuum detections). We adopt varying aperture sizes to account for the extended emissions seen towards several targets, noting that the minimum aperture diameter is wavelength dependent and is set by the broad MRS point-spread function (PSF), taken to be \(2.8 \times 1.22\,\lambda/D\) and is referred to here as the disk aperture \citep{2024ApJ...963..158P}.
As seen in the image cubes, there is noticeable background emission toward some disks, particularly for forbidden lines such as [Ne~II] and [Ar~II]. To try and best correct for this, we calculate the average background spectrum per pixel in an annulus beyond a \(2\farcs\) radius of the central source. This per-pixel background spectrum is then scaled by the number of pixels in the source aperture to produce the background spectrum expected within the source aperture. This scaled background spectrum is subtracted from the raw aperture spectrum to obtain the background-subtracted spectrum. This background subtraction may include weak extended emissions from the disks themselves but 2\farcs0 is beyond any clear extended line structures we see in the MIRI-MRS image slices. The resulting spectra are then locally continuum-subtracted, following the same procedure applied to the image cubes, assuming a linear baseline. To retrieve fluxes these local spectra are modeled with Gaussian profiles where the integrated line flux is taken as the area of the best-fit model and the uncertainties propagated from the fit covariance matrix. The minimum value for the full-width-half-maximum (FWHM) of these lines is set to be the native channel width where most lines have a FWHM of at least 2.0$\times$ the channel width. 
For non-detections, we report \(3\sigma\) upper limits assuming the line spans three channels, computed as: $F_{\rm ul} = 3\,\sigma_{\rm rms}\,\Delta v\,\sqrt{3}$, where \(\sigma_{\rm rms}\) is measured from line-free channels and \(\Delta v\) is the channel width.

\section{Results} 

\subsection{Overview of detected species}

\begin{figure*}
    \centering
    \includegraphics[width=0.9\hsize]{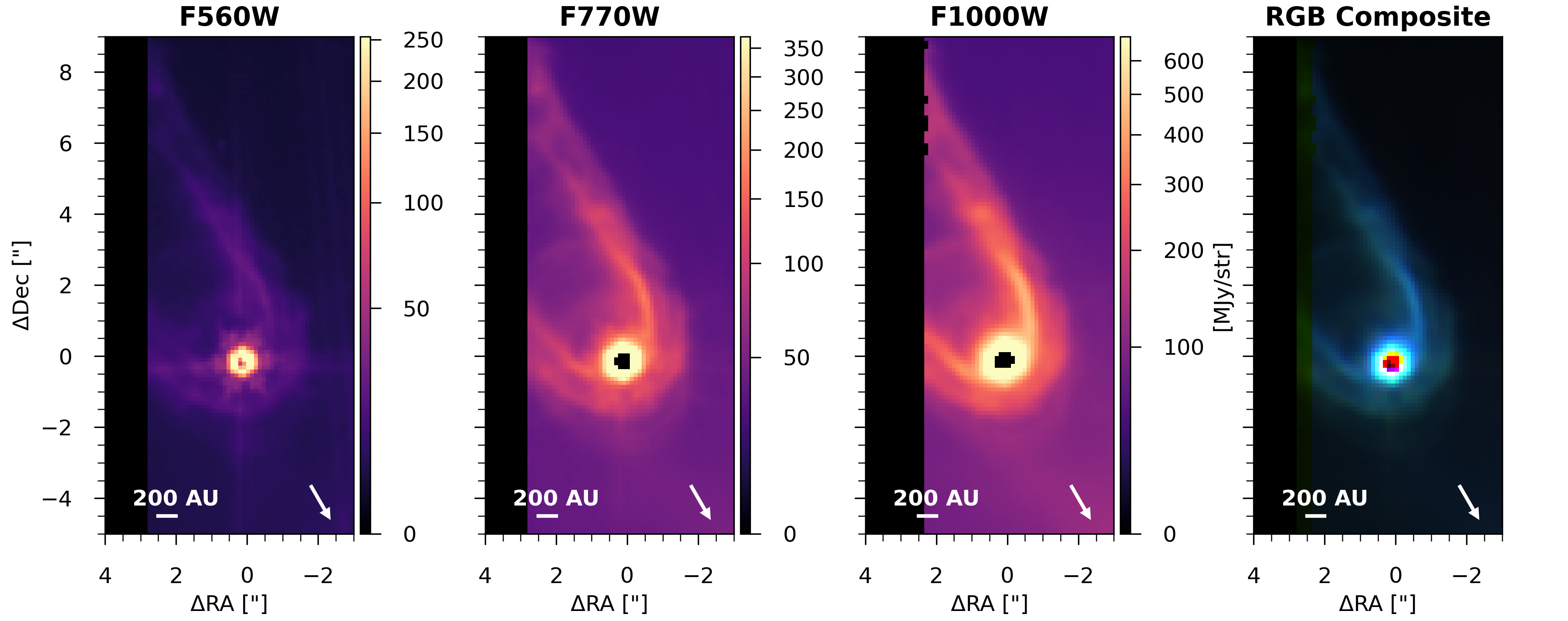}
    \caption{JWST/MIRI images of KCFF\#1 with three different filters alongside a RGB composite image. The arrow in the bottom right shows the direction towards 42~Ori.}
    \label{fig:disk1_miri_images}
\end{figure*}

\begin{figure*}[ht!]
    \centering
    \includegraphics[width=\hsize]{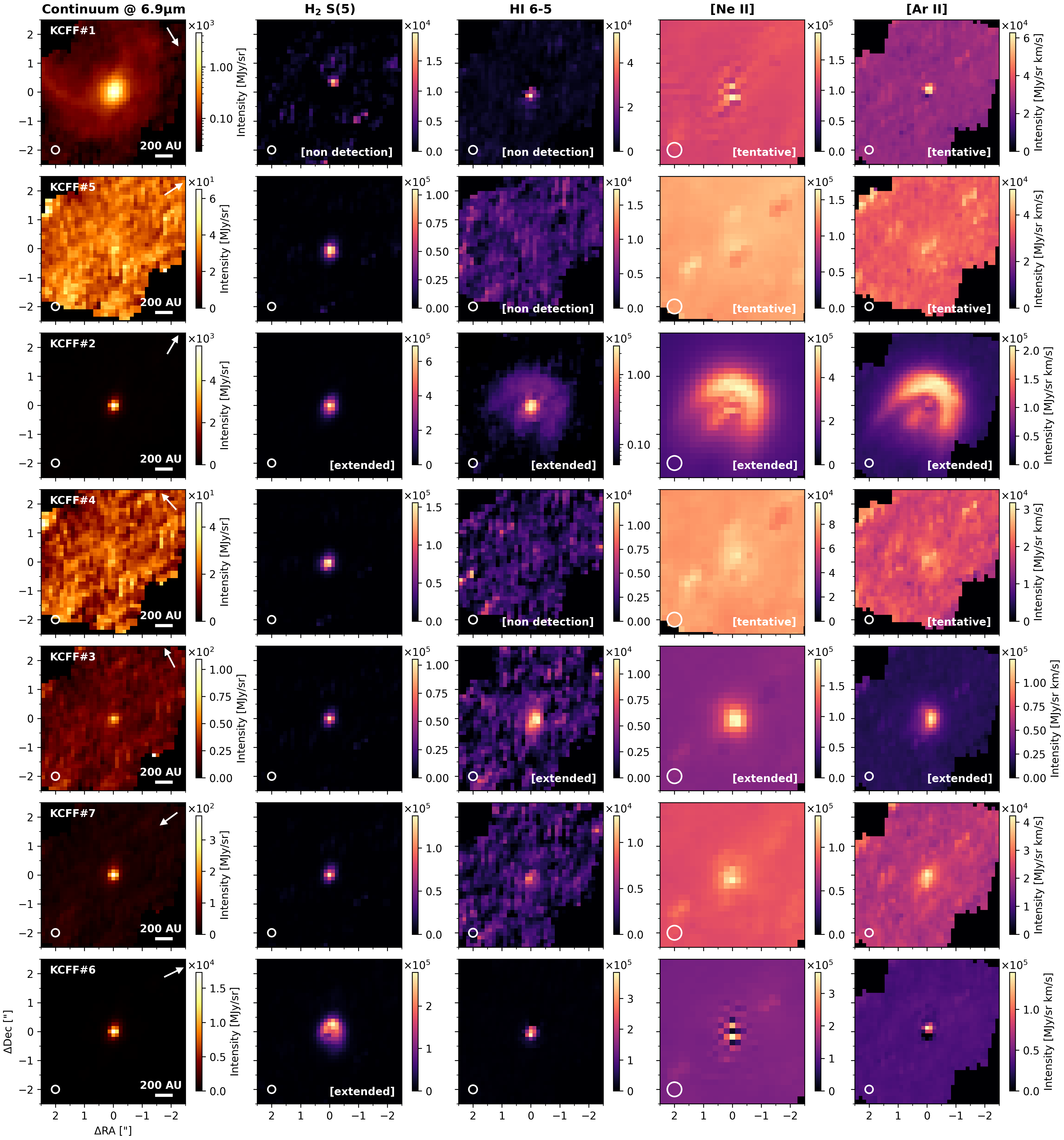}
    \caption{Integrated intensity maps of the \ce{H_2} S(5), HI (6-5), [Ne II], and [Ar II] emission lines, as well as the continuum at 6.9~$\mu$m, are shown for the full sample in order of increasing projected distance from 42~Ori. The circle in the bottom-left corner of each panel has a diameter of $1.22,\lambda/D$. All images are presented in relative RA and DEC coordinates, centered on the continuum peak for all disks except Disks 4 and 5, which are centered on the \ce{H2} line peak. The arrow in the top right shows the direction towards 42~Ori and sources are listed in order of increasing distance from 42~Ori.}
    \label{fig:moments}
\end{figure*}

In Table~\ref{tab:table1}, we summarize the detected emission lines considered in this study for each of the seven disks in the sample, noting whether the emission is spatially extended relative to the continuum. Some detections are tentative due to low signal-to-noise spectra, significant fringing, and/or significant background emission. The spectra for KCFF\#1 are generally line poor and  particularly difficult to interpret because of severe instrumental fringes in addition to the spatially extended continuum emission. This disk was also covered in the simultaneous imaging as shown in Figure~\ref{fig:disk1_miri_images} where a clear cometary cusp and extended tail is present in the continuum at 5.6, 7.7 and 10~$\mu$m. 
The rest of the sample all show compact and unresolved continuum emissions, shown in Figure~\ref{fig:moments}, as expected for Class II disks at these wavelengths and distances.The host star masses of KCFF\#4 and KCFF\#5 derived by \citet{2022MNRAS.512.2594H} are in the brown-dwarf regime, making weak continuum emission towards these sources not unsurprising due to their distance of 400~pc. 

We detect pure rotational \ce{H2} lines toward all sources except KCFF\#1. Additionally, in KCFF\#2 and KCFF\#6, this \ce{H2} emission is spatially extended beyond the continuum emission, suggestive of an externally driven photo-evaporative wind. Although the line-to-continuum ratios vary significantly across the sample, in the disks with notably weak continuum emission the lines are relatively brighter. For example, for the \ce{H_2} S(5) line the ratios of line-to-continuum are as follows: KCFF\#2$\approx$2, KCFF\#3$\approx$14, KCFF\#4$\approx$32, KCFF\#5$\approx$21, KCFF\#6$\approx$0.6, and KCFF\#7$\approx$4 with KCFF\#3,4 and 5 all thought to be $<$0.02$\mathrm{M_{\odot}}$ \citep{2022MNRAS.512.2594H}. 
These higher line fluxes from \ce{H_2} towards the lower mass stars may be expected due to the more significant impact of the external UV field has on heating the disk when the mass is lower \citep{2025ApJ...991...94C}. As explored in \citet{Zhou2026}, at first look when comparing to more nearby sources, e.g., \citet[][]{2025A&A...699A.194A}, the brightness of the \ce{H_2} lines relative to the continuum and other molecular features towards these very low mass stars appears to be distinct.

Atomic hydrogen recombination lines are present in four systems, with clear extended emission in KCFF\#2 and KCFF\#3. No significant compact detections of HI are found in KCFF\#1, 4 or 5 although significant background emission is present at these wavelengths. We also looked for forbidden line emission and see the [Ne II] and [Ar II] lines are detected (or tentatively detected) towards all disks. 
For KCFF\#2 in particular, the [HI], [Ne II] and [Ar II] appears to also have a blue-shifted velocity component possibly tracing a jet (as shown in Figure~\ref{fig:fig2} for [Ne II]]).  In addition, we detect [Ne III] in 4 sources and [Ar III] is not detected across the sample. We also checked for the presence of [S I], [S III], [Fe II], [Ni II] and [Cl II] lines and find that that some can be present in the background field but, compact detections associated with the disks are only found for [S III] in 3 (tentatively 1 more) sources. To synthesize these results we show the integrated intensity maps of the \ce{H_2} S(5), HI (6-5), [Ne II] and [Ar II] emission lines in Figure~\ref{fig:moments} alongside the average continuum around the \ce{H2} S(5) line at 6.9~$\mu$m for the full sample. 

\begin{table*}
\centering
\caption{Summary of targets, distance to 42~Ori and the detected emission lines and morphologies.}
\label{tab:table1}
\begin{tabular}{ccccclllllll} \hline \hline
KCFF ID & Name & $M_\star$ [$M_\odot$] & $\mathrm{d_{proj}}$ [pc] & $\mathrm{log_{10}(G_{FUV})}$ [$G_0$] & \ce{H2} & [\ce{H I}] & [Ne II] & [Ar II] & [Ne III] & [Ar III] & [S III] \\ 
\hline 

1 & 414-50092 & 0.20 & 0.036 & 4.8
& -- & -- & ? & ? & -- & -- & \checkmark{} \\ 
5 & 338-51180 & 0.015 & 0.116 & 3.8
& \checkmark{} & -- & ? & ? & -- & -- & --\\ 
2 & 551-51201 & 0.40 & 0.138 & 3.7 
& \checkmark{}E & \checkmark{}E & \checkmark{}E & \checkmark{}E & \checkmark{} & -- &  \checkmark{}\\ 
4 & 808-50020 & 0.015 & 0.148 & 3.6 
& \checkmark{} & -- & ? & ? & -- & --& --\\ 
3 & 881-50220 & 0.02 & 0.166 & 3.5
& \checkmark{} & \checkmark{}E & \checkmark{}E & \checkmark{}E & \checkmark{} & --& ?\\ 
7 & 313-48277 & 0.15 & 0.215 & 3.3 
& \checkmark{} & \checkmark{} & \checkmark{} & \checkmark{} & \checkmark{} & --& -- \\ 
6 & 252-52365 & 0.50 & 0.268 & 3.1 
& \checkmark{}E & \checkmark{} & \checkmark{} & \checkmark{} & \checkmark{} & -- & \checkmark{} \\

\hline
\end{tabular}
\tablecomments{The KCFF ID refers to the numerical identifier given in \cite{2016ApJ...826L..15K}. $M_\star$ is the stellar mass (taken from \citealt{2022MNRAS.512.2594H}), and $\mathrm{d_{proj}}$ is the projected distance to 42~Ori and the $\mathrm{G_{FUV}}$ values estimated as described in Section 2.
A tick denotes a detection, a question mark a tentative detection, and a dash a non-detection. ``E" refers to the presence of spatially extended emission.}
\end{table*}

\subsection{Extended continuum emission from KCFF\#1}

At a projected distance of $0.036~\mathrm{pc}$, KCFF\#~1 is the closest proplyd to 42~Ori and is experiencing an incident FUV field of $\approx 10^{5}\,G_0$ (see Table~1). We can see the impact of this irradiation at high angular resolution ($\approx0\farcs11$ or 44~au) and over a larger field of view (FOV), as this source was also covered in the MIRI simultaneous imaging, as shown in Figure~\ref{fig:disk1_miri_images}. The disk and proplyd tail are detected in all three filters, with the tail extending to over $3000~\mathrm{au}$ from the central star and beyond the MIRI FOV. Whether or not these MIRI filters are primarily tracing thermal continuum and/or scattered light and, if there is significant contributions from PAH features is unclear but, the F1000W filter is the brightest of the three which, covers the silicate feature, and when comparing MRS spectra extracted from the disk and the tail there is no clear enhancement in the PAH features along these streamlines leading us to believe that dust is responsible for this extended emission. 

\begin{figure*}
    \centering
    \includegraphics[width=0.9\hsize]{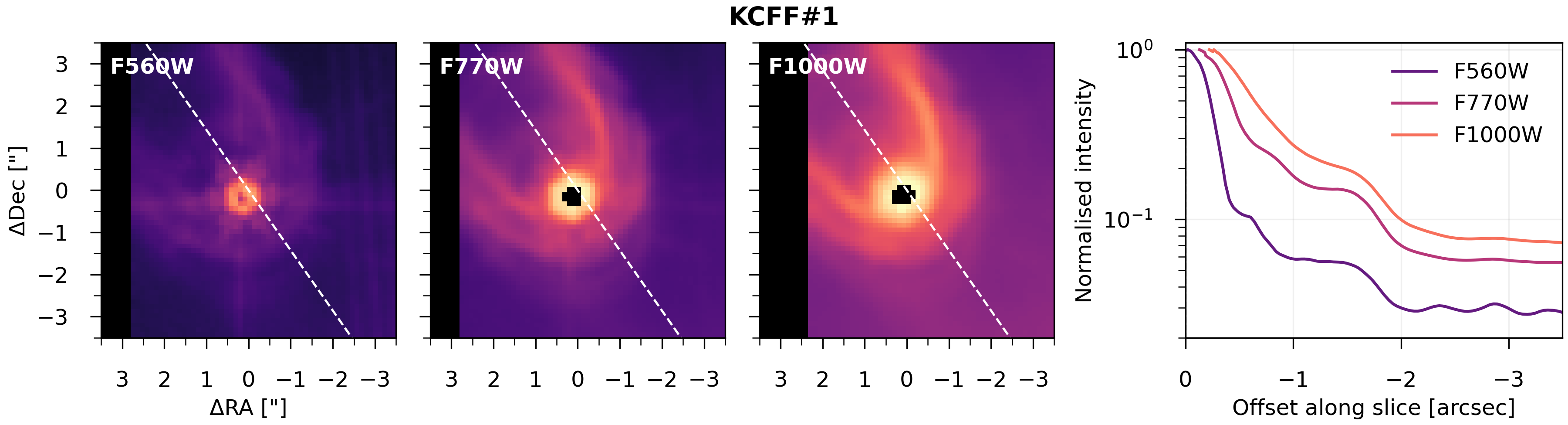}
    \caption{Broadband MIRI continuum images of KCFF\#1 with three different filters alongside intensity slices which show the average flux over a width of 3 pixels along the dashed line along the direction towards 42~Ori.}
    \label{fig:disk1_slices}
\end{figure*}

This proposed dusty tail was previously shown by \citet{2016ApJ...826L..15K} with \textit{Spitzer} data and extends to $\approx$8000~au, but with these MIRI images we can also assess the distance from the star to the proplyd cusp or ionization front (IF). Mass-loss rates for proplyds can be estimated from ionization balance by comparing the radius of the IF and the incident ionizing flux \citep{1998ApJ...499..758J}. If we take the IF to be at $1.5-2.0\arcsec$, as shown in Figure~\ref{fig:disk1_slices} where we take slices of the intensity maps, and adopt the same assumptions as \citet{2022MNRAS.512.2594H} did for KCFF\#2-7, we estimate a mass-loss via 

{\footnotesize
\begin{equation}
\frac{\dot{M}_{w}}{10^{-8}\ M_\odot\ \mathrm{yr}^{-1}} =
\left(\frac{1}{1200}\right)^{3/2}
\left(\frac{R_{\mathrm{IF}}}{\mathrm{au}}\right)^{3/2}
\left(\frac{d_{\mathrm{sep}}}{\mathrm{pc}}\right)^{-1}
\left(\frac{\dot{N}_{\mathrm{Ly}}}{10^{45}\ \mathrm{s}^{-1}}\right)^{1/2}.
\end{equation}
}

where the ionizing photon output from 42
Ori $\dot{N}_{\mathrm{Ly}}$ is taken to be $2\times10^{45}~s^{-1}$ \citep{2012ApJ...756..137B} and the distance 0.036~pc \citep{2016ApJ...826L..15K}. 
This results in a rate of $\approx 1-2 \times 10^{-7}~M_\odot~\mathrm{yr^{-1}}$, which is similar to that as derived by \citet{2022MNRAS.512.2594H} for KCFF\#2. 

Although there is a clear IF in the MIRI images of KCFF\#1, the brighter streams (most prominent in F770W and F1000W) appear to originate directly from the central source, indicating that dust is being stripped from the disk itself and the external UV is penetrating deep into the system. These streams and the lack of any significant ($\times$10 less than in KCFF\#2 and 6) molecular (\ce{H_2}) or atomic (HI) gas associated with the central source or ionization front indicating that this disk it likely at a very evolved state of dispersal. The compact emission seen in [S III], [Ne II] and [Ar II] would be inline with this scenario. That being said, \citet{Zhou2026} do detect \ce{CO_2} and \ce{C_2H_2} emission from the central disk indicating that there is some gas (and likely small dust) reservoir on the scales of a few au. 

\subsection{Extended atomic and molecular emissions}

Both extended atomic and molecular gas emissions are detected across a sub-set of the disks, in particular KCFF\#2, \#3 and \#6. Here we characterize these structures and compare them to the radii of the known proplyd ionization fronts and direction towards the ionizing source 42~Ori. 

\begin{figure*}
\centering
\includegraphics[width=0.9\hsize]{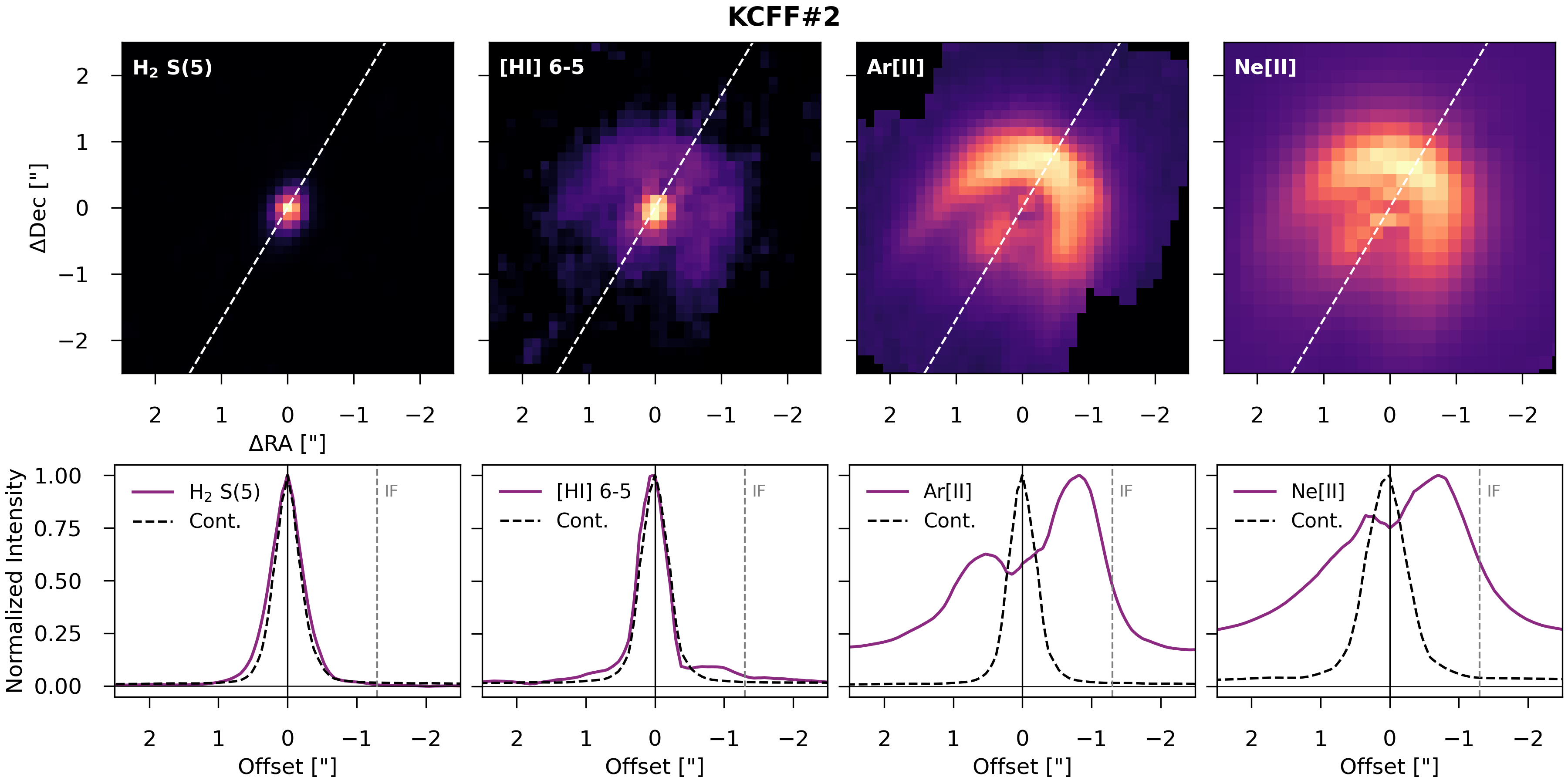}
\caption{Top: Integrated intensity maps of selected emission lines toward KCFF\#2, where the dashed line indicates the direction to 42~Ori.
Bottom: Intensity profiles extracted along the disk axis in the direction toward 42~Ori, where each profile represents the average flux over a width of 3 pixels perpendicular to the dashed line. The colored curve shows the line emission corresponding to the map above, the dashed black curve is the local continuum around the line, and the vertical dashed line marks the location of the ionization front (IF) as derived by \citet{2016ApJ...826L..15K}.}
\label{fig:disk2_slices}
\end{figure*}

The IF of KCFF\#2 is located at 1\farcs35 and was the largest measured for this sample in \citet{2016ApJ...826L..15K}. We detect clear extended emission in [Ne II], [Ar II], and HI, which appear to trace the PDR between the ionization front and the disk as shown in Figure~\ref{fig:disk2_summary}, whereas the \ce{H_2} emission, although extended relative to the continuum (shown more clearly in Figure~\ref{fig:channels_h2_disk2}), lies interior to these atomic lines. In Figure~\ref{fig:disk2_slices}, we show integrated intensity maps of the \ce{H_2} S(5), HI (6–5), [Ar II] and [Ne II] lines toward KCFF\#2, along with emission profiles extracted perpendicular to the IF along the direction to the ionizing source 42 Ori. We select the \ce{H_2} S(5) and H I (6–5) lines for comparison because they are both bright and close in frequency space, so the data slices have approximately the same pixel size (0\farcs13). In contrast, the [Ne II] and [Ar II] data have pixel sizes of 0\farcs2 and 0\farcs13, respectively. As a check, we re-gridded the [Ar II] data to 0\farcs2 and found that the emission peaks at the same radius as the [Ne II] emission. For this source in particular, we also see evidence for a jet traced in [HI], [Ne II] and [Ar II] that is somewhat aligned along the direction to 42~Ori indicating, contributions to these emission lines from both internal and external mass loss processes. 

From the values presented in \citet{2016ApJ...826L..15K}, the rest of the sample have IFs on $<$1\farcs0 scales. That being said, we detect clear extended gas emission in KCFF\#3 and KCFF\#6. KCFF\#3 is similar to KCFF\#2, with extended HI, [Ne II], and [Ar II], and these emissions peak within the IF of 0\farcs4. Interestingly though, the extended emission in KCFF\#3, in particular the [Ar II] and HI, do not peak in full accordance with the projected location of 42~Ori as shown in Figure~\ref{fig:disk3_slices}. KCFF\#6 is distinct to the other two systems, with extended emission seen only in \ce{H_2} as shown in Figure~\ref{fig:disk6_slices} and Figure~\ref{fig:channels_h2_disk6}) where, the extension in \ce{H_2} is directed towards 42~Ori but within the IF radius as measured by \citet{2016ApJ...826L..15K}. 
KCFF\#6 is in an $\approx4\times$ lower irradiation regime than KCFF\#2 and KCFF\#3 with a similar host star mass to KCFF\#2 therefore, the lack of extended HI (or [Ne II] and [Ar II]) may be reflective of these environmental differences where, the gas is heated by the FUV (and EUV) but the impact is not as destructive in terms of photo-ionization and photo-dissociation. Additionally, extended \ce{H_2} emission from  KCFF\#6  may also be linked to the mass/size of the disk \citep[see][]{2019MNRAS.485.3895H} given KCFF\#6 has the highest host star mass across the sample. 

\subsection{Molecular hydrogen as a tracer of the disk surface and/or mass loss}

\begin{figure*}
\centering
    \includegraphics[width=0.95\hsize]{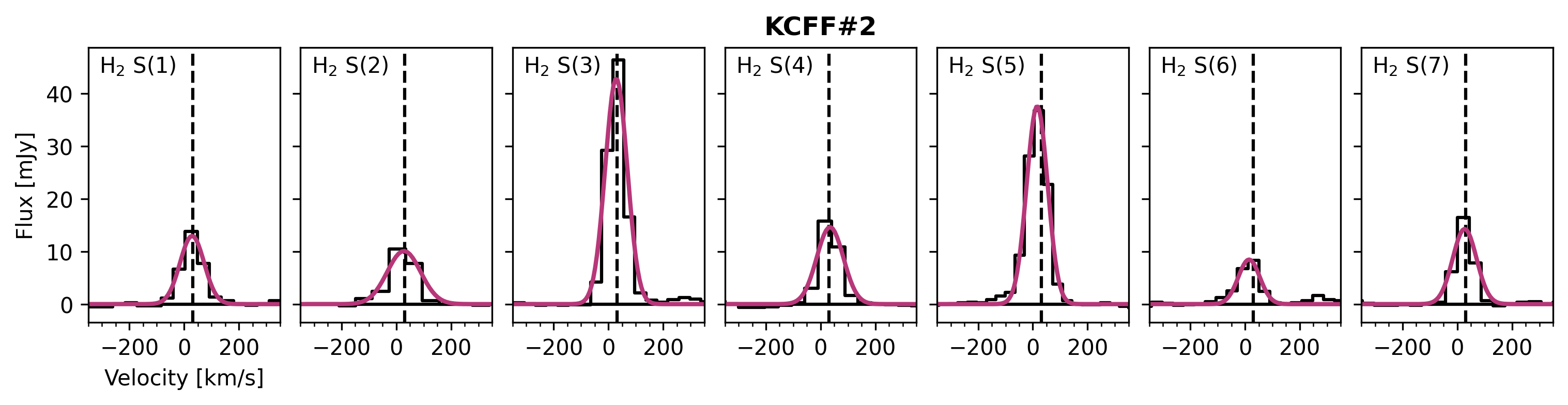}
    \caption{Spectra (black) extracted from a 1\farcs0 aperture towards Disk 2  of the \ce{H_2} lines and corresponding gaussian fits (purple). The x-axis is velocity relative to the rest wavelength of the line and the dashed vertical line highlights the approximate heliocentric velocity of 31~km~$\mathrm{s^{-1}}$.}
    \label{fig:disk_2_spectra}
\end{figure*}

The pure rotational lines of \ce{H_2} accessible with MIRI trace hot gas in the upper disk atmosphere and/or disk winds \citep[e.g.;][]{2024ApJ...965L..13A, 2024A&A...687A..96F, 2025ApJ...980..148S}. In Figure~\ref{fig:disk_2_spectra} we show the \ce{H_2} (0,0) spectra for KCFF\#2 extracted uniformly from a 1\farcs0 aperture after a local continuum and background subtraction. Although the S(8) line is present in the brightest disks it is typically blended with CO ro-vibrational emission from the inner disk and therefore excluded from our analysis. The spectra for the rest of the sample are shown in Appendix Figure~\ref{fig:spectra_H2} where in all cases the lines are spectrally unresolved with FWHM $\approx$2.4$\times \Delta$v and there are no indications of any significant blue or red velocity shifts in the lines. 

\begin{figure*}
    \centering
    \includegraphics[width=0.975\hsize]{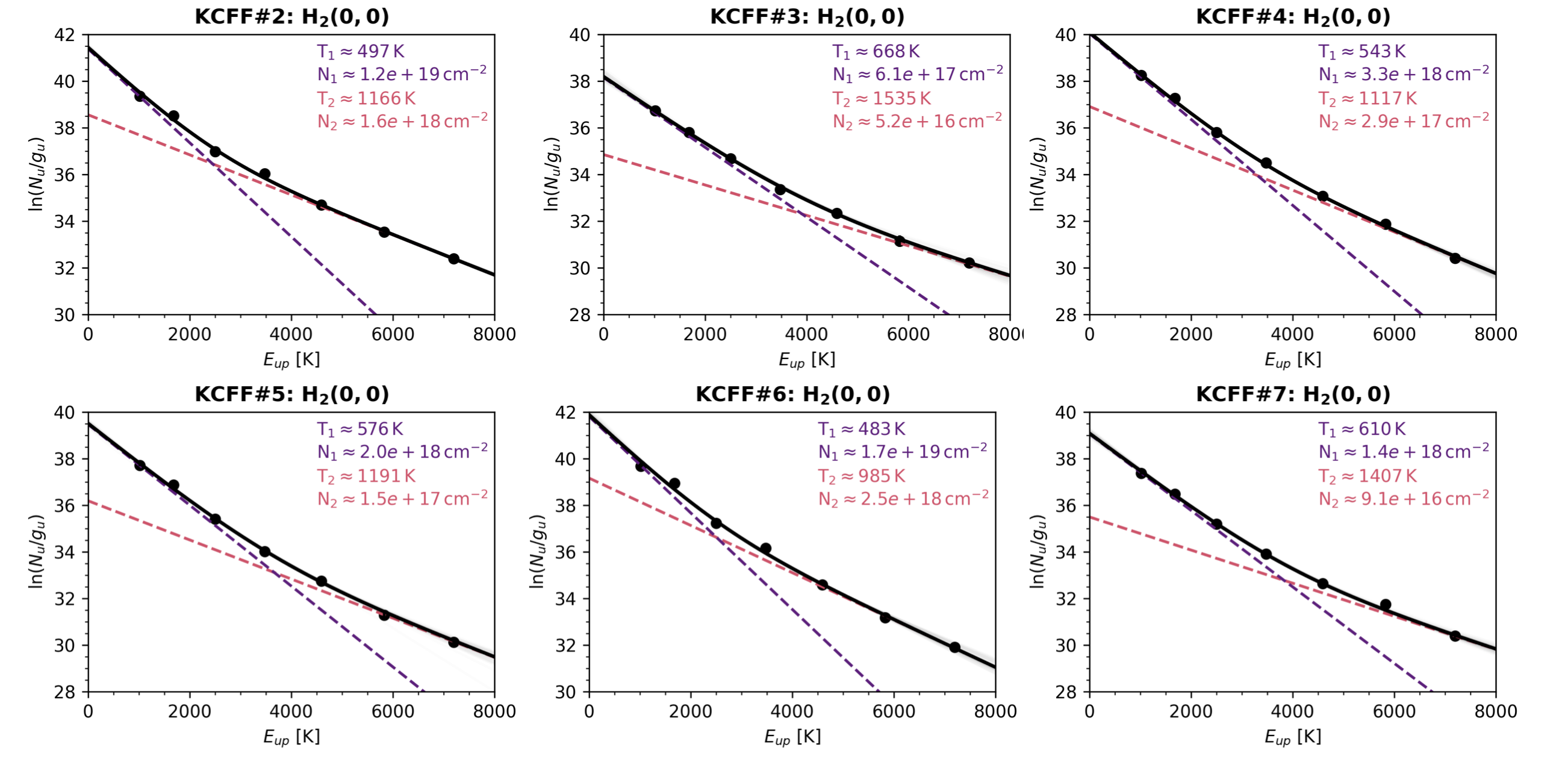}
    \caption{Rotational diagrams for the \ce{H_2} emission detected toward KCFF\#2–7. Dashed colored lines show the best-fit warm and hot components, the solid black line shows the combined model, and gray lines indicate random draws from the posterior distributions.  }
    \label{fig:rotation_diagrams}
\end{figure*}

Using the extracted line fluxes for transitions S(1) through S(7), we calculate the column density and rotational temperature of the \ce{H_2} for KCFF\#2-7 using a rotational diagram \cite[e.g.,][]{1999ApJ...517..209G, 2025ApJ...980..148S}. 
This procedure relied on the assumption that these lines are in LTE, as was seen for MIRI \ce{H_2} lines in the irradiated disk d203-506 \citep{2023Natur.621...56B}, but, we note that the \ce{H_2} excitation may be more complicated when considering the vibrational transitions in the JWST/NIRSpec range. 
We collect the molecular data for these calculations from the high-resolution transmission molecular absorption database \citep[HITRAN;][]{HITRAN2020}, which provides the partition function and transition properties such as the upper-level degeneracies, upper-level energies, and Einstein coefficients. Additionally, because these transitions have extremely low probabilities, the lines can be assumed to be optically thin. Inspection of the data shows that a two-component temperature and density model is required to reproduce the observations, similar to that seen in other disks observed with MIRI \citep[e.g.,][]{2025ApJ...980..148S,2025ApJ...991..128R}. We show the resulting rotational diagrams and best-fit models in Figure~\ref{fig:rotation_diagrams}, and summarize the derived parameters in Table~\ref{table:H2} where these disks all consistently require a warm ($\approx$500-600~K) component of gas as well as a hotter ($\approx$1000-1500~K), more rarefied component. 

\begin{figure*}
    \centering
    \includegraphics[width=0.9\hsize]{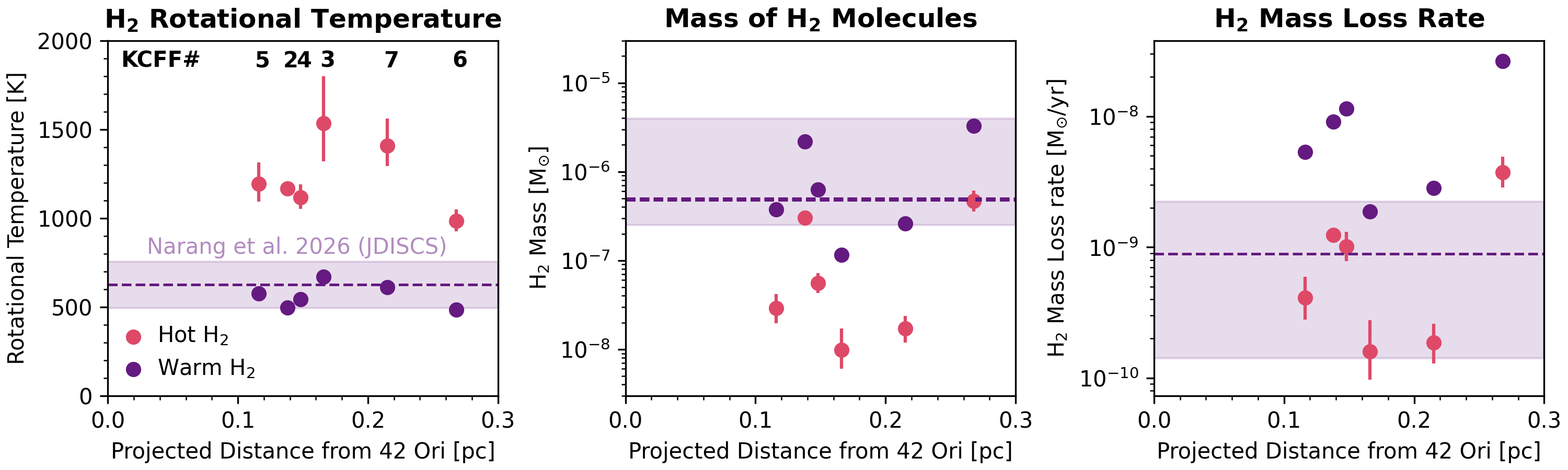}
    \caption{Properties of the warm and hot \ce{H_2} reservoir for KCFF\#2-7 as a function of projected distance to 42~Ori. The dashed purple line and shaded region show the median and range for warm \ce{H_2} as reported by \citet{2026arXiv260507016N} for a sample of isolated/nearby Class II disks.}
    \label{fig:h2_distance}
\end{figure*}

We assume a circular emitting area with a radius of 1\farcs0 for this analysis however, because the \ce{H_2} lines are optically thin, the column density and emitting area are degenerate. We therefore calculate the total number and mass of \ce{H_2} molecules which in this regime is independent of the chosen emitting area. The masses are listed in Table~\ref{table:H2}, assuming an emitting area of 1\farcs0 and a distance of 400~pc. Notably, the disks with the highest \ce{H_2} masses - KCFF\#2 and \#6 also have extended \ce{H_2} emissions. Overall there is no trend in \ce{H_2} mass and temperature in relation to the incident FUV, as highlighted in Figure~\ref{fig:h2_distance}, indicating that the gas properties are likely shaped by the host star properties and the disk mass/size. These two temperature components may represent the gas in the warm molecular layer and a more diffuse and hotter wind and/or may both trace wind components. The potential mass loss rates of such winds will be discussed in Section 4.1.

\subsection{Atomic hydrogen as a tracer of accretion}

Across the MIRI wavelength range, there are a number of hydrogen recombination lines that are commonly used to measure accretion luminosity and mass accretion rates, with a past focus on the HI (7–6) line at 12.37~$\mu$m in particular \citep{2015ApJ...801...31R}. In addition to tracing accretion, these HI lines are also expected to trace the photo-dissociation of \ce{H_2} between the proplyd ionization front and the disk surface \citep[e.g.,][]{1998ApJ...502L..71S}. A further complication is that these lines can be contaminated by molecular emission, particularly \ce{H_2O} from the inner disk \citep[see studies from][]{2015ApJ...801...31R, 2025AJ....169..165B, 2025ApJ...985..224T, 2025arXiv251203456S} but, as shown by \citet{Zhou2026}, only KCFF\#2 and KCFF\#6 have clear detections of \ce{H_2O}. We therefore focus our analysis on a subset of HI lines: both the expected bright transitions (6–5) and (7–6), which may be affected by molecular contamination in some sources, as well as detectable transitions identified as good alternatives by \citet{2025ApJ...985..224T, 2025arXiv251203456S}, namely (8–6), (10–7) and (8-7).

For KCFF\#1, similar to \ce{H_2}, the fringing in the spectra and extended emissions make it difficult to confidently show the presence of compact HI emission in this source. For KCFF\#2, we detect both compact and extended HI emission with the (6-5) transition shown in Figure~\ref{fig:moments} and Figure~\ref{fig:HI_disk2}, where the extended emission appears to trace both the ionization front and, tentatively, a jet (as indicated in this source from free-free emission \citealt{2024ApJ...967..103B}). Similarly, KCFF\#~3 shows extended emission in the HI (6–5) line. There are no clear detections of HI towards KCFF\#4 and KCFF\#5 and compact HI is detected towards KCFF\#6 and KCFF\#7. The channel maps for the HI (6-5) emissions towards KCFF\#1,2,5,6 and 7 are shown in Appendix Figures~\ref{fig:HI_disk1}, \ref{fig:HI_disk2}, \ref{fig:HI_disk3}, \ref{fig:HI_disk6}, and \ref{fig:HI_disk7} with PSF and/or continuum subtraction where applicable. 

Using the refined empirical relations from \citet{2025arXiv251203456S} (and \citet{2025ApJ...985..224T} for the 8-7 transition, note the relations in both studies are slightly different for the same transitions), we calculate the accretion luminosities across the sample. For this analysis we use the line luminosity calculated from the line flux from the disk-sized aperture and assume a source distance of 400~pc.
This small aperture size will limit contamination from the PDR in the case of KCFF\#2 in particular. The spectra for KCFF\#2 are shown in Figure~\ref{fig:disk2_HI_spectra} with the rest of the sample shown in Appendix Figure~\ref{fig:spectra_HI}. 
Similar to the \ce{H_2} there is no strong evidence that these lines are spectrally resolved but for some of the well detected lines the fits on the FWHM can be up to $\approx$4-5$\times$dv.  
We then convert these accretion luminosities to accretion rates using the stellar masses from \citet{2022MNRAS.512.2594H} and the 1~Myr isochrones from \citet{2015A&A...577A..42B}\footnote{Available here: \url{https://perso.ens-lyon.fr/isabelle.baraffe/BHAC15dir/BHAC15_iso.2mass}} to determine the stellar radii. We also assume $R_\mathrm{in}/R_\star = 5$ across the sample \citep[e.g.;][]{2015ApJ...801...31R}. These values in Table~3 should be viewed as upper-limits due to potential unresolved contributions from the proplyd ionization front and possible jets, as seen in KCFF\#2 in particular \citep[see jet towards HH 30][]{2025AJ....169..296B}. 

The inferred accretion rates are listed in Table~\ref{tab:accretion}, along with upper limits where applicable. For KCFF\#2 and KCFF\#6 we recommend the rates from the lines not overlapping with \ce{H_2O}, which are the (8-6) and (10-7). Although not contaminated by \ce{H_2O}, the (8-7) is at much longer wavelengths where, for KCFF\#2, we cannot spatially disentangle the ionization front from the inner disk emission, and this is likely why the derived rate here is notably higher.
Overall, for KCFF\#2 and KCFF\#6 the values for the (8-6) and (10-7) lines, and the upper limits for KCFF\#1, are consistent with expectations for Class II disks albeit on the higher end \citep[e.g.;][]{2023ASPC..534..539M} and the irradiated disk d204-504 \citep{2025NatAs.tmp..147S}. 
Furthermore, when comparing to the sample of externally irradiated disks in NGC 3603 presented by 
\citet{2025A&A...698A.172R} although they report higher mass accretion rates for the irradiated disks in their sample this is only really prevalent for host stars above 1~$\mathrm{M_{\odot}}$ which is beyond the masses present in our sample.
For KCFF\#7 the rates are higher than expected for a 0.15$M_{\odot}$ host star, which may be due to contamination from the ionization front and/or an unresolved jet contribution. For the lower mass stars in the sample, the accretion rate expectations from the literature are less clear and, in the case of KCFF\#3 in particular, the ionization front is likely inflating the estimates.

\begin{table*}
\centering
\caption{Warm and Hot \ce{H_2} properties for each source derived from a rotational diagram analysis.}
\begin{tabular}{c ccc ccc}
\hline\hline
\multirow{2}{*}{Source ID} 
    & \multicolumn{3}{c|}{Warm \ce{H_2}} 
    & \multicolumn{3}{c}{Hot \ce{H_2}} \\
& $N_{\mathrm{col}}$ [$\mathrm{cm^{-2}}$] 
& $T_{\mathrm{ex}}$ [K] 
& Mass [$\mathrm{M_{\odot}}$]
& $N_{\mathrm{col}}$ [$\mathrm{cm^{-2}}$] 
& $T_{\mathrm{ex}}$ [K]
& Mass [$\mathrm{M_{\odot}}$] \\
\hline

5 &
$2.24^{+0.12}_{-0.11}\times10^{18}$ &
$576^{+21}_{-22}$ &
$3.77^{+0.20}_{-0.19}\times10^{-7}$ &
$1.72^{+0.77}_{-0.55}\times10^{17}$ &
$1191^{+123}_{-98}$ &
$2.90^{+0.13}_{-0.09}\times10^{-8}$ \\

2 &
$1.31^{+0.06}_{-0.05}\times10^{19}$ &
$497^{+11}_{-10}$ &
$2.20^{+0.10}_{-0.09}\times10^{-6}$ &
$1.79^{+0.11}_{-0.11}\times10^{18}$ &
$1166^{+20}_{-19}$ &
$3.01^{+0.02}_{-0.02}\times10^{-7}$ \\

4 &
$3.74^{+0.11}_{-0.11}\times10^{18}$ &
$543^{+13}_{-14}$ &
$6.29^{+0.02}_{-0.02}\times10^{-7}$ &
$3.31^{+0.96}_{-0.74}\times10^{17}$ &
$1117^{+73}_{-64}$ &
$5.57^{+0.16}_{-0.13}\times10^{-8}$ \\

3 &
$6.90^{+0.62}_{-0.55}\times10^{17}$ &
$668^{+41}_{-49}$ &
$1.16^{+0.10}_{-0.09}\times10^{-7}$ &
$5.86^{+4.26}_{-2.30}\times10^{16}$ &
$1535^{+262}_{-215}$ &
$9.86^{+7.18}_{-3.86}\times10^{-9}$ \\

7 &
$1.55^{+0.10}_{-0.09}\times10^{18}$ &
$610^{+27}_{-27}$ &
$2.61^{+0.17}_{-0.16}\times10^{-7}$ &
$1.02^{+0.40}_{-0.32}\times10^{17}$ &
$1407^{+152}_{-114}$ &
$1.72^{+0.07}_{-0.05}\times10^{-8}$ \\

6 &
$1.96^{+0.16}_{-0.17}\times10^{19}$ &
$483^{+27}_{-23}$ &
$3.29^{+0.27}_{-0.29}\times10^{-6}$ &
$2.77^{+0.88}_{-0.65}\times10^{18}$ &
$985^{+64}_{-61}$ &
$4.66^{+0.15}_{-0.11}\times10^{-7}$ \\

\hline
\end{tabular}
\label{table:H2}
\end{table*}

\begin{figure*}
\centering
    \includegraphics[width=0.75\hsize,trim={0cm 0 7cm 0},clip]{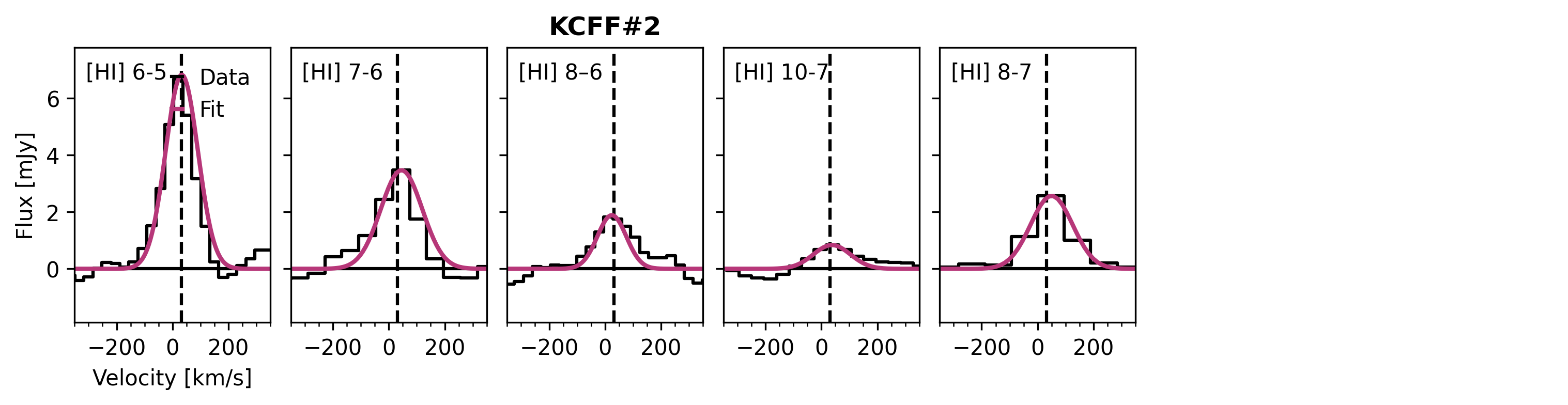}
    \caption{Spectra (black) extracted from a the disk-sized aperture towards Disk 2 of select \ce{HI} lines and corresponding gaussian fits (purple). The x-axis is velocity relative to the rest wavelength of the line and the dashed vertical line highlights the approximate heliocentric velocity of 31~km~$\mathrm{s^{-1}}$.}
    \label{fig:disk2_HI_spectra}
\end{figure*}

\begin{sidewaystable*}
\centering
\caption{Derived mass accretion rates from multiple HI lines, and mass loss rates derived from proplyd ionisation fronts ($\dot{M}_{\mathrm{loss, IF}}$; tracing the total mass loss: \citealt[][]{2022MNRAS.512.2594H} and this work) free–free emission ($\dot{M}_{\mathrm{loss, ff}}$: \citealt{2024ApJ...967..103B}), hot \ce{H2} emission ($\dot{M}_{\mathrm{loss, H_2}}$: this work), and warm \ce{H2} emission ($\dot{M}_{\mathrm{loss, warm\,H_2}}$: this work).}
\begin{tabular}{c c c c c c c c c c}
\hline\hline
Source ID 
& $\dot{M}_\mathrm{acc}$ (6--5) 
& $\dot{M}_\mathrm{acc}$ (7--6)
& $\dot{M}_\mathrm{acc}$ (8--6)
& $\dot{M}_\mathrm{acc}$ (10--7)
& $\dot{M}_\mathrm{acc}$ (8--7)
& $\dot{M}_\mathrm{loss, IF}$
& $\dot{M}_\mathrm{loss, ff}$
& $\dot{M}_\mathrm{loss, hot\,H_2}$
& $\dot{M}_\mathrm{loss, warm\,H_2}$ \\
& [$10^{-8}\,M_\odot\,\mathrm{yr^{-1}}$] 
& [$10^{-8}\,M_\odot\,\mathrm{yr^{-1}}$] 
& [$10^{-8}\,M_\odot\,\mathrm{yr^{-1}}$] 
& [$10^{-8}\,M_\odot\,\mathrm{yr^{-1}}$] 
& [$10^{-8}\,M_\odot\,\mathrm{yr^{-1}}$] 
& [$10^{-8}\,M_\odot\,\mathrm{yr^{-1}}$]
& [$10^{-8}\,M_\odot\,\mathrm{yr^{-1}}$]
& [$10^{-10}\,M_\odot\,\mathrm{yr^{-1}}$]
& [$10^{-9}\,M_\odot\,\mathrm{yr^{-1}}$] \\
\hline

1 
& $< 0.38$ 
& $< 1.49$ 
& $< 0.50$ 
& $< 2.04$ 
& $< 4.28$ 
& 10--20 
& $0.295$
& -- 
& -- \\

5 
& $< 0.20$ 
& $< 0.97$ 
& $< 0.27$ 
& $< 0.96$ 
& $< 0.82$ 
& 4.6 
& $<0.357$
& $4.1^{+1.8}_{-1.3}$ 
& $5.37^{+0.27}_{-0.28}$ \\

2 
& $7.11 \pm 0.30$ 
& $6.74 \pm 0.34$ 
& $1.99 \pm 0.33$ 
& $2.38 \pm 0.53$ 
& $12.06 \pm 4.21$ 
& 17.8 
& $1.77$
& $12.4^{+0.8}_{-0.8}$ 
& $9.08^{+0.38}_{-0.40}$ \\

4 
& $< 0.20$ 
& $< 1.00$ 
& $< 0.28$ 
& $< 1.19$ 
& $< 0.46$ 
& 3.3 
& $<0.337$
& $10.1^{+2.9}_{-2.3}$ 
& $11.42^{+0.35}_{-0.35}$ \\

3 
& $1.70 \pm 0.09$ 
& $2.28 \pm 0.11$ 
& $0.43 \pm 0.13$ 
& $< 0.40$ 
& $0.38 \pm 0.22$ 
& 2.4 
& $0.532$
& $1.6^{+1.2}_{-0.6}$ 
& $1.88^{+0.15}_{-0.17}$ \\

6 
& $7.49 \pm 0.63$ 
& $2.85 \pm 0.83$ 
& $3.17 \pm 0.54$ 
& $5.85 \pm 1.24$ 
& $< 88.59$ 
& 3.6 
& $<0.662$
& $37.5^{+11.9}_{-8.8}$ 
& $26.51^{+2.30}_{-2.13}$ \\

7 
& $14.36 \pm 0.61$ 
& $13.60 \pm 0.68$ 
& $4.02 \pm 0.67$ 
& $4.81 \pm 1.07$ 
& $24.34 \pm 8.49$ 
& 3.1 
& $<0.489$
& $1.9^{+0.7}_{-0.6}$ 
& $2.84^{+0.17}_{-0.19}$ \\

\hline
\end{tabular}
\label{tab:accretion}
\end{sidewaystable*}

\begin{figure}
    \centering
    \includegraphics[width=0.99\linewidth]{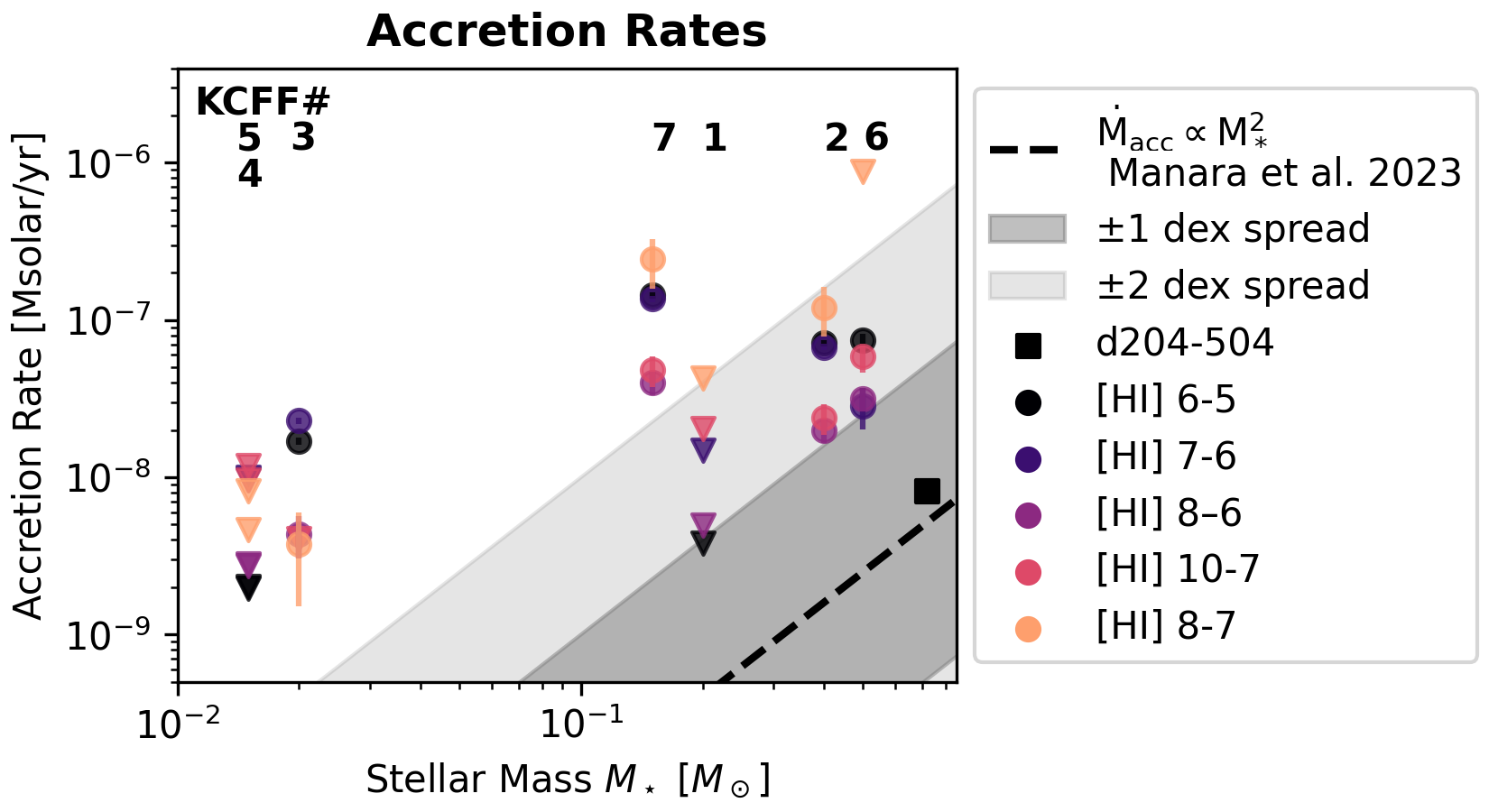}
    \caption{Accretion rate estimates from HI line luminosities across the sample. The dashed line shows the qualitative correlation from \citet{2023ASPC..534..539M}, with the observed 1–2~dex scatter seen in nearby Class II disks. The square point shows the irradiated disk d204-504 for comparison \citep{2025NatAs.tmp..147S}.}
    \label{fig:placeholder}
\end{figure}

\subsection{Forbidden line emissions}

UV radiation can ionize gas in disks and generate forbidden-line emission from various atomic species. In particular, these lines are anticipated to trace disk dispersal due to both internal and external photo-evaporative winds, in addition to MHD-driven winds (for a pre-JWST review see \citealt[][]{2023ASPC..534..567P} and more recently \citealt{2024AJ....167..127B} and \citealt{2025MNRAS.541.2917P}). Across the sample, we (tentatively in the case of KCFF\#1, \#4, and \#5) detect [Ne II] and [Ar II] toward all seven disks (see Figure~\ref{fig:moments}), with the channel maps shown in Figure~2 and in Appendix Section C, where in some disks the background emission is clearly significant. We additionally detect [Ne III] and [S III] in a subset of systems, but do not detect [Ar III]. Interestingly, only [S III] is convincingly detected in KCFF\#1 (see Figure~\ref{fig:disk1_SIII}) where, similar to the [Ne II] and [Ar II], the emission is compact. Note that [S III] has higher ionization energy than both [Ne II] and [Ar II]. 

\begin{figure*}
\centering
    \includegraphics[width=0.7\hsize]{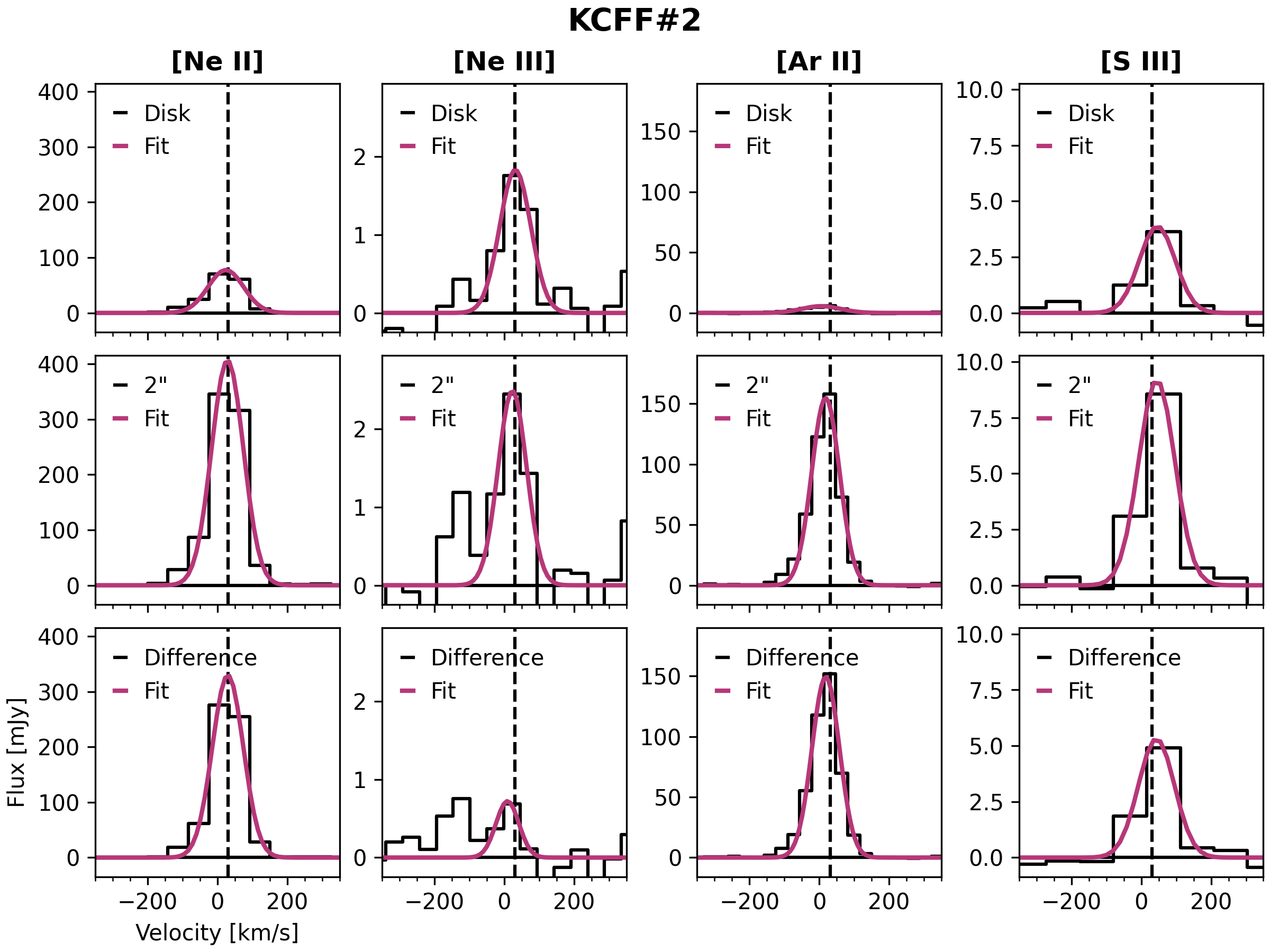}
    \caption{Spectra (black) extracted from a disk and 2\farcs aperture towards Disk 2 of forbidden emission lines and corresponding gaussian fits (purple). The x-axis is velocity relative to the rest wavelength of the line and the dashed vertical line highlights the approximate heliocentric velocity of 31~km~$\mathrm{s^{-1}}$.}
    \label{fig:disk2_forbidden}
\end{figure*}

Due to the presence of extended emission, we extract spectra from both a disk-sized aperture and a larger 2\farcs diameter aperture, with the resulting line fluxes listed in Table~\ref{table:forbidden_fluxes}.  The spectra for KCFF\#2 is shown in Figure~\ref{fig:disk2_forbidden}, where, for example, in [Ar II] the flux is mostly from larger scales than the disk, whereas [Ne III] appears to be emitted from a more compact region than [Ne II]. For KCFF\#2 in particular there may also be contributions to these line emission from a jet. Spectra for KCFF\#1, \#3, \#6, and \#7 are shown in Appendix Figures~\ref{fig:disk1_forbidden}, \ref{fig:disk3_forbidden}, \ref{fig:disk6_forbidden} and \ref{fig:disk7_forbidden}.

The ratios of these lines can inform on the hardness of the radiation field and if X-rays or UV photons dominate, as initially explored by \citet{2012ApJ...759...47S} and more recently by \citet{2024AJ....167..127B} for the wind from the T Cha disk. The relative line ratios [Ne II]/[Ne III], [Ar II]/[Ar III], and [Ne II]/[Ar II] are shown in Figure~\ref{fig:line_ratios} for the various aperture combinations for the four disks with detections (not tentative detections) of [Ne II] and [Ar II].
The difference in the [Ne II] and [Ar II] emission in KCFF\#2 on different scales is highlighted by a [Ne II]/[Ar II] of 13 from the disk but $\approx$3 in the extended PDR; there is a similar trend for the ratio of [Ne II] with [Ne III] signaling different irradiation conditions in the inner disk and the IF. Looking at the lower limits on [Ar II]/[Ar III] line ratios, in almost all cases this is in line with irradiation from EUV radiation rather than hard X-rays \citet{2009ApJ...703.1203H}. 
Unlike for isolated disks where the [Ne II] ionization is from hard X-rays stellar photons  e.g., Fig. 7 in \citet{2014ApJ...795....1P}, here the primary ionization is from external EUV which is shown by the different line ratios. Furthermore, qualitatively, it appears that the disks KCFF\#2 and \#3, which are in an $\approx2-4\times$ higher irradiation regime than KCFF\#6 and \#7, have higher [Ne II]/[Ne III] line ratios. We will discuss comparisons to other Class II disks in more detail in Section 4.2.

\begin{table*}[ht]
\centering
\footnotesize
\caption{Forbidden line fluxes.}
\begin{tabular}{c c c c c c c c c c c}
\hline\hline
Source ID
& \multicolumn{2}{c}{[Ne\,II] (12.81~$\mu$m)}
& \multicolumn{2}{c}{[Ne\,III] (15.55~$\mu$m)}
& \multicolumn{2}{c}{[Ar\,II] (6.99~$\mu$m)}
& \multicolumn{2}{c}{[Ar\,III] (8.99~$\mu$m)}
& \multicolumn{2}{c}{[S\,III] (18.71~$\mu$m)} \\
\cline{2-3} \cline{4-5} \cline{6-7} \cline{8-9} \cline{10-11}
& Disk
& 2\arcsec
& Disk
& 2\arcsec
& Disk
& 2\arcsec
& Disk
& 2\arcsec
& Disk
& 2\arcsec \\
\hline
1 & 0.18$\pm$0.02 & 0.86$\pm$0.03 & 0.08$\pm$0.02 & 0.12$\pm$0.04 & 0.07$\pm$0.01 & $<$0.12 & $<$0.04 & $<$0.13 & 0.80$\pm$0.03 & 0.88$\pm$0.07 \\

5 & 0.202$\pm$0.003 & 1.38$\pm$0.02 & $<$0.02 & $<$0.08 & 0.044$\pm$0.002 & $<$0.08 & $<$0.01 & $<$0.09 & $<$0.46 & $<$1.71 \\

2 & 9.90$\pm$0.03 & 46.57$\pm$0.04 & 0.21$\pm$0.03 & 0.26$\pm$0.03 & 0.77$\pm$0.03 & 15.47$\pm$0.06 & $<$0.04 & $<$0.13 & 0.50$\pm$0.10 & 1.19$\pm$0.14 \\

4 & 0.257$\pm$0.003 & 1.01$\pm$0.02 & $<$0.02 & $<$0.07 & 0.042$\pm$0.002 & $<$0.08 & $<$0.008 & $<$0.090 & $<$0.30 & $<$1.1 \\


3 & 1.915$\pm$0.003 & 2.810$\pm$0.009 & 0.039$\pm$0.002 & 0.052$\pm$0.008 & 0.585$\pm$0.002 & 0.95$\pm$0.04 & $<$0.01 & $<$0.01 & 0.08$\pm$0.03 & 0.17$\pm$0.05 \\

7 & 0.613$\pm$0.003 & 1.32$\pm$0.01 & 0.06$\pm$0.00 & 0.08$\pm$0.01 & 0.102$\pm$0.002 & $<$0.11 & $<$0.01 & $<$0.08 & $<$0.14 & $<$0.20 \\

6 & 0.87$\pm$0.05 & 2.47$\pm$0.03 & 0.50$\pm$0.10 & 0.55$\pm$0.11 & $<$0.12 & 0.25$\pm$0.05 & $<$0.11 & 0.43$\pm$0.09 & 0.58$\pm$0.06 & 0.80$\pm$0.09 \\

\hline
\end{tabular}
\begin{tablenotes}
\item{In units of Jy km\,s$^{-1}$.}
\end{tablenotes}
\label{table:forbidden_fluxes}

\end{table*}

\section{Discussion} 

\subsection{Disk evolution and lifetimes in NGC 1977}

Here we bring together different sources of mass loss seen towards these disks (as listed in Table~\ref{tab:accretion}) with the goal of gaining insight into the evolution and lifetimes of the systems. These are the mass loss from a potential molecular wind traced in \ce{H_2}, the mass accretion rate onto the central stars estimated from the [HI] lines and, the mass loss rates determined form the IF radius and distance to 42~Ori \citep{2022MNRAS.512.2594H} in addition to that inferred from the presence of free-free emission \cite{2024ApJ...967..103B}.

Our inferred temperature components for \ce{H_2} gas are consistent with that seen towards other disks with JWST, namely the transition
disks GM~Aur and SY Cha \citep{2025ApJ...991..128R,2025ApJ...980..148S}, the hydrocarbon-rich M-dwarf disk J160532 where a warm component is present only \citep{2024A&A...687A..96F} 
and, the irradiated disks in Orion where a \ce{H_2} winds with a rotational temperature of $\approx$1000~K are reported \citep{2024Sci...383..988B, 2025NatAs...9.1326S}.
The hot \ce{H_2} detected towards SY Cha is extended up to 1\farcs0 from the central source and is seen to be tracing the upper disk atmosphere on the near side of disk and \citet{2025ApJ...980..148S} propose this emission is tracing a photo-evaporative disk wind. 
In contrast, the extended $\approx$600~K \ce{H_2} emission traces winds in the sample of Class II disks collated by \citet{2026arXiv260507016N}.

Given the distance of 400~pc to our sources and, the spatial and spectral resolution of MIRI-MRS, disentangling \ce{H_2} winds on the expected $<$500~au scales (see Model A from \citealt{2019MNRAS.485.3895H}) expected for external photo-evaporation from the bound \ce{H_2} in the upper layers of the disk is challenging. That being said, we see clear extended emission beyond the continuum disk in \ce{H_2} for both KCFF\#2 and KCFF\#6. For KCFF\#6 in particular, this emission extends out to $>$300~au suggesting that material is flowing from the outer disk. If we therefore interpret the \ce{H_2} components across our sample as a photo-evaporative winds, we can estimate mass loss rates in a similar way as done by \citet{2024ApJ...965L..13A} and \citet{2025ApJ...980..148S}.
These estimates assume a steady mass flux from the outer edge of the disk to the ionisation front (IF) at a velocity of 10~km/s. For the disk outer radius and IF location we adopt the values from \citet{2016ApJ...826L..15K}. Where the disk is unresolved, we assume a characteristic radius of 30~au, noting that a smaller assumed radius would increase the separation between the disk edge and the IF and therefore result in a lower inferred mass-loss rate. 
From this we estimate a range of mass loss rates from 10$^{-10}$ to 10$^{-9}$ $M_{\odot}/yr$ for the hot component of \ce{H_2} and 10$^{-9}$ to 10$^{-8}$ for the warm component of \ce{H_2} - as listed in Table~\ref{tab:accretion} with no clear trend with distance to 42~Ori, as shown in Figure~\ref{fig:h2_distance} where instead there is likely a link between mass loss and host star and/or disk mass. 
Specifically, the hot \ce{H_2} rates are 10-150 times lower than those measured from the IF radii, while the warm \ce{H_2} rates are closer, 1.4-20 times lower, with KCFF\#6 showing the closest agreement.
Although these rates are all notably lower than those measured from the IF radii; depending on the assumed \ce{H_2}/H ratio in the wind, the true total mass loss could be significantly higher than what the \ce{H_2} emission alone traces.
These \ce{H_2} mass loss rates are also lower than those inferred from the free-free emission \citep{2024ApJ...967..103B}.
Our rates are lower than those measured for the more irradiated disks d203-506 and d203-504, also assessed from \ce{H_2} lines albeit via different methods; \citet{2024Sci...383..988B, 2025NatAs...9.1326S} where they report gas temperatures of $\approx$1000~K and mass loss rates of $10^{-7}-10^{-6}~\mathrm{M_{\odot}yr^{-1}}$.
When considering more local and isolated disks, \citet{2026arXiv260507016N} report mass loss rates of order $10^{-9}~\mathrm{M_{\odot}yr^{-1}}$, with a 
single-component gas temperature of $\approx$624$\pm$130~K derived from extended \ce{H_2} emission (as highlighted in Figure~9). If mass loss is traced by the warm gas, our rates are notably higher than in these isolated systems, although absolute disk masses are needed for our sample to put these mass loss rates in better context.

\begin{figure}
    \includegraphics[width=0.85\linewidth]{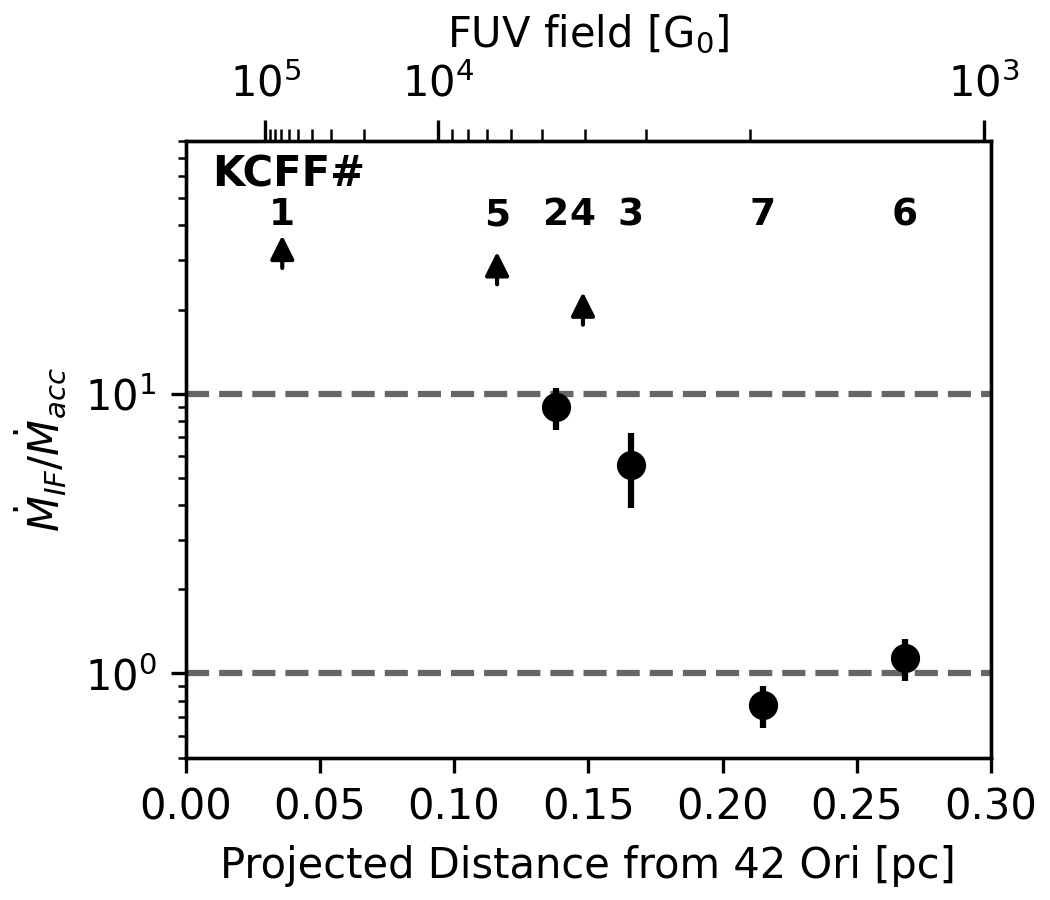}
    \caption{Ratio of the mass loss rates to mass accretion rates as a function of distance from 42~Ori and estimated FUV exposure.}
    \label{fig:loss_ratios}
\end{figure}

We also estimate mass accretion rates across the sample from HI emission lines. Looking at rates (mostly upper limits) derived from the (6–5) for KCFF\#1, \#3, \#4, and \#5 and the rates from the (8–6) for KCFF\#2, \#6, and \#7, we estimate the ratio of mass loss measured from the IF to mass loss from accretion and show this in Figure~\ref{fig:loss_ratios}. For KCFF\#1, \#2, \#3, \#4, and \#5 these values are $>$25, $9.0\pm1.5$, $5.6\pm1.7$, $>$17, and $>$23, receptively indicating that disk dispersal is dominated by the externally driven mass loss. These are conservative estimates, as the mass accretion rates are upper limits. Nevertheless, the two mass-loss estimates agree within an order of magnitude for four systems. Our values are similar or lower to the ratio of $\approx$37 
measured for the irradiated disk d204-504 which is in a more extreme FUV environment \citep{2025MNRAS.tmp.1907C}. Interestingly, for the two more distant disks from 42~Ori KCFF\#6 and \#7 the mass loss ratios are $1.1\pm0.2$ and $0.8\pm0.1$, indicating that there is approximately equal contribution between inside-out and outside-in disk clearing.

We currently lack any estimates of the masses (and in most cases sizes) of the disks in NGC~1977, which, coupled with the mass loss rates, would yield a more accurate estimate of the potential lifetimes of these systems. If we assume a favorable current disk mass of 10~Jupiter masses of gas, as seen towards disks in the ONC \citet{2023ApJ...947....7B}, then all disks should disperse within $<$0.3~Myr. We show an estimate of the disk lifetimes assuming disk masses of 1, 3, 10, and 30~Jupiter masses in Table~\ref{tab:disk_depletion_timescales}, where in almost all cases the disks are expected to disperse within $<$1~Myr. KCFF\#1, as seen in Figure~\ref{fig:disk1_miri_images}, appears to be at a more evolved stage of dispersal when compared to KCFF\#2 and KCFF\#6 which both have, at first look, relativity typical mid-IR spectra for Class~II disks \citep[see][and Figure~\ref{fig:disk2_summary}]{Zhou2026}. Even given the small number of systems in this survey, KCFF\#1, \#2 and \#6 provide anchor points for looking at the relative roles of internal versus external processes in disk evolution across similar host stars where future work, e.g, thermochemical disk modeling of such systems \citep[e.g.;][]{2025ApJ...991...94C}, and sub-millimeter continuum and line data will help unravel these effects further. 

\begin{table*}
\centering
\caption{Approximate total disk depletion timescales for disk masses of 1, 3, 10, and 30~$M_\mathrm{Jup}$. Timescales include all mass loss sources: external ionization fronts, H$_2$ winds, and accretion.}
\begin{tabular}{c c c c c}
\hline\hline
Source ID & 1~$M_\mathrm{Jup}$ [Myr] & 3~$M_\mathrm{Jup}$ [Myr] & 10~$M_\mathrm{Jup}$ [Myr] & 30~$M_\mathrm{Jup}$ [Myr] \\
\hline
1 & $\approx 0.012$ & $\approx 0.03$ & $\approx 0.09$ & $\approx 0.28$ \\
5 & $\approx 0.020$ & $\approx 0.06$ & $\approx 0.20$ & $\approx 0.59$ \\
2 & $\approx 0.009$ & $\approx 0.014$ & $\approx 0.05$ & $\approx 0.14$ \\
4 & $\approx 0.027$ & $\approx 0.08$ & $\approx 0.27$ & $\approx 0.81$ \\
3 & $\approx 0.034$ & $\approx 0.10$ & $\approx 0.34$ & $\approx 1.01$ \\
7 & $\approx 0.013$ & $\approx 0.04$ & $\approx 0.13$ & $\approx 0.40$ \\
6 & $\approx 0.014$ & $\approx 0.04$ & $\approx 0.14$ & $\approx 0.41$ \\
\hline
\end{tabular}
\label{tab:disk_depletion_timescales}
\end{table*}

\subsection{Interpreting the atomic line ratios in the context of isolated Class II disks}

\begin{figure*}[]
    \centering 
    \includegraphics[width=0.95\hsize]{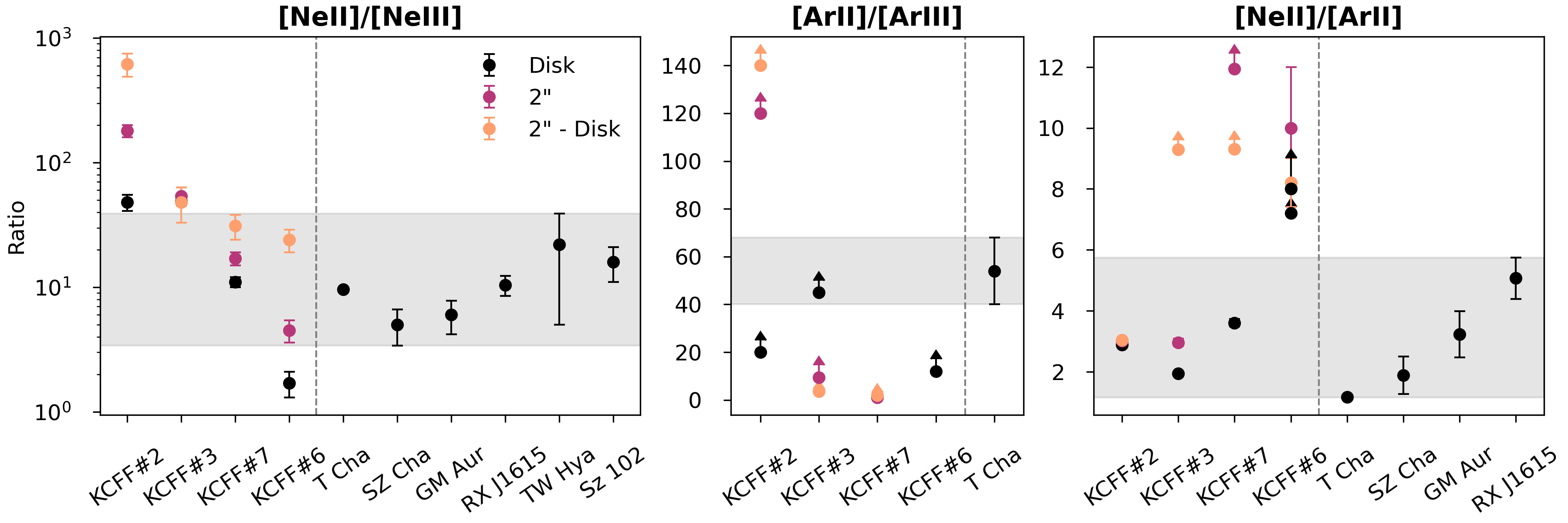}
    \caption{Line flux ratios of different atomic lines toward our sample (left of the dashed line) and other isolated Class II systems (right of the dashed line). The different colors mark the different aperture sizes, where the “Disk” apertures are set by the wavelength of the line. In the case of the [Ne II]/[Ar II] ratio, where these lines are well separated in wavelength space, we recalculate the [Ar II] line flux using the same aperture size as the [Ne II] line. The data for the isolated Class II disks are taken from \citet{2007ApJ...665..492L, 2010ApJ...712..274N, 2023ApJ...958L...4E, 2024AJ....167..127B, 2026AJ....171...39V} where the range in these values is shown with the gray shading. }
    \label{fig:line_ratios}
\end{figure*}

As highlighted by \citet{2025MNRAS.541.2917P}, and as is evident across our sample, the cometary morphology of the proplyd IF as traced by forbidden emission lines is not always going to be spatially resolvable. This means we need to look for diagnostics, e.g., specific line ratios, that are distinct from expectations for internally driven UV and X-ray mass loss processes. In Figure~\ref{fig:line_ratios} we compare the line ratios on different spatial scales of [Ne II]/[Ne III], [Ar II]/[Ar III] and [Ne II]/[Ar II] of our sample to a collection of Class II disks with data taken from \citet{2007ApJ...665..492L, 2010ApJ...712..274N, 2023ApJ...958L...4E, 2024AJ....167..127B, 2026AJ....171...39V}. 

There is a legacy of data from \textit{Spitzer}, and now JWST, measuring atomic line ratios in nearby Class II disks, where typically detections of both [Ne II] and [Ne III] result in line ratios of order $\approx$10–20 \citep{2007ApJ...665..492L, 2010ApJ...712..274N, 2023ApJ...958L...4E, 2024AJ....167..127B} - as shown in Figure~\ref{fig:line_ratios}. In KCFF\#2 we see that this ratio is significantly higher in the IF region, with a [Ne II]-dominated wind where [Ne II]/[Ne III]$>100$. With a smaller aperture size this ratio decreases, possibly indicating a more significant contribution from [Ne III] relative to [Ne II] in the inner disk gas, but given the spatial resolution of the data it is not possible to fully exclude contamination from the extended flow. Even in the disk-integrated case for KCFF\#3, which is similarly irradiated, this ratio is $\approx$50 which is elevated when compared to the most isolated systems. Furthermore, for the disks exposed to FUV field of $\approx10^{3}$~$G_0$ these ratios drop down to $\approx$1-10 on disk scales but are still slightly higher when considering the 2\farcs aperture. The variation across our sources, which tends towards the values in more isolated systems, may indicate that between $\approx10^{3}$ and $\approx10^{4}$~$G_0$ the gas in the IF overtakes the contribution from the inner disk gas. The overall high ratios of [Ne II]/[Ne III] are consistent with an impinging radiation field dominated by the softer UV rather than hard EUV models of \citet{2009ApJ...703.1203H}. 

With [Ar III] only detected towards the wind in T~Cha system, there are fewer comparison points available than with [Ne III] but, the non-detection of [Ar III] in the extended wind of KCFF\#2 does provide an interesting upper limit. Specifically, [Ar II]/[Ar III]$>$100, which is distinct from the value of $\approx$50 reported by \citep{2024AJ....167..127B} where their result is consistent with the EUV models of \citet{2009ApJ...703.1203H}. Similar to the under-abundance of [Ne III] the lack of [Ar III] shows that the external UV field is softer than the 27.6 eV needed to produce [Ar III]. We also detect [S III] in a sub-set of disks (KCFF\#1,2,6) which has an ionization energy of 23.3~eV which is between that of [Ar II] and [Ar III]. 

We can also investigate the relative ratios of [Ne II] to [Ar II]. Towards Class II disks with JWST these ratios range from $\approx$1-5 \citep{2023ApJ...958L...4E,2024AJ....167..127B, 2026AJ....171...39V}. For KCFF\#6 and 7 our ratios are generally higher but this trend does not hold for the disks closer to 42~Ori KCFF\#2 and 3 which have lower ratios making this, at first look, a more complicated diagnostic interpret. 
Overall, it seems from this small sample that the excess [Ne II] and [Ar II] in the proplyd ionization fronts can lead to line ratios that are distinguishable from internal photo-evaporation in more isolated systems when comparing to the higher ionization states, namely [Ne III] and [Ar III]. 
This hypothesis can be tested with more observations of disks across an external FUV range of $\approx10^{3}-10^{5}$ in combination with models accounting for both internal and external forbidden line generation. This will be particularly useful for smaller and/or more distant systems where directly resolving the IF is not possible \citep[as already explored by][]{2025MNRAS.541.2917P}.

\section{Conclusions} 

We have presented JWST/MIRI observations of seven protoplanetary disks undergoing varying levels of external irradiation from the B star 42~Ori. These data reveal a wealth of molecular and atomic features (see also \citealt{Zhou2026}), in addition to an array of extended emission structures. The main conclusions of this work are as follows:

\begin{itemize}

\item The closest proplyd to 42~Ori, KCFF\#1, is experiencing extreme mass loss, as traced by an $\approx$8000~au dusty tail anchored on disk scales. The lack of any clear compact molecular \ce{H_2} or atomic HI gas associated with the central source or ionization front and the compact emissions from forbidden lines indicates that this disk is likely at a very evolved stage of dispersal. Nonetheless, the presence of molecular gas (\ce{CO_2} and \ce{C_2H_2}; see \citealt{Zhou2026}) on disk scales suggests that even under these harsh conditions a small inner disk persists.   

\item Extended molecular and atomic emission is detected across multiple disks, where the morphologies are consistent with stratified PDRs shaped by external irradiation. KCFF\#2 is particularly striking, where extended HI, [Ne II], and [Ar II] trace the gas at/between the proplyd ionization front and the molecular disk. Interestingly, KCFF\#6, which is in a $\approx$4$\times$ lower FUV environment, only shows extended \ce{H_2} emission, consistent with a disk that is comparatively less affected by external UV irradiation from 42~Ori.
    
\item KCFF\#2–7 all exhibit two-temperature-component pure rotational \ce{H_2} emission, with characteristic temperatures of $\approx$500–700~K and $\approx$1000–1500~K. With no clear trend in excitation temperature or column density with distance to 42~Ori, it is likely that the internal radiation field from the host stars, as well as stellar spectral type and disk mass, are the primary drivers setting the bulk molecular gas properties in these systems. That being said, the line-to-continuum ratios of \ce{H_2} for the lowest-mass hosts in the sample are $>$10, indicating potentially larger reservoirs of warm gas and winds due to the elevated irradiation environment. 

\item Interpreting the hot \ce{H_2} emission as disk winds results in mass-loss rates of order $10^{-10}-10^{-9}~M_{\odot}~yr^{-1}$. If the warm \ce{H_2} instead traces a wind these rates are an order of magnitude higher. Comparing these values to the measured mass-loss rates from the ionization fronts and our inferred mass accretion rates from the HI lines indicates that in KCFF\#1,2,3,4 and 5 external mass loss as traced by the IF dominates, whereas in the less irradiated KCFF\#6 and 7 mass accretion and mass loss from winds are contributing equally. 

\item The detections of forbidden-line emission across the sample are broadly consistent with ionization by a relatively soft external UV radiation field. In particular, the ionization front in KCFF\#2 shows especially distinct ratios (a high [Ne II]/[Ne III] and strong lower limits on [Ar II]/[Ar III]) compared to typical isolated Class II disks, highlighting the contribution from externally irradiated gas on larger scales. Overall, the measured line ratios differ from those seen in more isolated systems, suggesting that emission from the proplyd ionization fronts can dominate over internally driven mass-loss signatures.
\end{itemize}

Overall, this work demonstrates how JWST/MIRI can be used to assess the impact of environment on disk evolution, with NGC~1977 providing an anchor point between the more strongly irradiated disks in the ONC and the more local disk population. Upcoming surveys of the ONC (e.g., GO 7534) will enable a more quantitative approach to disentangling internal and external processes on the physics and chemistry of planet formation.

\begin{acknowledgments}

The JWST data presented in this article were obtained from the Mikulski Archive for Space Telescopes (MAST) at the Space Telescope Science Institute. The specific observations analyzed can be accessed via \dataset[doi:10.17909/kwgg-yy54]{https://doi.org/10.17909/kwgg-yy54}
We thank Carlos E. Romero-Mirza for his work in leading the JWST proposal. 
We also thank Sierra Grant and Nicole Arulanantham for very useful discussions in the preparation of this manuscript in addition to helpful comments from Olivier Berné, Giovanni Rosotti and Adolfo Carvalho. 
A.S.B. is supported by a Clay Postdoctoral Fellowship from the Smithsonian Astrophysical Observatory.
Q. Z. is supported by the National Science Foundation Graduate Research Fellowship Program under Grant No. DGE 2140743. Any opinions, findings, and conclusions or recommendations expressed in this material are those of the authors and do not necessarily reflect the views of the National Science Foundation.
J.S.K. acknowledges NASA/Space Telescope Science Institute grant JWST-GO-06130 and NASA’s Nexus for Exoplanet System Science (NExSS) research coordination network sponsored by NASA’s Science Mission Directorate and project “Alien Earths” funded under Agreement No. 80NSSC21K0593. 
J.K.C. is supported by the Kavli-Laukien Origins of Life Consortium at Harvard.
TJH acknowledges UKRI guaranteed funding for a Horizon Europe ERC consolidator grant (EP/Y024710/1) and a Royal Society Dorothy Hodgkin Fellowship.
R. Boyden acknowledges support from the the Virginia Initiative on Cosmic Origins (VICO), the National Radio Astronomy Observatory, and NASA/Space Telescope Science Institute grant JWST-GO-06130. The National Radio Astronomy Observatory is a facility of the National Science Foundation operated under cooperative agreement by Associated Universities, Inc. 
Part of this research was carried out at the Jet Propulsion Laboratory, California Institute of Technology, under a contract with the National Aeronautics and Space Administration (80NM0018D0004). 

\end{acknowledgments}


%



\clearpage
\newpage
\begin{appendix}

\section{Channel maps of \ce{H_2} emission}

\begin{figure*}[th!]
    \centering
    \includegraphics[width=0.95\hsize]{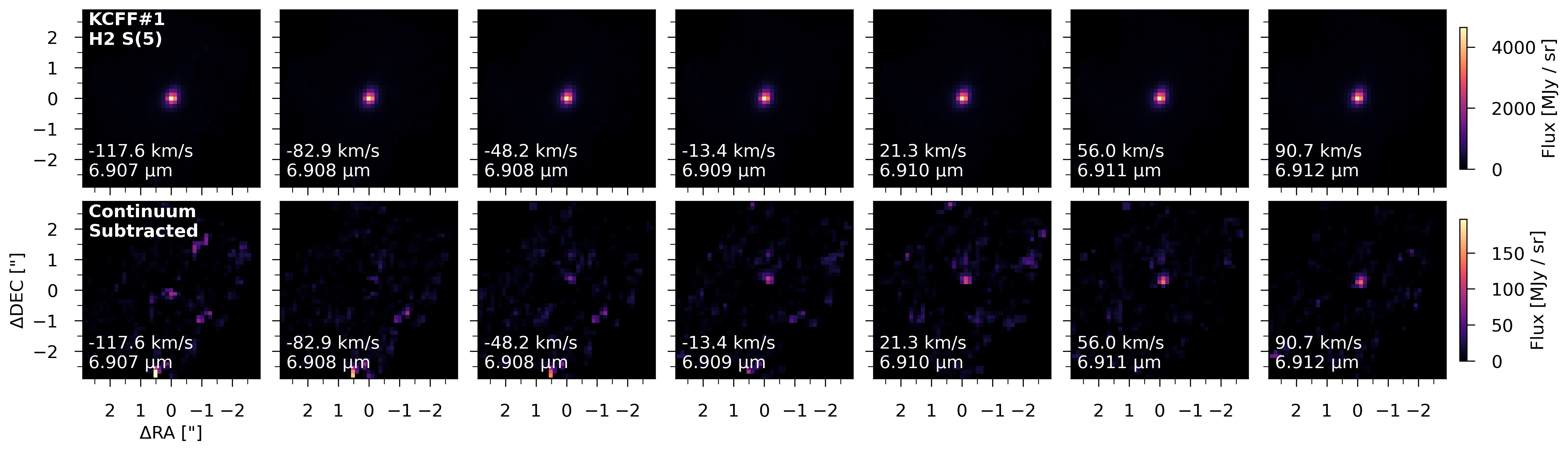}
    \caption{Same as Figure~\ref{fig:placeholder} for the \ce{H_2} (0,0) S(5) line towards KCFF\#1.}
    \label{fig:channels_h2_disk1}
\end{figure*}

\begin{figure*}[th!]
    \centering
    \includegraphics[width=0.95\hsize]{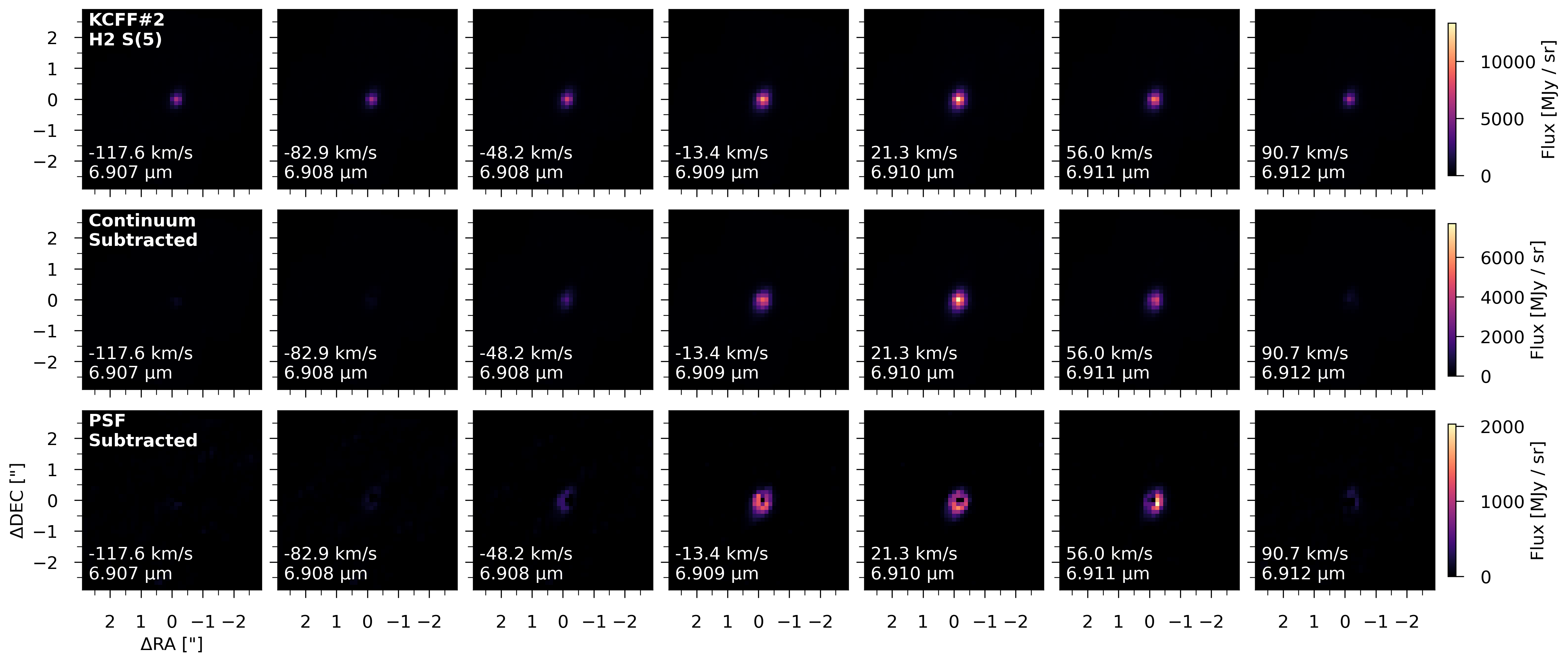}
    \caption{Same as Figure~\ref{fig:placeholder} for the \ce{H_2} (0,0) S(5) line towards KCFF\#2.}
    \label{fig:channels_h2_disk2}
\end{figure*}

\begin{figure*}[h!]
    \centering
    \includegraphics[width=0.95\hsize]{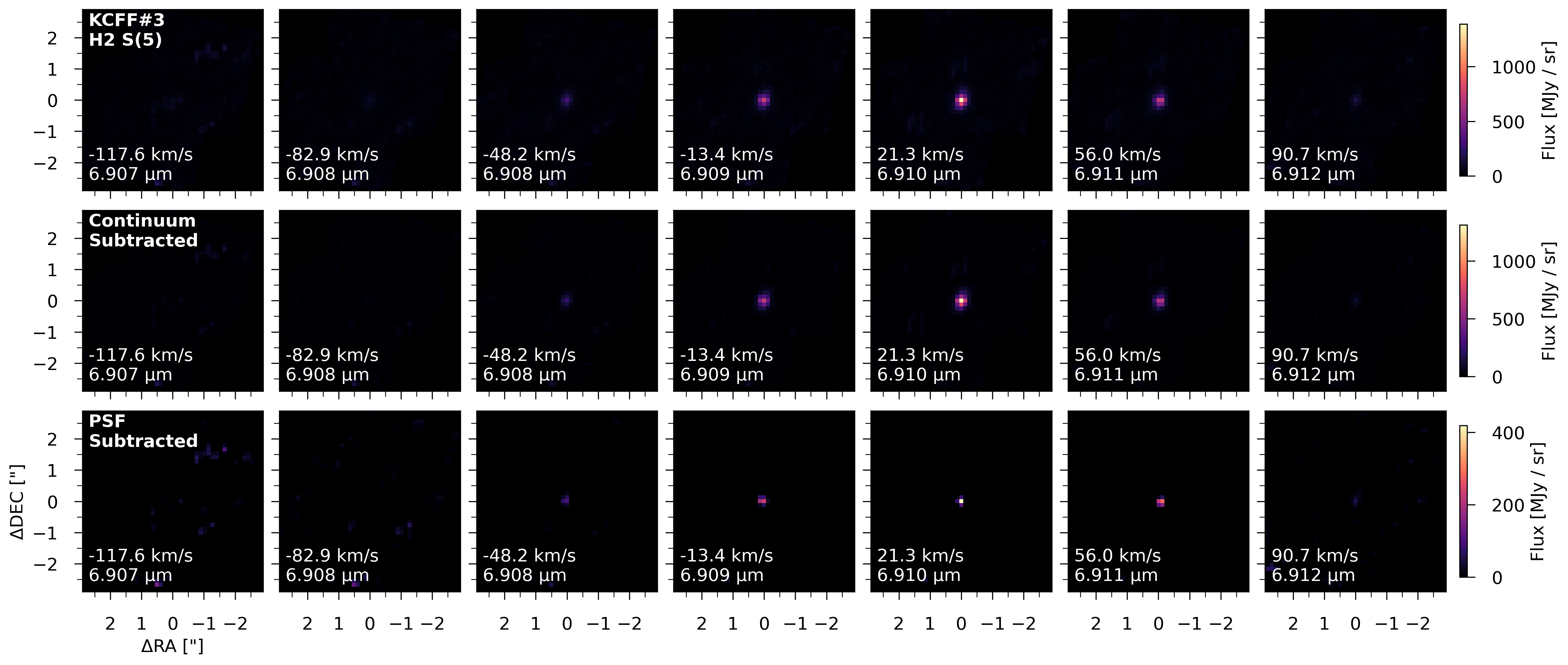}
    \caption{Same as Figure~\ref{fig:placeholder} for the \ce{H_2} (0,0) S(5) line towards KCFF\#3. }
    \label{fig:channels_h2_disk3}
\end{figure*}

\begin{figure*}[h!]
    \centering
    \includegraphics[width=0.95\hsize]{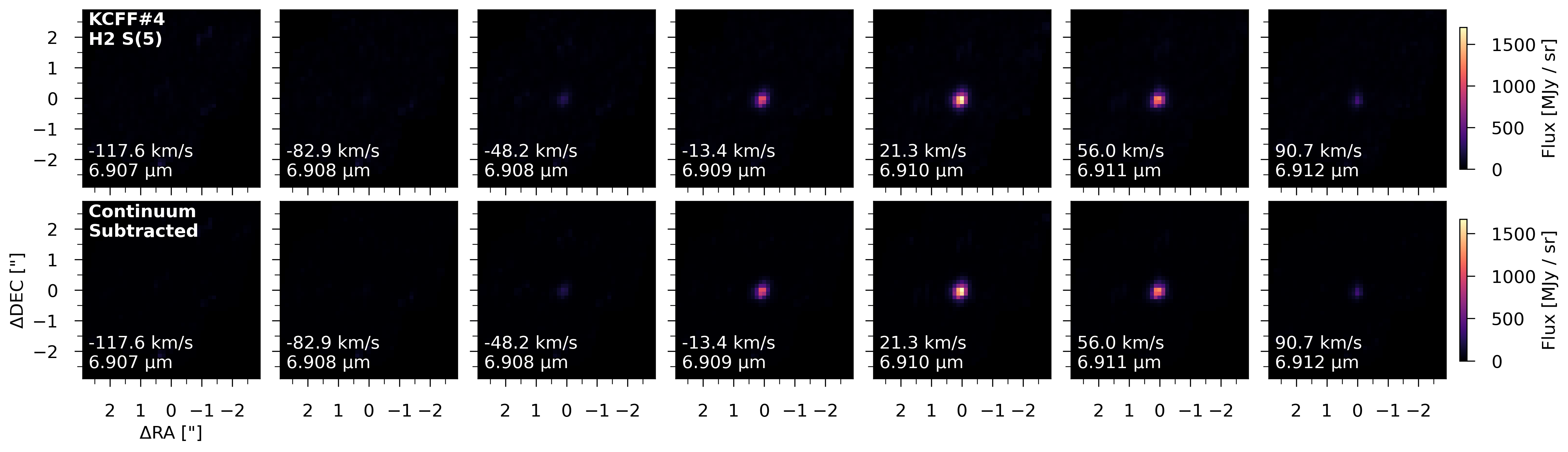}
    \caption{Same as Figure~\ref{fig:placeholder} for the \ce{H_2} (0,0) S(5) line towards KCFF\#4. }
    \label{fig:channels_h2_disk4}
\end{figure*}

\begin{figure*}[h!]
    \centering
    \includegraphics[width=0.95\hsize]{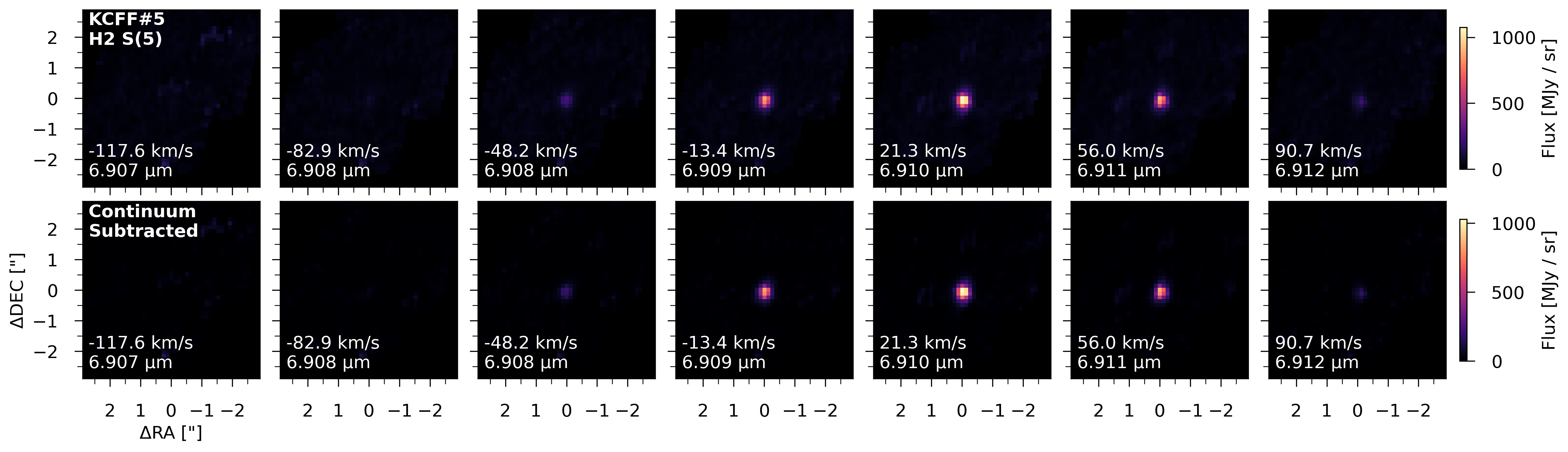}
    \caption{Same as Figure~\ref{fig:placeholder} for the \ce{H_2} (0,0) S(5) line towards KCFF\#5. }
    \label{fig:channels_h2_disk5}
\end{figure*}

\begin{figure*}[h!]
    \centering
    \includegraphics[width=0.95\hsize]{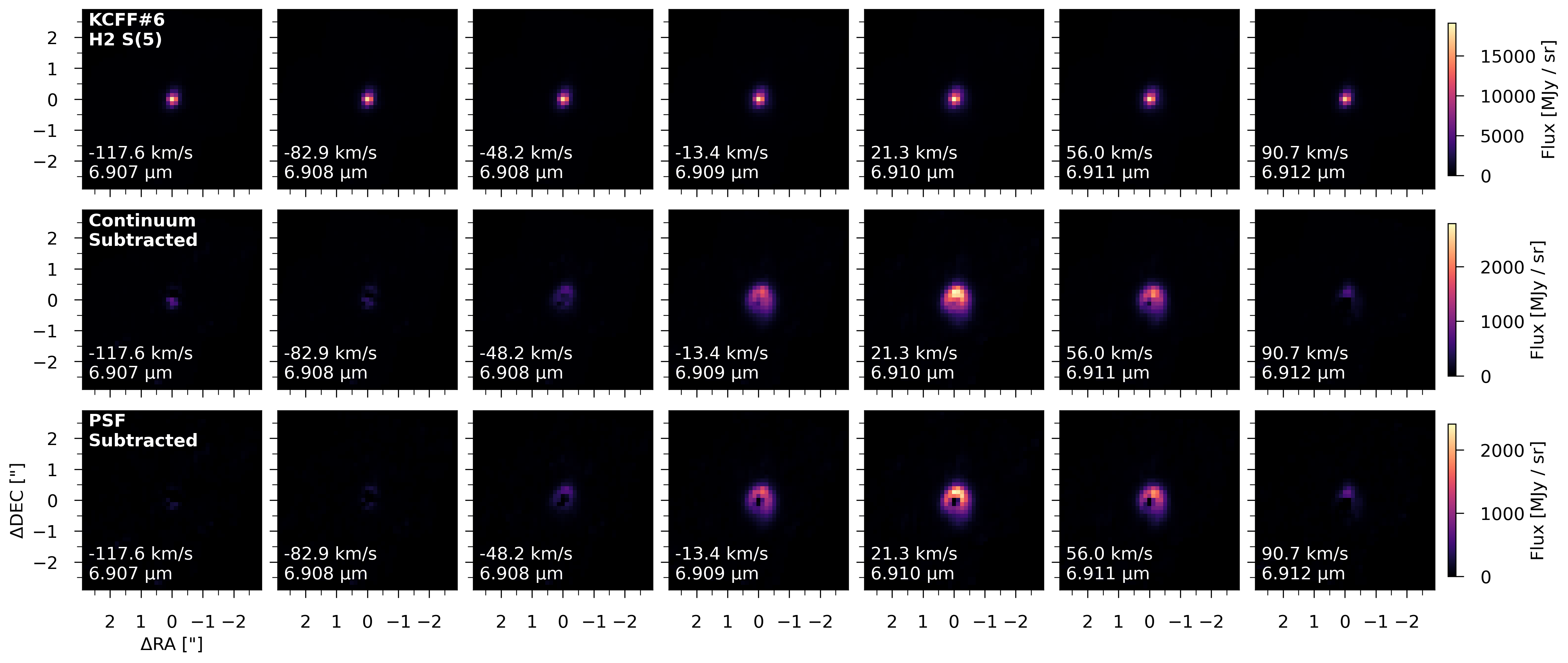}
    \caption{Same as Figure~\ref{fig:placeholder} for the \ce{H_2} (0,0) S(5) line towards KCFF\#6. }
    \label{fig:channels_h2_disk6}
\end{figure*}

\begin{figure*}[h!]
    \centering
    \includegraphics[width=0.95\hsize]{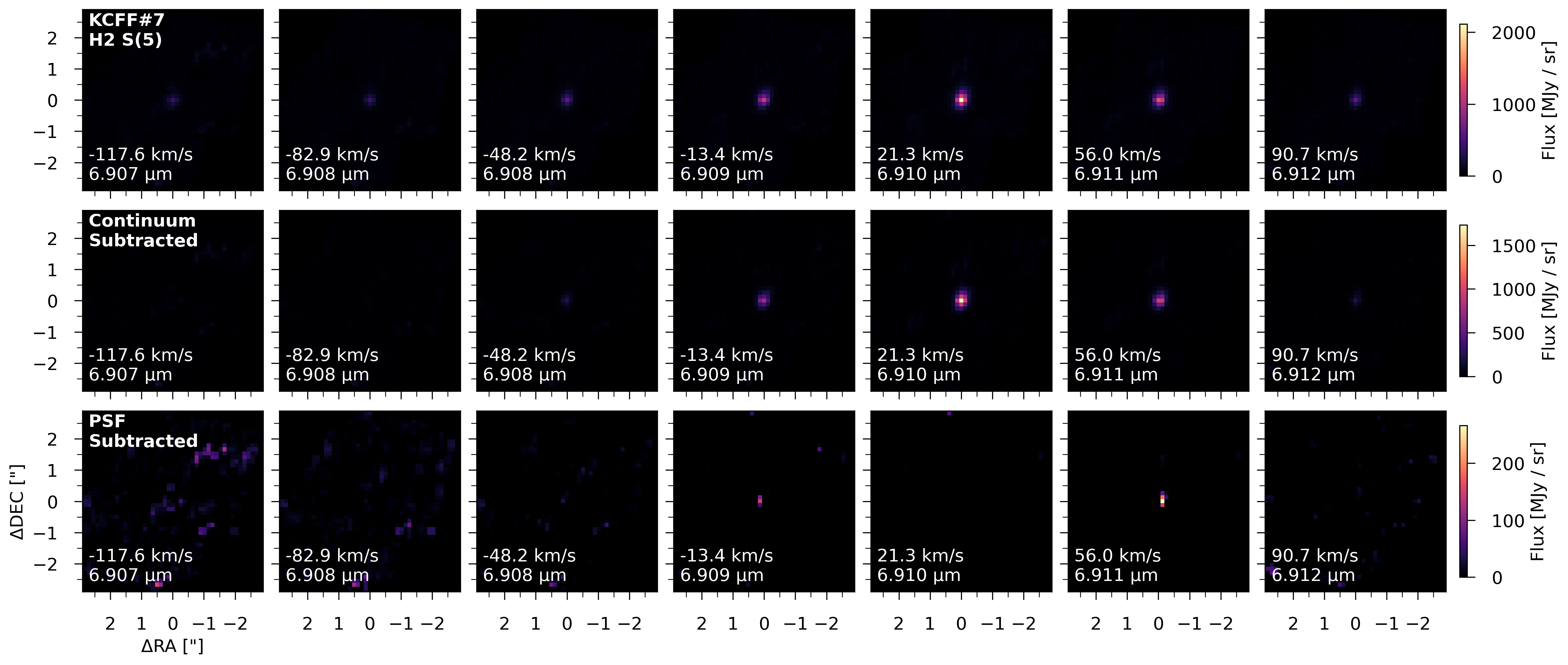}
    \caption{Same as Figure~\ref{fig:placeholder} for the \ce{H_2} (0,0) S(5) line towards KCFF\#7.}
    \label{fig:channels_h2_disk7}
\end{figure*}

\clearpage
\newpage
\section{Channel maps of \ce{HI} emission}

\begin{figure*}[th!]
    \centering
    \includegraphics[width=0.95\hsize]{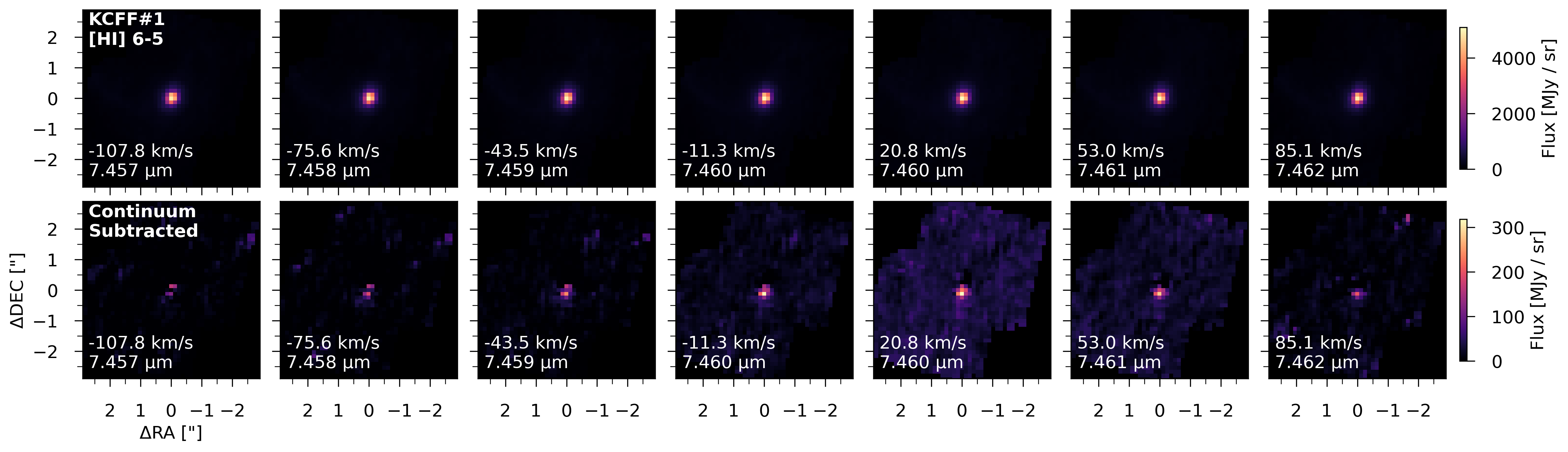}
    \caption{Same as Figure~\ref{fig:placeholder} for the \ce{HI} 6-5 line towards KCFF\#1.}
    \label{fig:HI_disk1}
\end{figure*}

\begin{figure*}[]
    \centering
    \includegraphics[width=0.95\hsize]{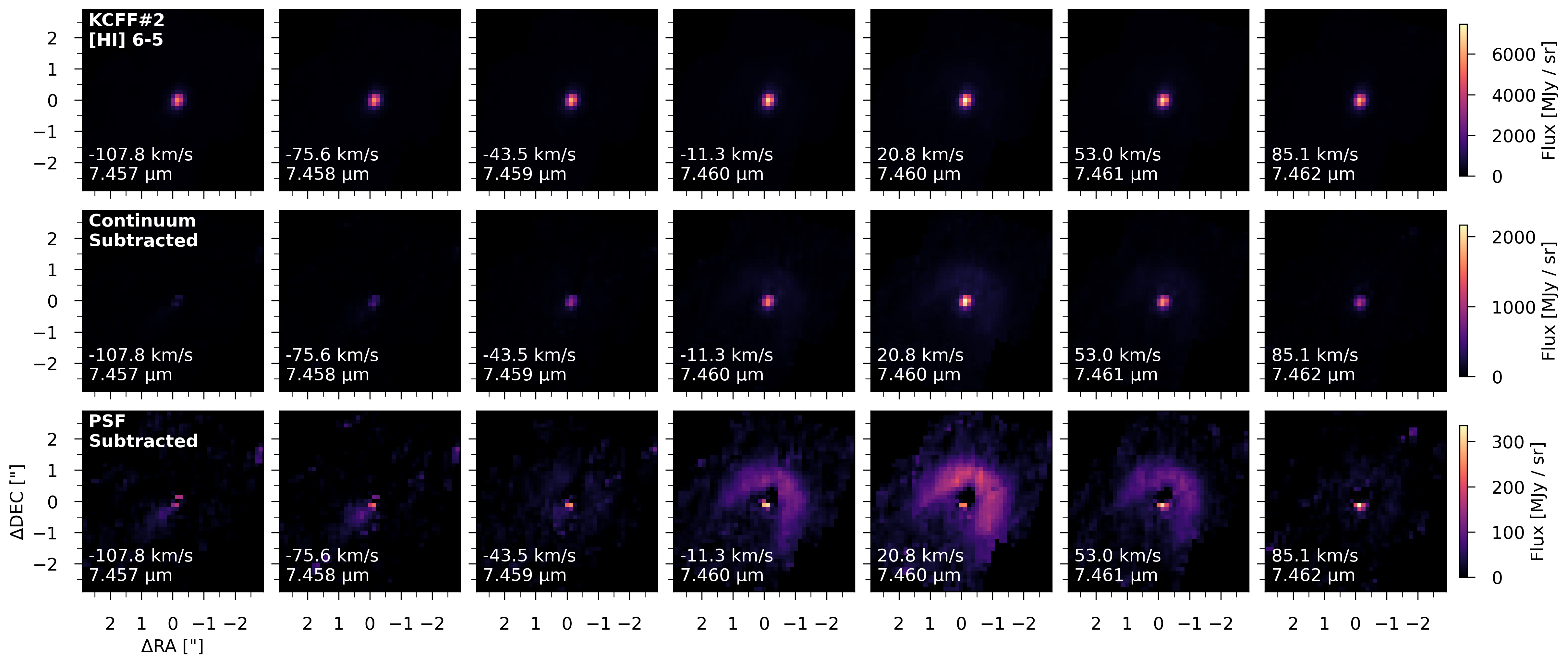}
    \caption{Same as Figure~\ref{fig:placeholder} for the detection of HI (6-5) in KCFF\#2}
    \label{fig:HI_disk2}
\end{figure*}

\begin{figure*}[]
    \centering
    \includegraphics[width=0.95\hsize]{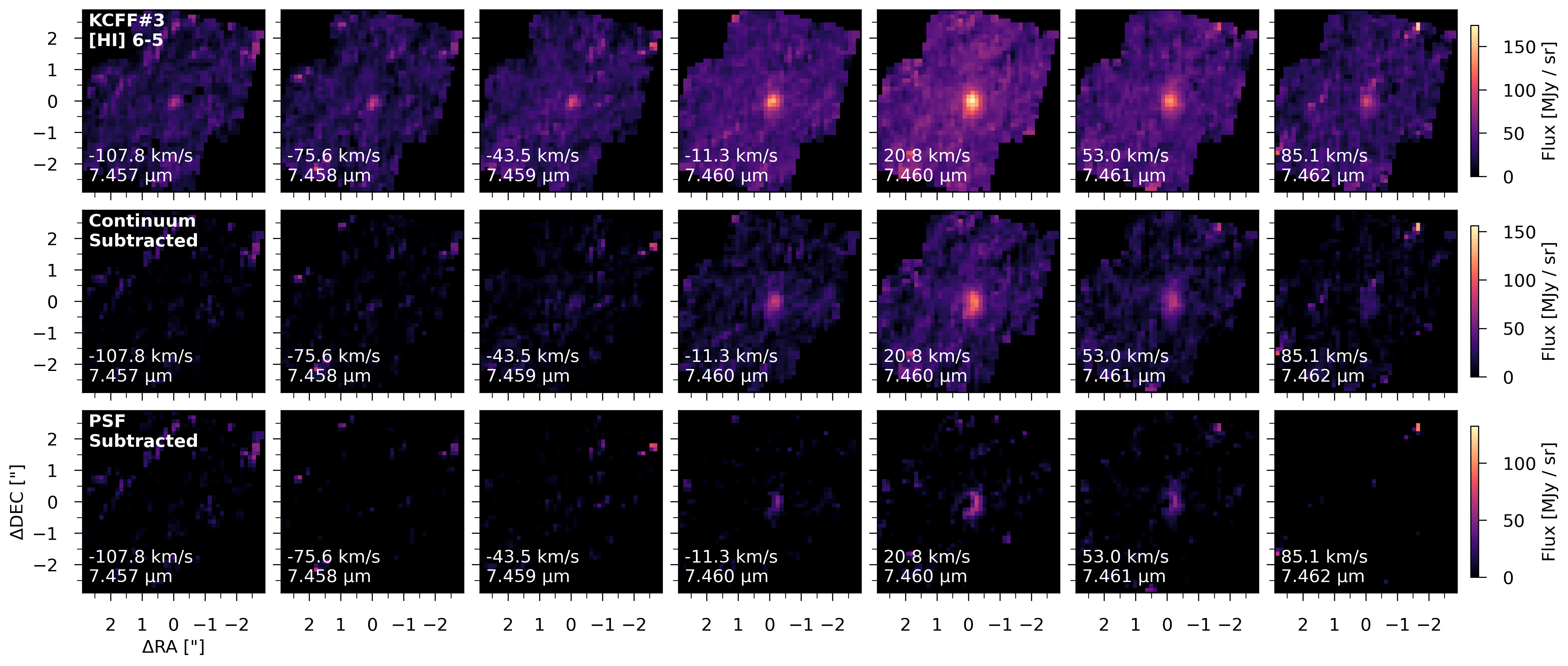}
    \caption{Same as Figure~\ref{fig:placeholder} for the detection of HI (6-5) in KCFF\#3}
    \label{fig:HI_disk3}
\end{figure*}



\begin{figure*}[]
    \centering
    \includegraphics[width=0.95\hsize]{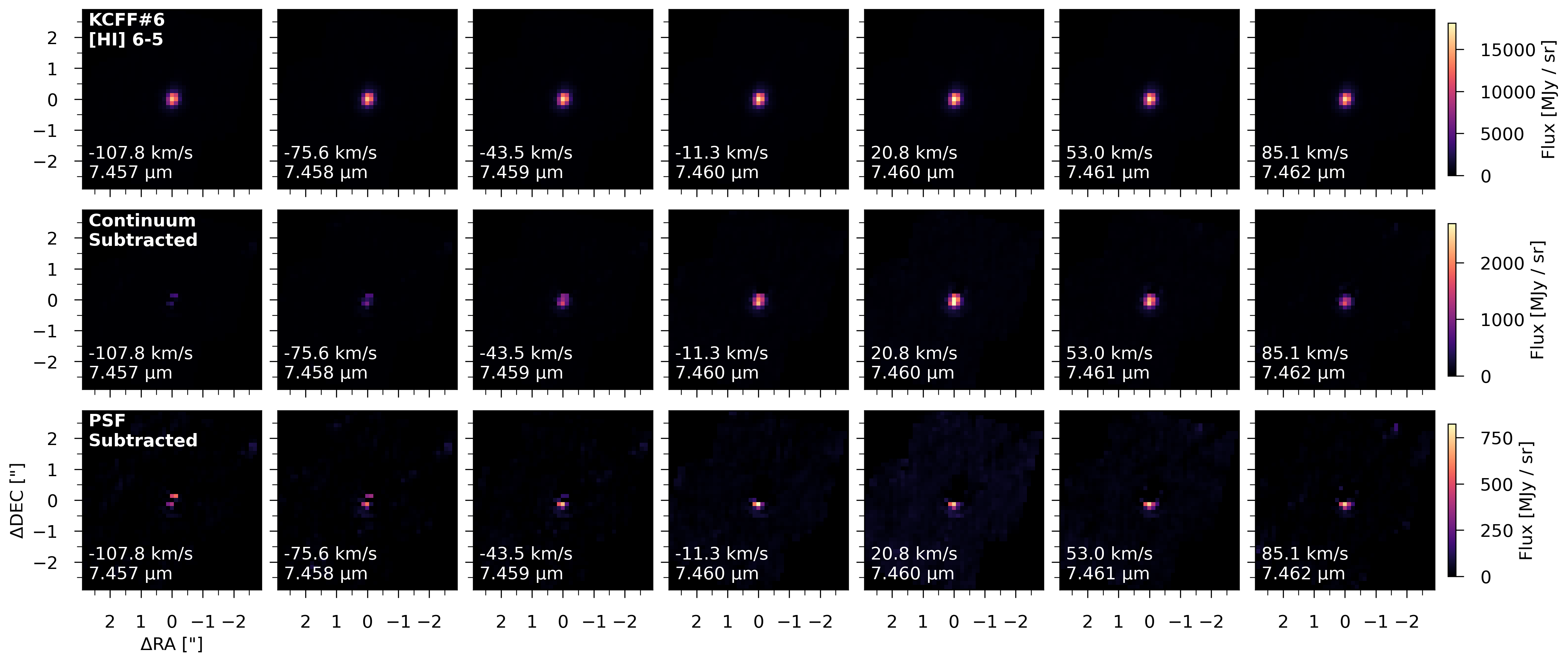}
    \caption{As in Figure~\ref{fig:placeholder} for the detection of HI (6-5) in KCFF\#6}
    \label{fig:HI_disk6}
\end{figure*}

\begin{figure*}[]
    \centering
    \includegraphics[width=0.95\hsize]{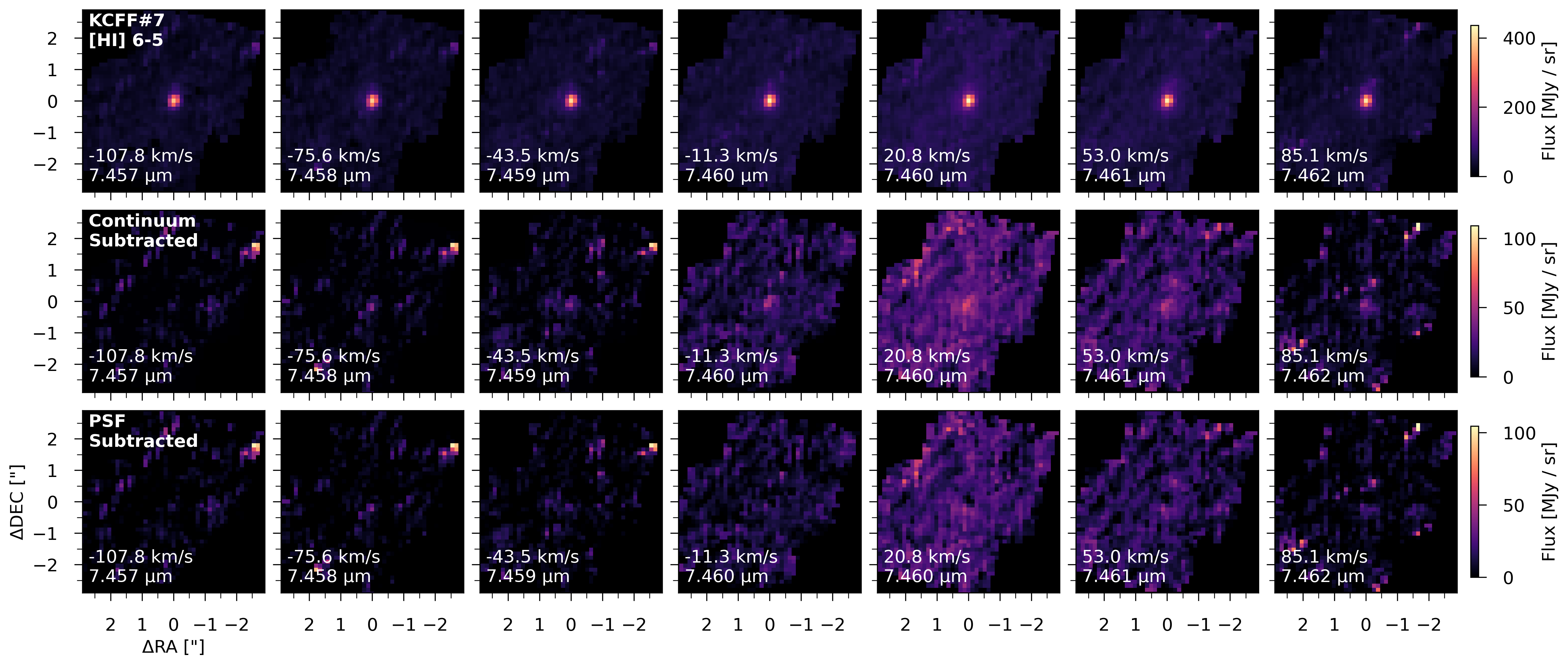}
    \caption{Same as Figure~\ref{fig:placeholder} for the detection of HI (6-5) in KCFF\#7}
    \label{fig:HI_disk7}
\end{figure*}

\clearpage
\newpage
\section{Channel maps of forbidden line emission}

\begin{figure*}[h!]
    \centering
    \includegraphics[width=0.95\hsize]{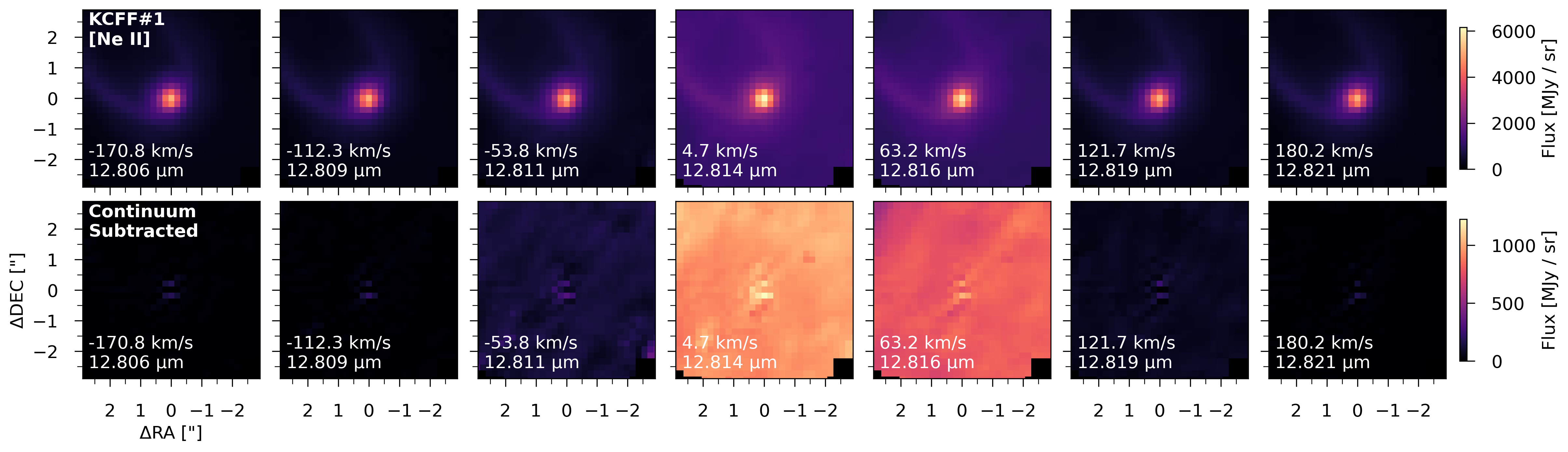}
    \caption{Same as Figure~\ref{fig:placeholder} for the detection of [Ne II] in KCFF\#1}
    \label{fig:NeII_disk1}
\end{figure*}

\begin{figure*}[h!]
    \centering
    \includegraphics[width=0.95\hsize]{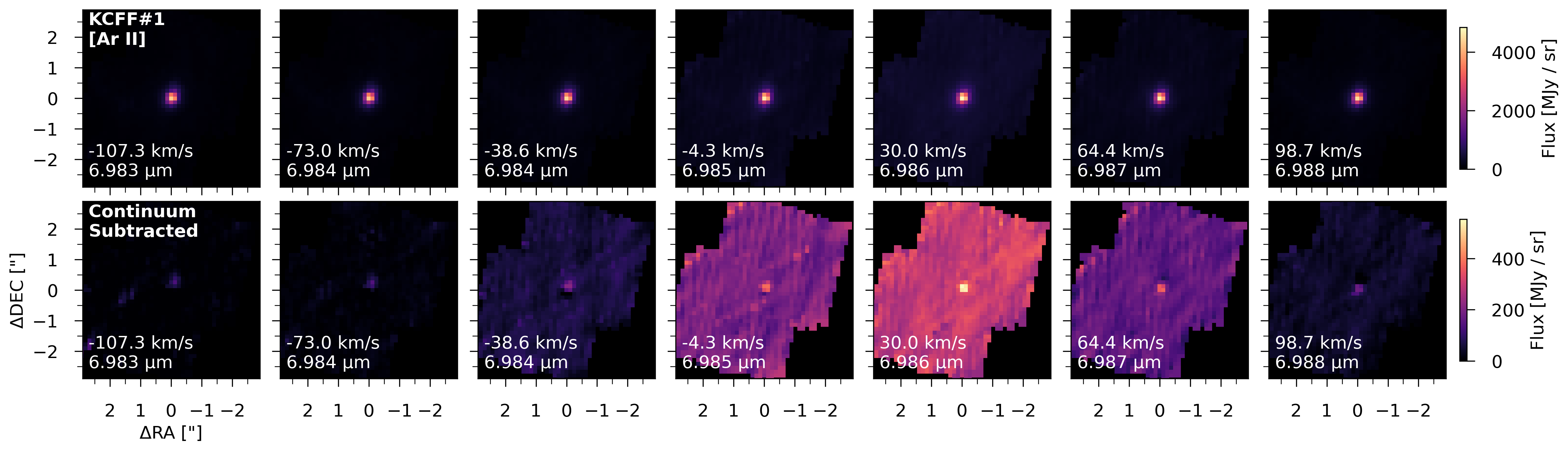}
    \caption{Same as Figure~\ref{fig:placeholder} for the detection of [Ar II] in KCFF\#1}
    \label{fig:ArII_disk1}
\end{figure*}

\begin{figure*}[]
    \centering
    \includegraphics[width=0.95\hsize]{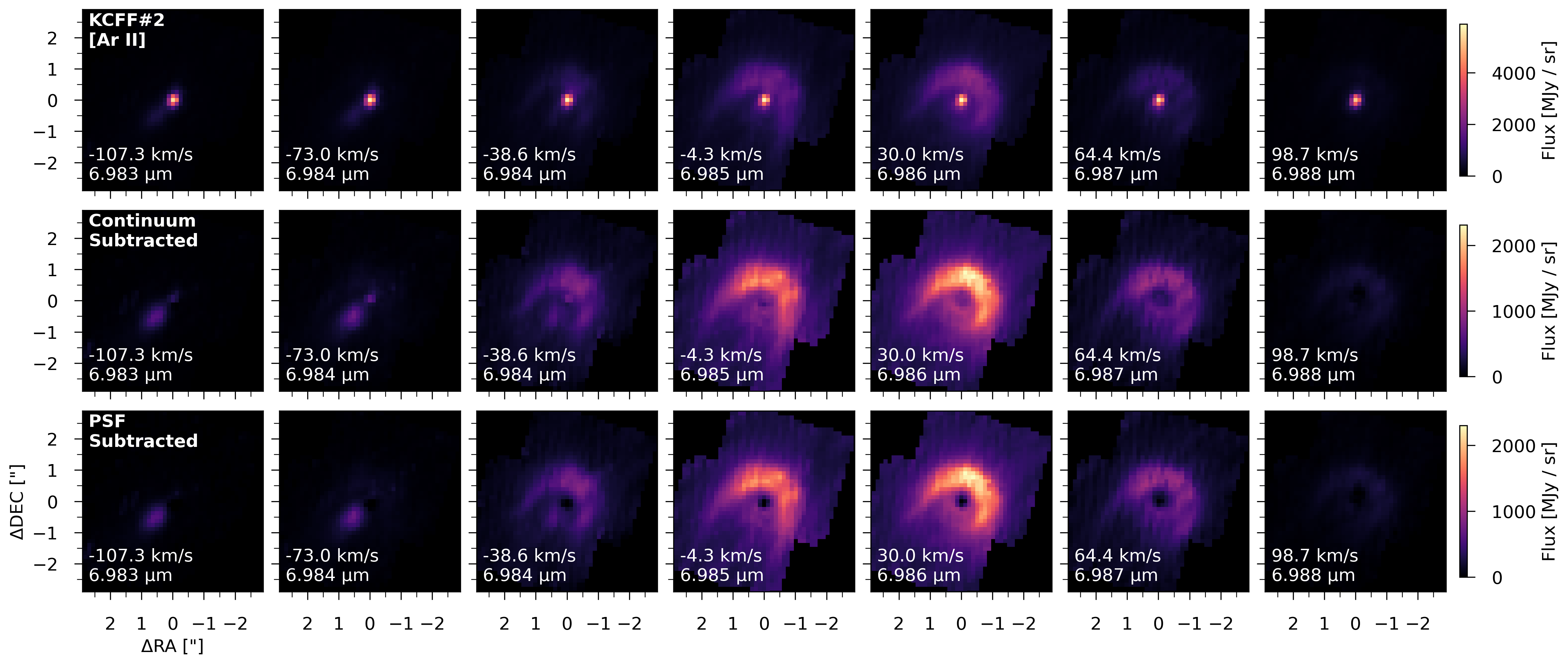}
    \caption{Same as Figure~\ref{fig:placeholder} for the detection of [Ar II] in KCFF\#2}
    \label{fig:ArII_disk2}
\end{figure*}

\begin{figure*}[]
    \centering
    \includegraphics[width=0.95\hsize]{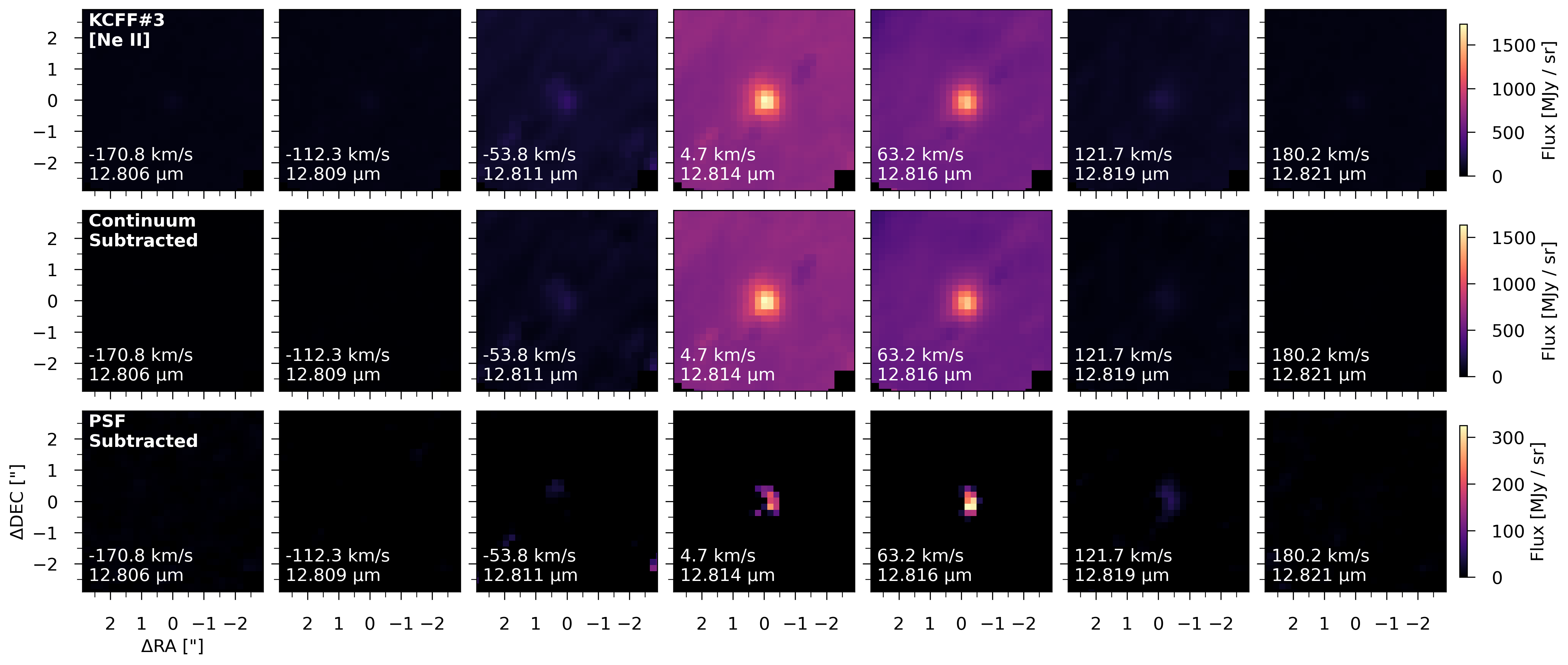}
    \caption{Same as Figure~\ref{fig:placeholder} for the detection of [Ne II] in KCFF\#3}
    \label{fig:NeII_disk3}
\end{figure*}

\begin{figure*}[]
    \centering
    \includegraphics[width=0.95\hsize]{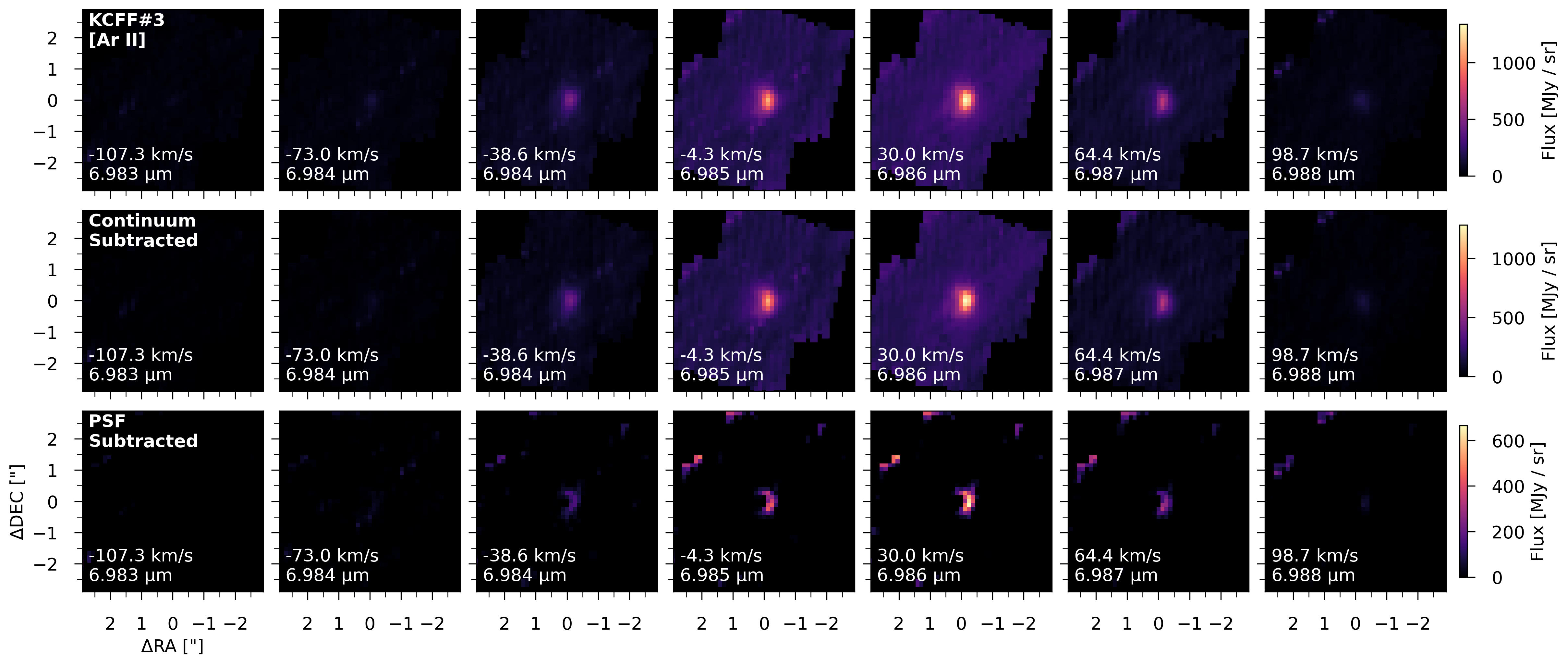}
    \caption{Same as Figure~\ref{fig:placeholder} for the detection of [Ar II] in KCFF\#3}
    \label{fig:ArII_disk3}
\end{figure*}

\begin{figure*}[]
    \centering
    \includegraphics[width=0.95\hsize]{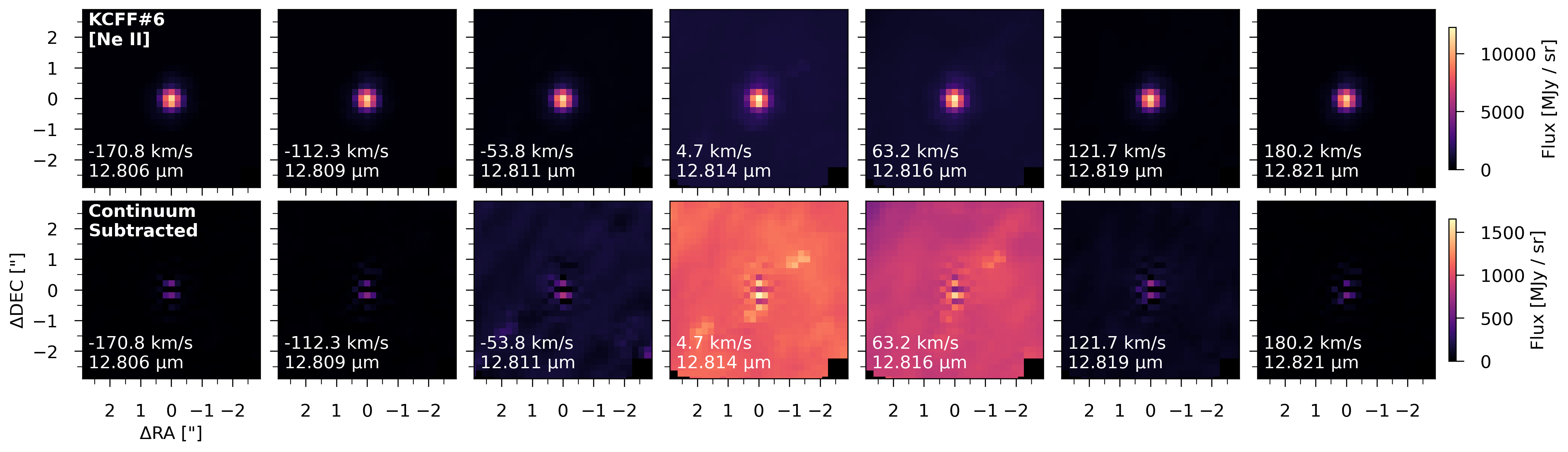}
    \caption{Same as Figure~\ref{fig:placeholder} for the detection of [Ne II] in KCFF\#6}
    \label{fig:NeII_disk6}
\end{figure*}

\begin{figure*}[]
    \centering
    \includegraphics[width=0.95\hsize]{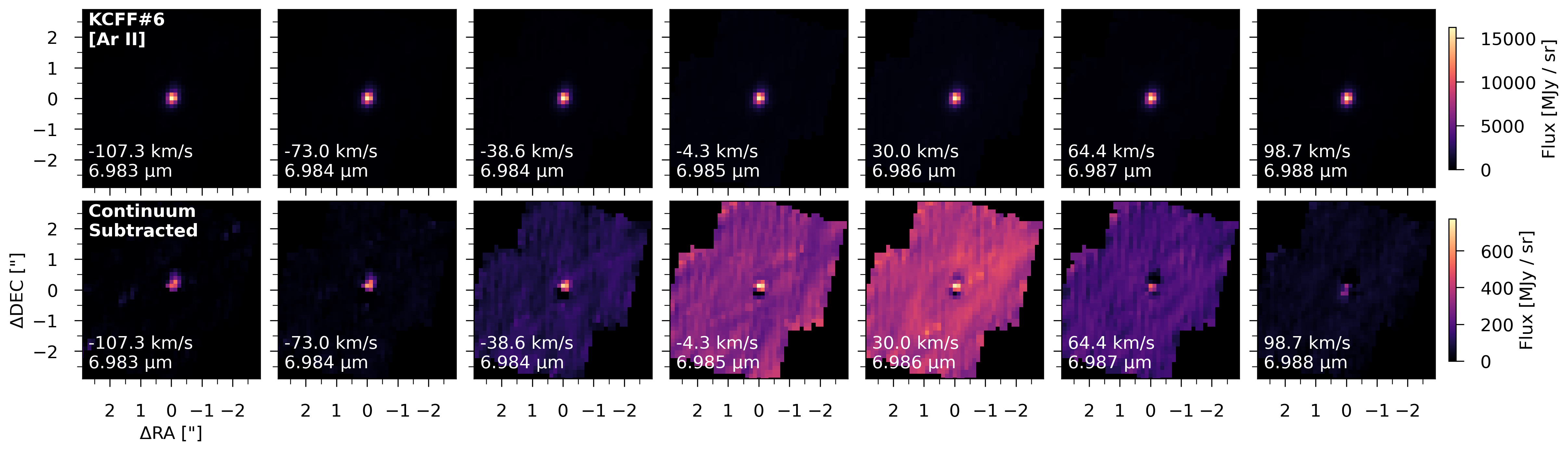}
    \caption{Same as Figure~\ref{fig:placeholder} for the detection of [Ar II] in KCFF\#6}
    \label{fig:ArII_disk6}
\end{figure*}

\begin{figure*}[]
    \centering
    \includegraphics[width=0.95\hsize]{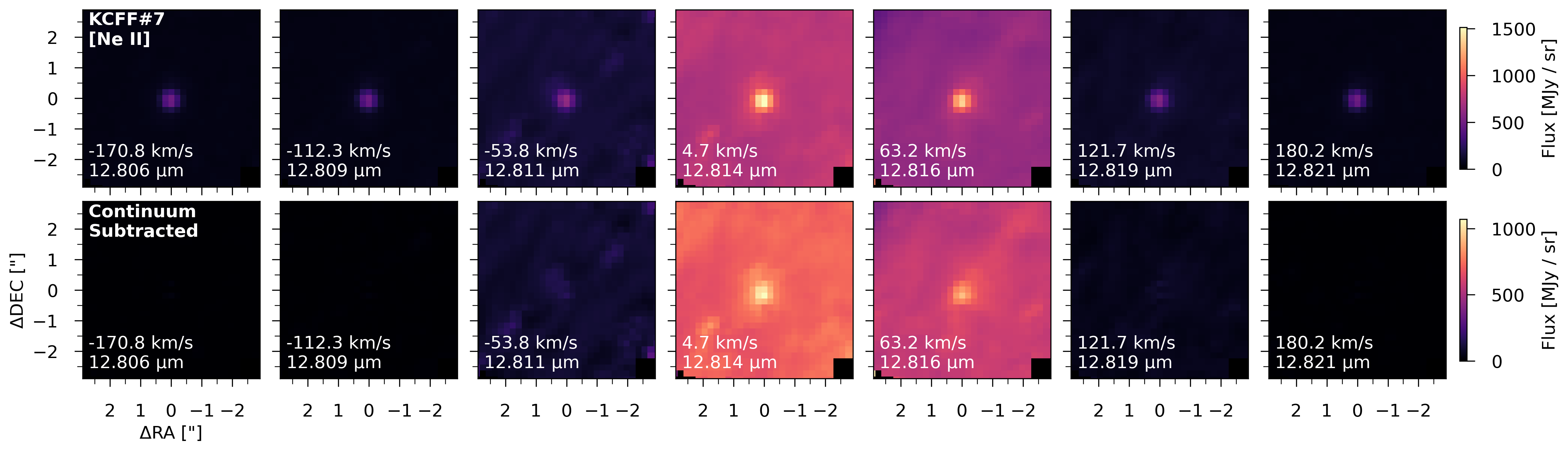}
    \caption{Same as Figure~\ref{fig:placeholder} for the detection of [Ne II] in KCFF\#7}
    \label{fig:NeII_disk7}
\end{figure*}

\begin{figure*}[]
    \centering
    \includegraphics[width=0.95\hsize]{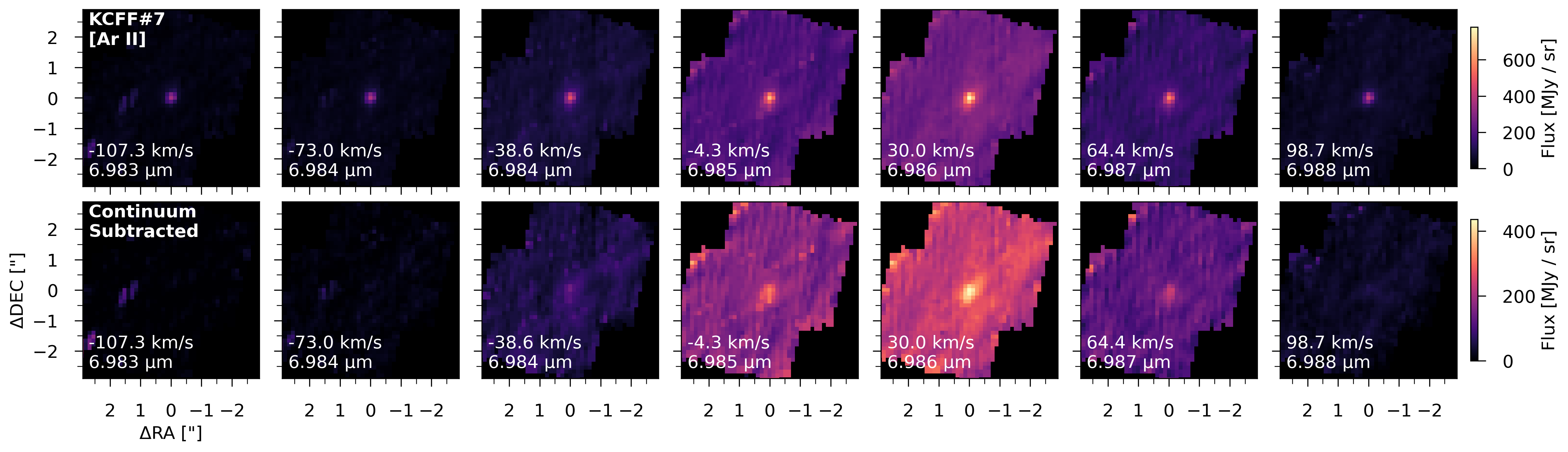}
    \caption{Same as Figure~\ref{fig:placeholder} for the detection of [Ar II] in KCFF\#7}
    \label{fig:ArII_disk7}
\end{figure*}

\begin{figure*}[h!]
    \centering
    \includegraphics[width=0.95\hsize]{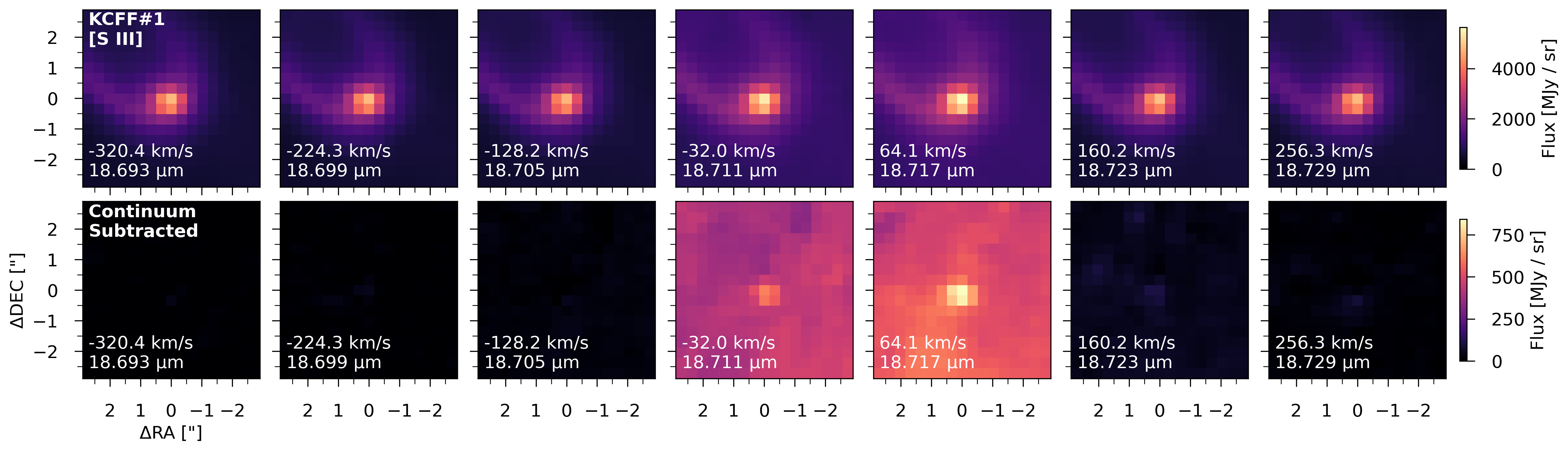}
    \caption{Same as Figure~\ref{fig:placeholder} for the detection of [S III] in KCFF\#1}
    \label{fig:disk1_SIII}
\end{figure*}

\begin{figure*}[h!]
    \centering
    \includegraphics[width=0.95\hsize]{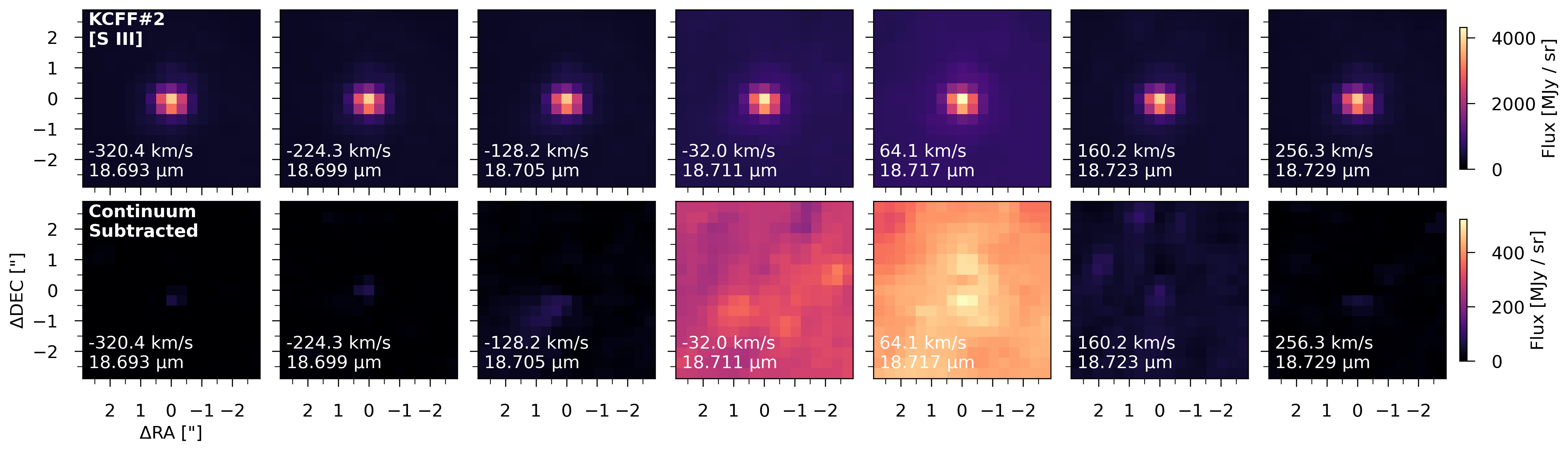}
    \caption{Same as Figure~\ref{fig:placeholder} for the detection of [S III] in KCFF\#2}   
    \label{fig:disk2_SIII}
\end{figure*}

\begin{figure*}[h!]
    \centering
    \includegraphics[width=0.95\hsize]{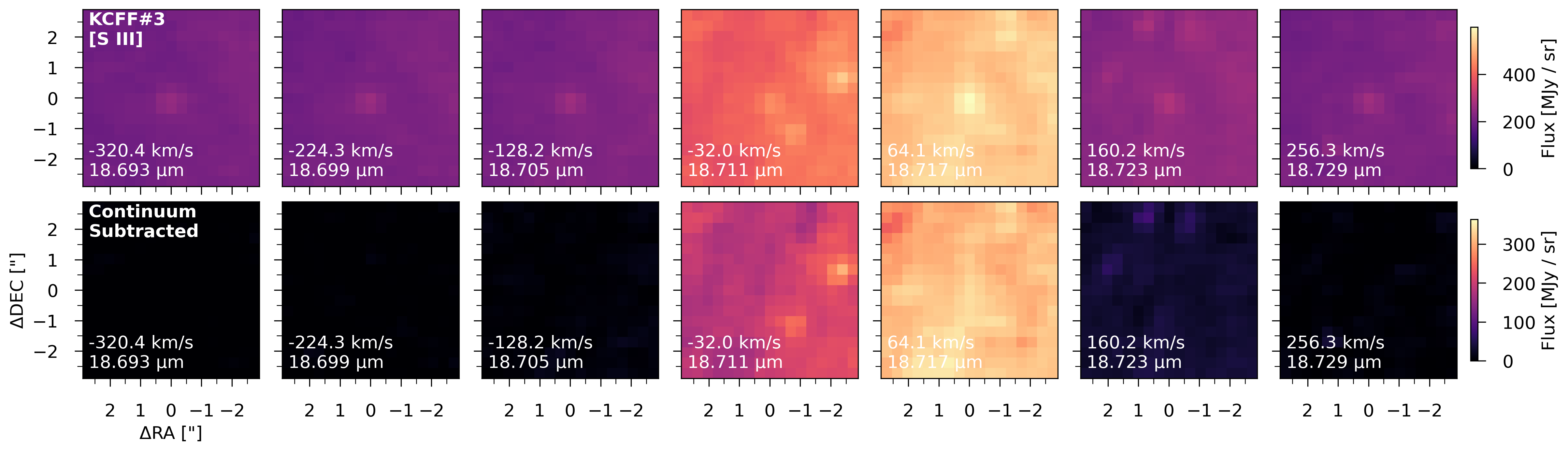}
    \caption{Same as Figure~\ref{fig:placeholder} for the detection of [S III] in KCFF\#3}
    \label{fig:disk3_SIII}
\end{figure*}

\begin{figure*}[h!]
    \centering
    \includegraphics[width=0.95\hsize]{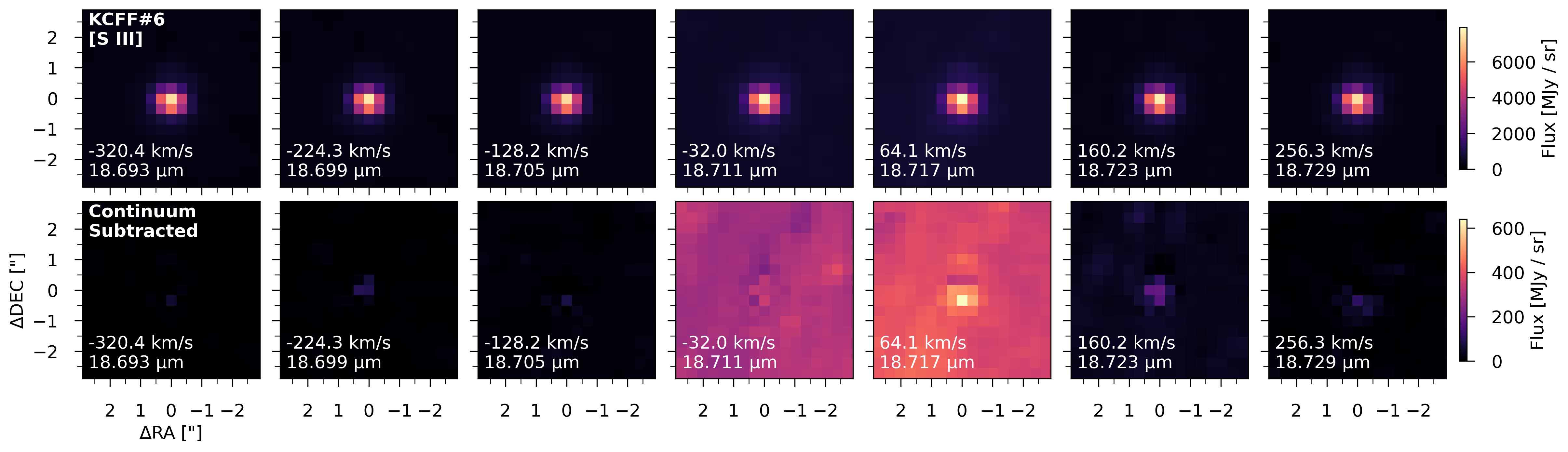}
    \caption{Same as Figure~\ref{fig:placeholder} for the detection of [S III] in KCFF\#6}
    \label{fig:disk6_SIII}
\end{figure*}


\clearpage
\newpage
\section{Emission slices across KCFF\#3 and KCFF\#6 towards the direction of 42 Ori}

\begin{figure*}[h!]
\centering
\includegraphics[width=0.9\hsize]{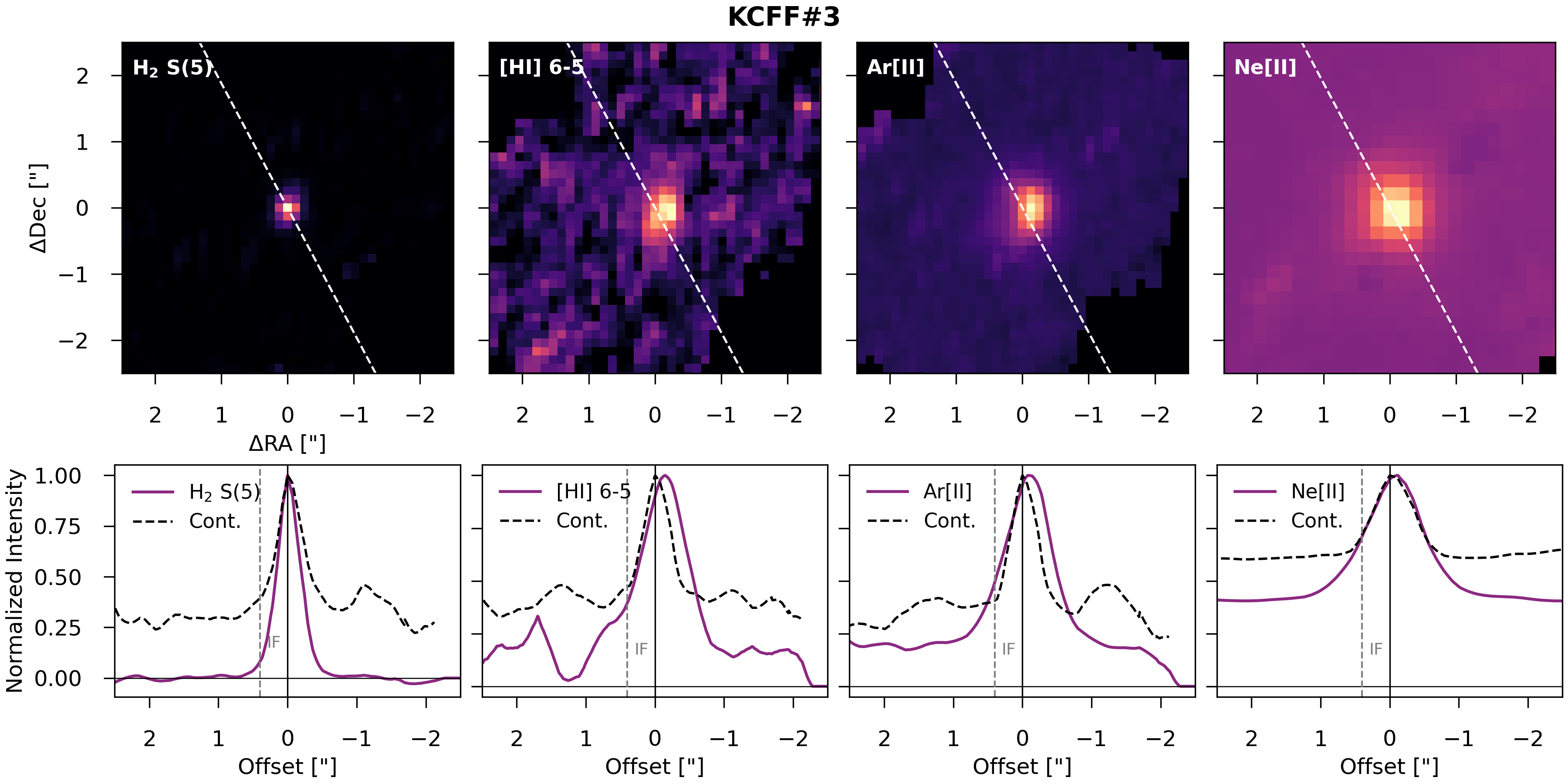}
\caption{Same as Figure~\ref{fig:disk2_slices} for KCFF\#3.}
\label{fig:disk3_slices}
\end{figure*}

\begin{figure*}[h!]
\centering
\includegraphics[width=0.9\hsize]{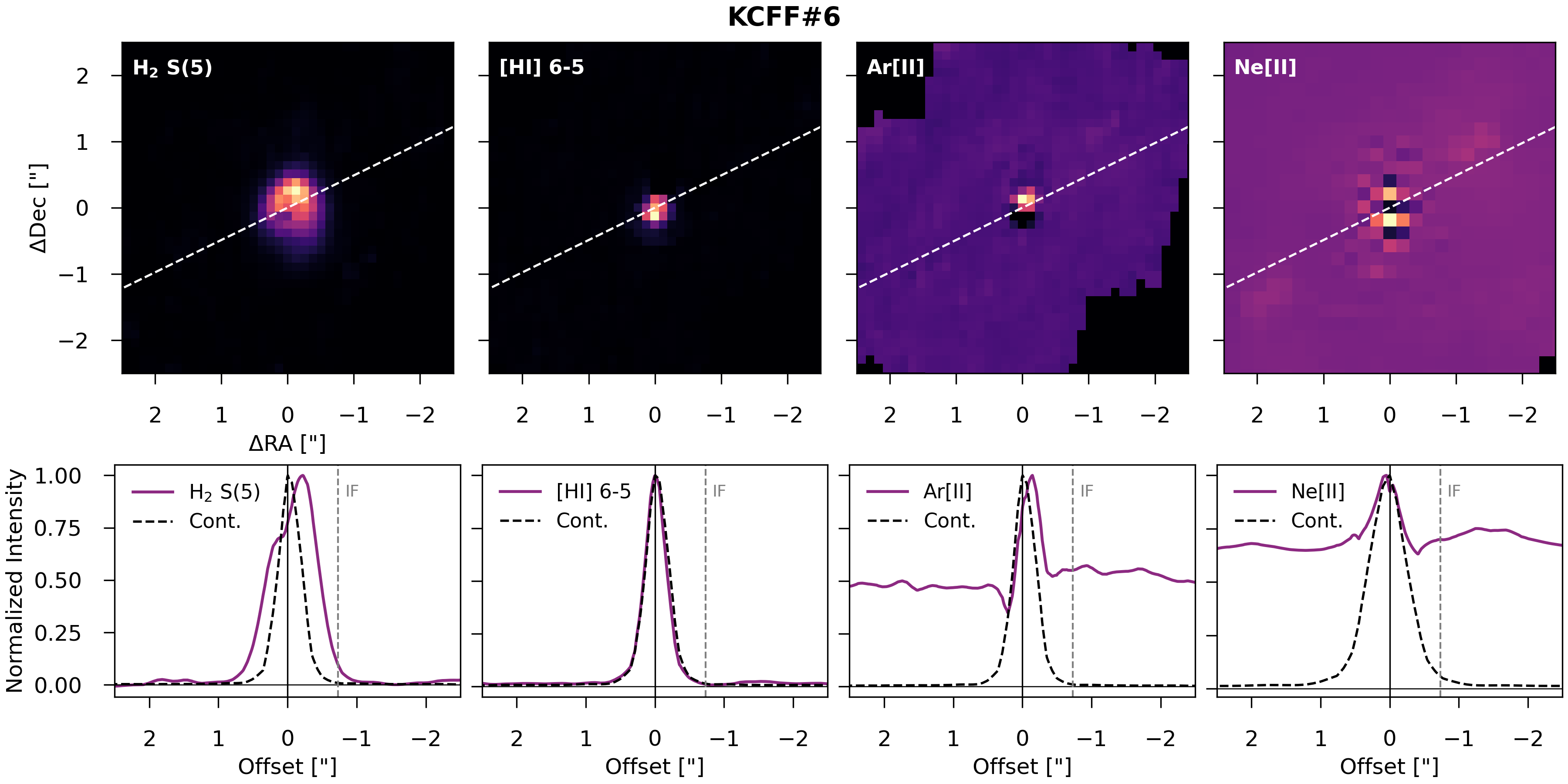}
\caption{Same as Figure~\ref{fig:disk2_slices} for KCFF\#6.}
\label{fig:disk6_slices}
\end{figure*}

\newpage
\subsection{\ce{H_2} spectra and model fits}

\begin{figure*}[h!]
    \centering
    \includegraphics[width=0.78\hsize]{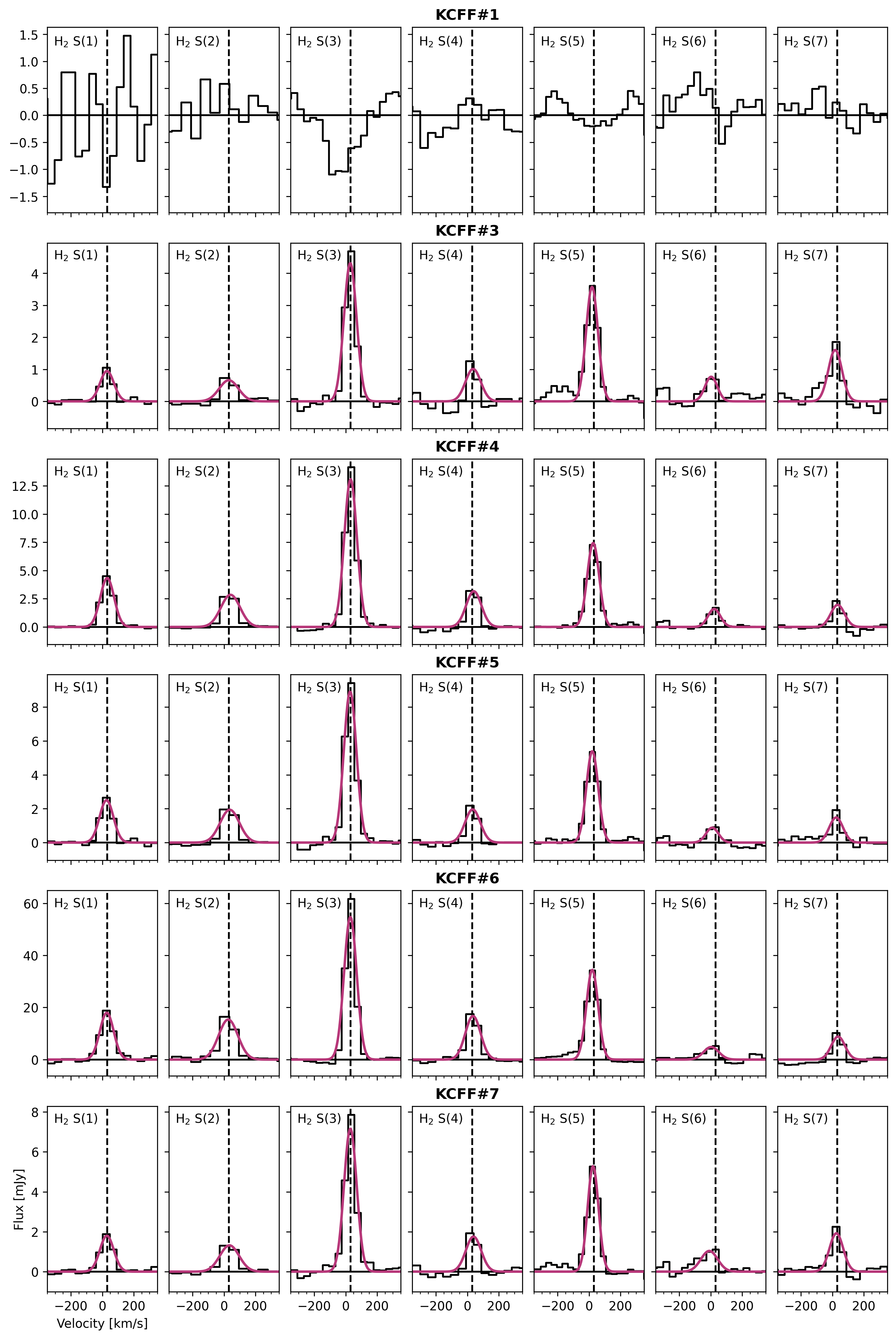}
    \caption{Same as Figure~\ref{fig:disk_2_spectra} for KCFF\#1,3,4,5,6 and 7.}
    \label{fig:spectra_H2}
\end{figure*}

\newpage
\subsection{\ce{HI} spectra and model fits}

\begin{figure*}[h!]
    \centering
    \includegraphics[width=0.6\hsize,trim={0 0 6cm 0},clip]{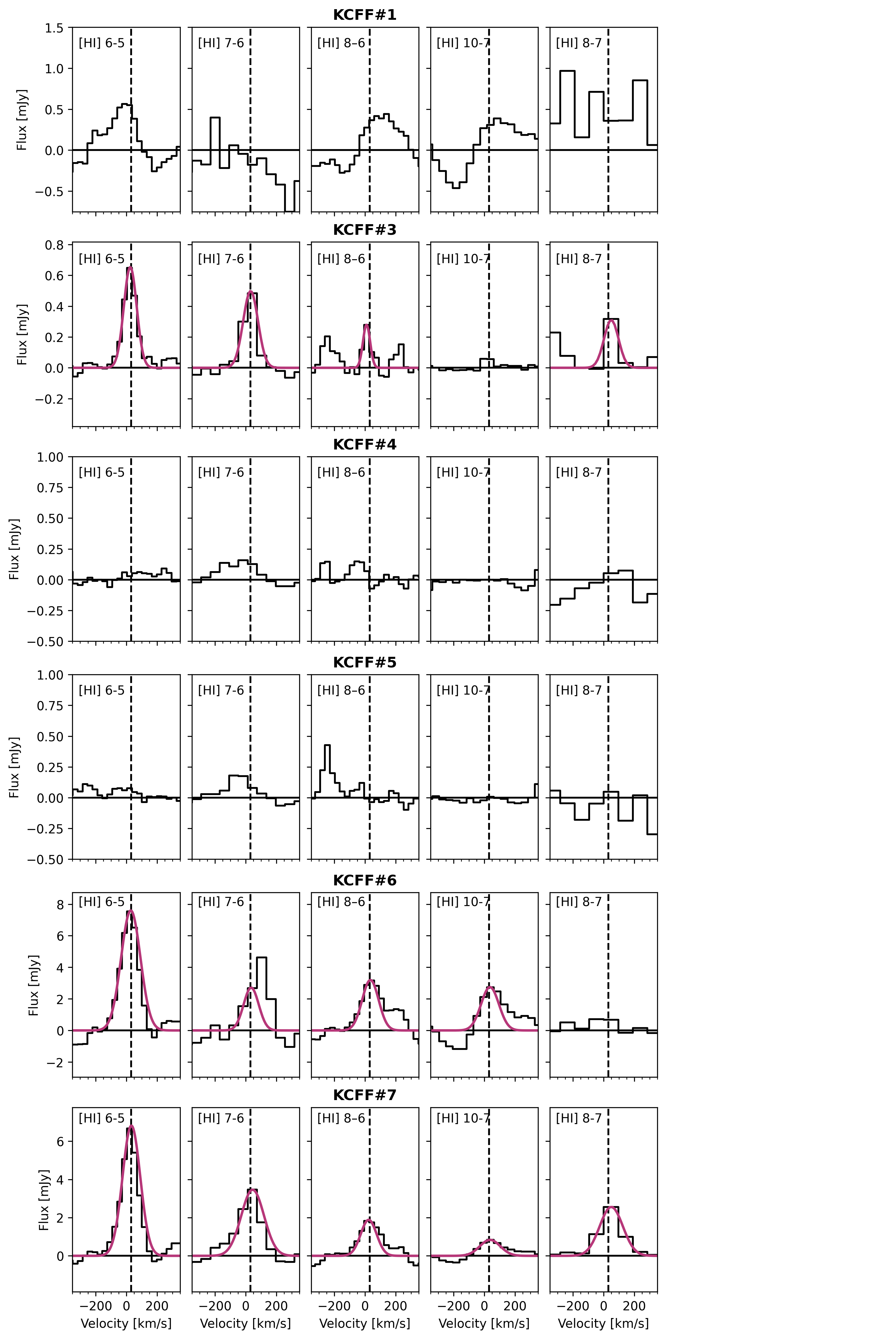}
    \caption{Same as Figure~\ref{fig:disk2_HI_spectra} for KCFF\#1,3,4,5,6 and 7.}
    \label{fig:spectra_HI}
\end{figure*}

\newpage
\subsection{Forbidden emission line spectra and model fits}

\begin{figure*}[h!]
\centering
    \includegraphics[width=0.65\hsize]{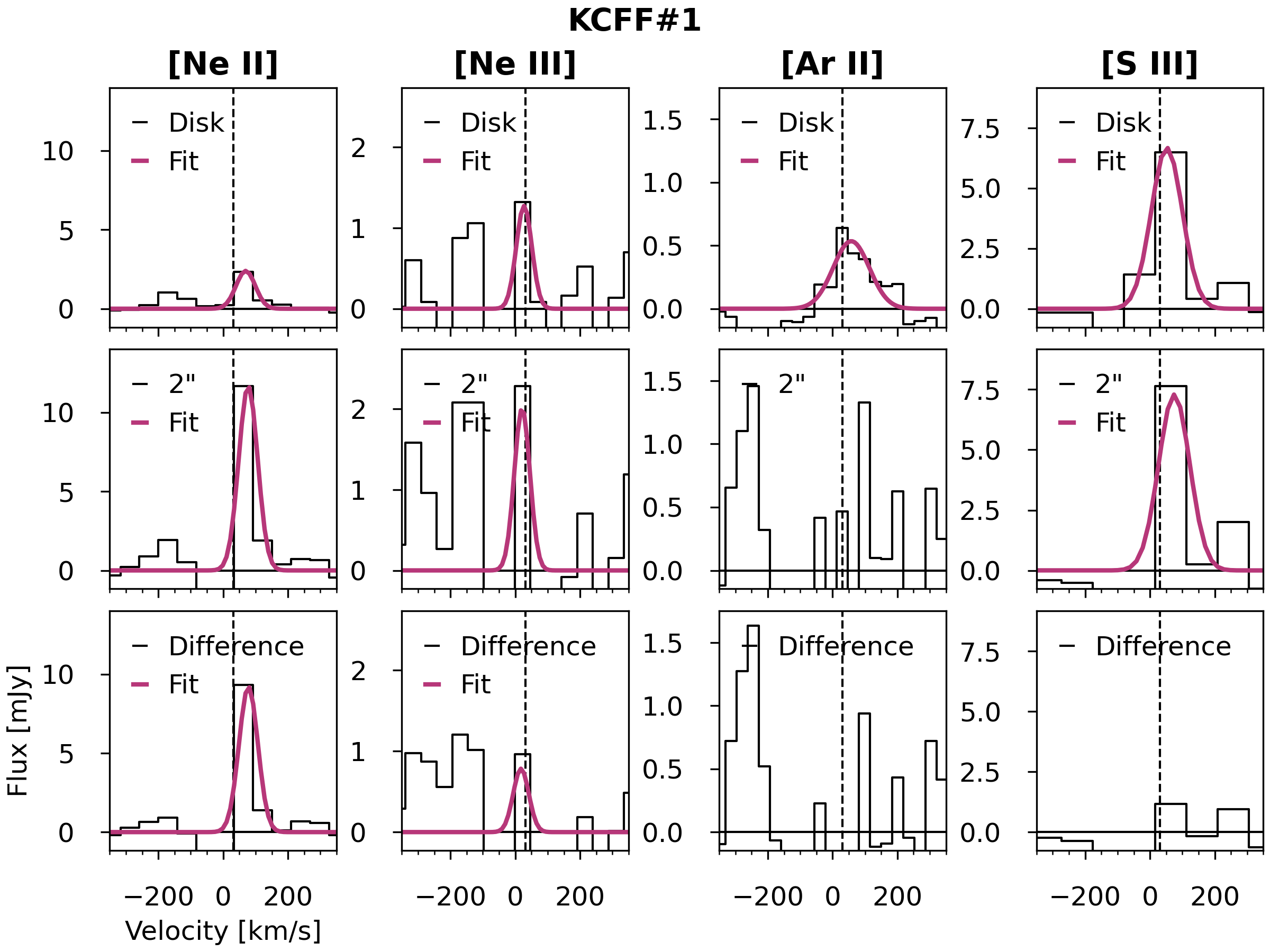}
    \caption{Spectra (black) extracted from a disk and 2\farcs aperture towards KCFF\#1 of forbidden emission lines and corresponding gaussian fits (purple). The x-axis is velocity relative to the rest wavelength of the line and the dashed vertical line highlights the approximate heliocentric velocity of 31~km~$\mathrm{s^{-1}}$.}
    \label{fig:disk1_forbidden}
\end{figure*}

\begin{figure*}[h!]
\centering
    \includegraphics[width=0.65\hsize]{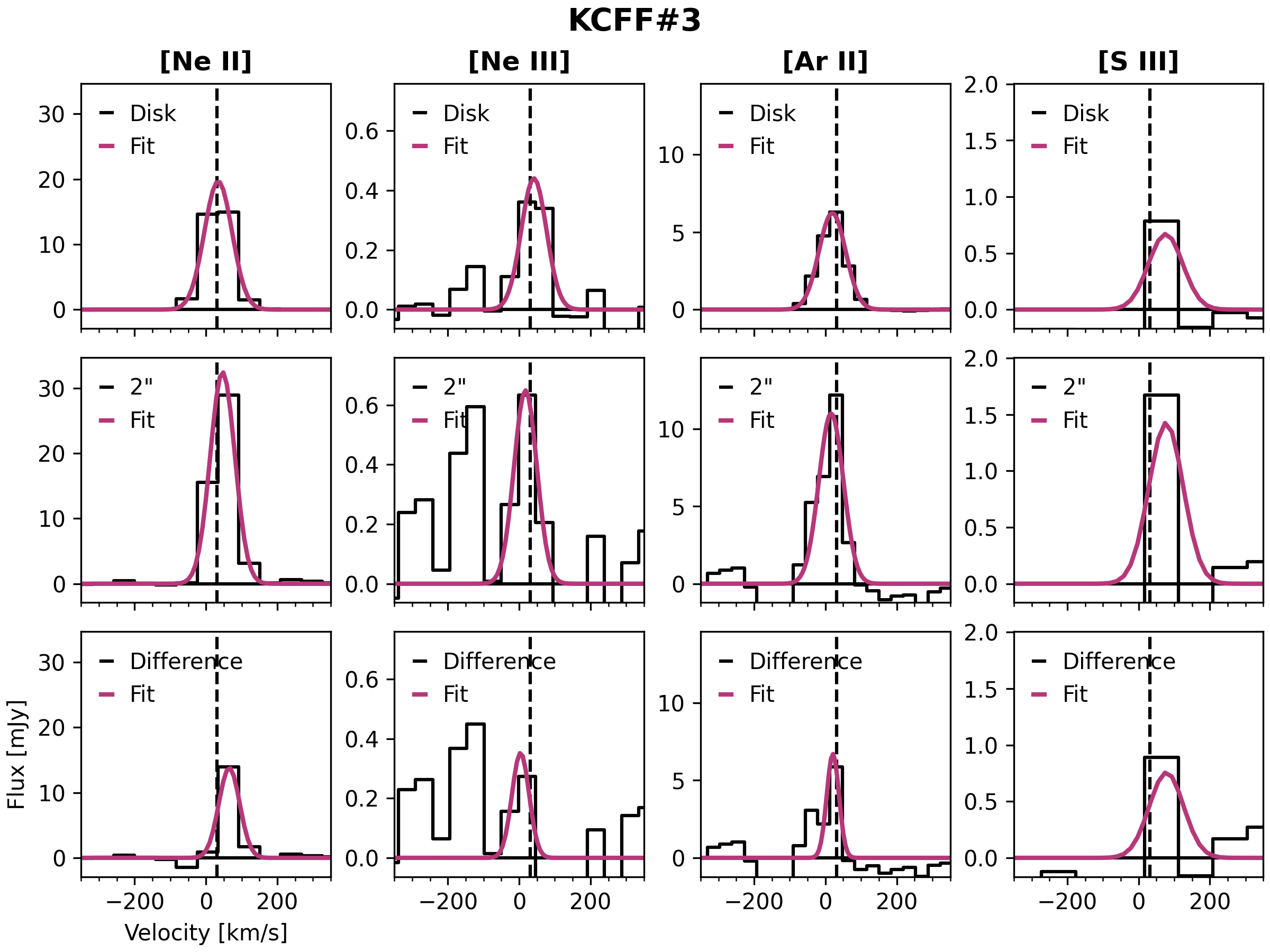}
    \caption{Spectra (black) extracted from a disk and 2\farcs aperture towards KCFF\#3 of forbidden emission lines and corresponding gaussian fits (purple). The x-axis is velocity relative to the rest wavelength of the line and the dashed vertical line highlights the approximate heliocentric velocity of 31~km~$\mathrm{s^{-1}}$.}
    \label{fig:disk3_forbidden}
\end{figure*}

\begin{figure*}[h!]
\centering
    \includegraphics[width=0.65\hsize]{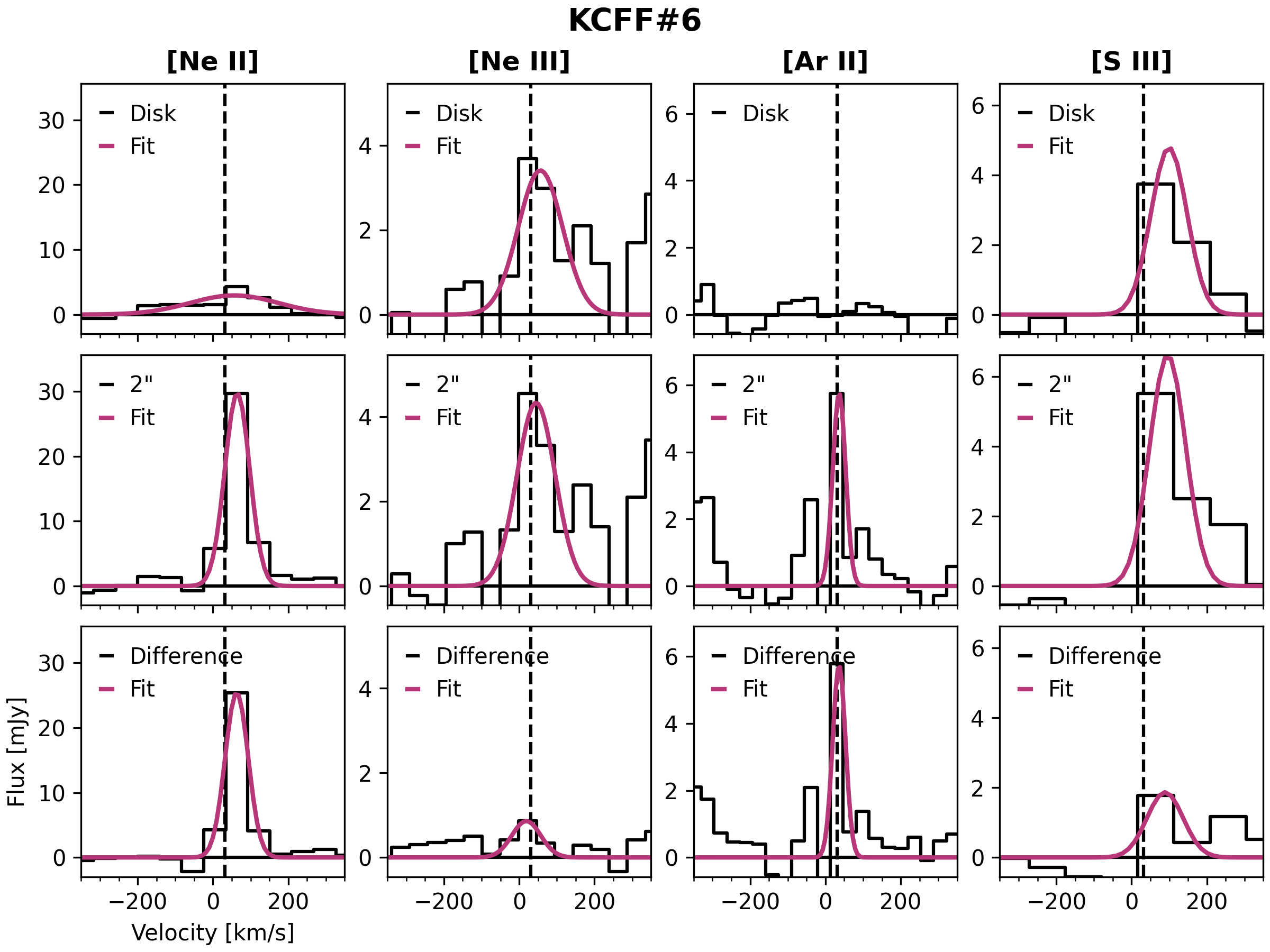}
    \caption{Spectra (black) extracted from a disk and 2\farcs aperture towards KCFF\#6 of forbidden emission lines and corresponding gaussian fits (purple). The x-axis is velocity relative to the rest wavelength of the line and the dashed vertical line highlights the approximate heliocentric velocity of 31~km~$\mathrm{s^{-1}}$.}
    \label{fig:disk6_forbidden}
\end{figure*}

\begin{figure*}[h!]
\centering
    \includegraphics[width=0.65\hsize]{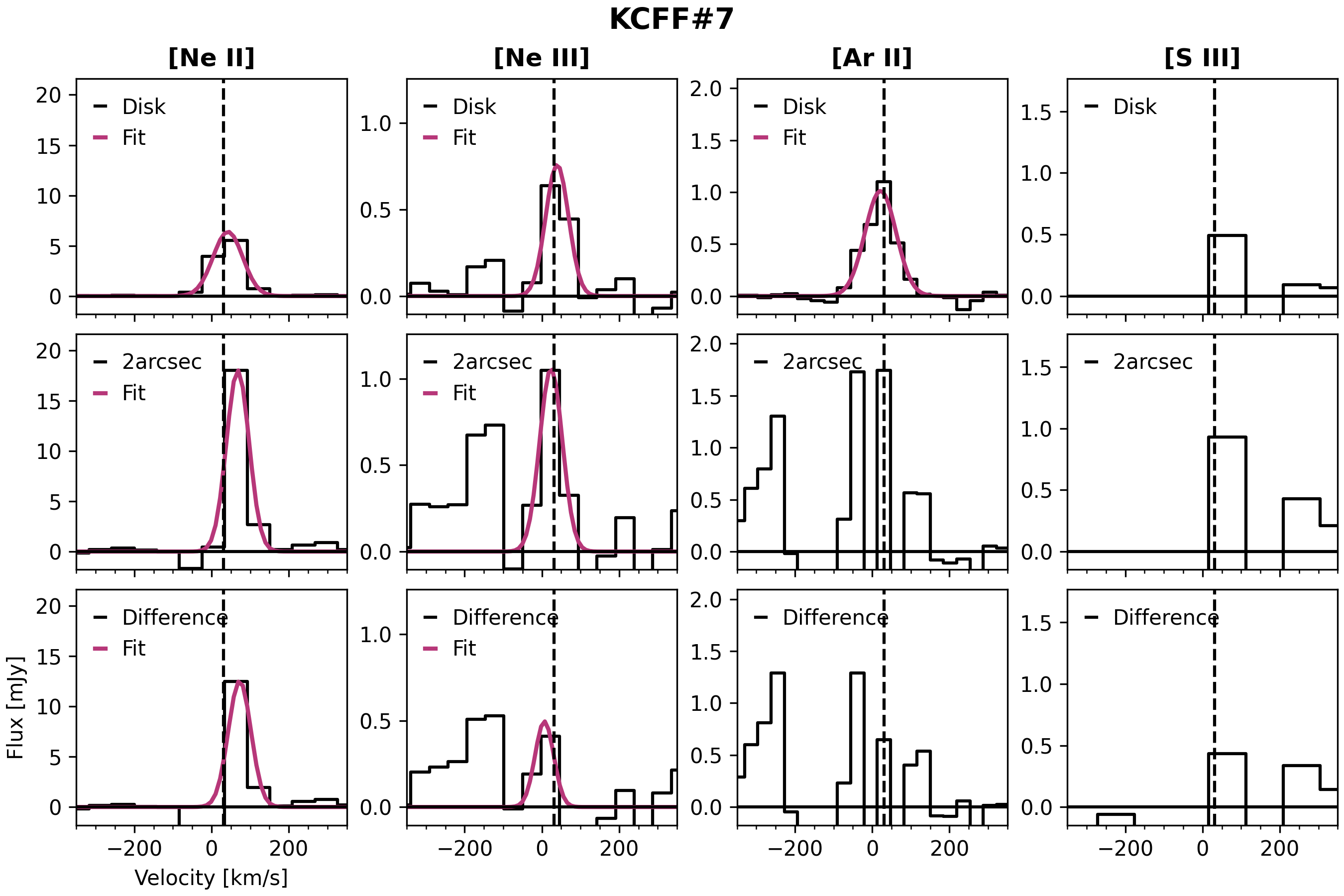}
    \caption{Spectra (black) extracted from a disk and 2\farcs aperture towards KCFF\#7 of forbidden emission lines and corresponding gaussian fits (purple). The x-axis is velocity relative to the rest wavelength of the line and the dashed vertical line highlights the approximate heliocentric velocity of 31~km~$\mathrm{s^{-1}}$.}
    \label{fig:disk7_forbidden}
\end{figure*}

\end{appendix}

\clearpage
\newpage
\bibliography{sample701}{}

@ARTICLE{1987ApJ...314..535G,
       author = {{Garay}, Guido and {Moran}, James M. and {Reid}, Mark J.},
        title = "{Compact Continuum Radio Sources in the Orion Nebula}",
      journal = {\apj},
         year = 1987,
        month = mar,
       volume = {314},
        pages = {535},
          doi = {10.1086/165084},
       adsurl = {https://ui.adsabs.harvard.edu/abs/1987ApJ...314..535G}
}

@ARTICLE{2025A&A...695A..74A,
       author = {{Anania}, Rossella and {Winter}, Andrew J. and {Rosotti}, Giovanni and {Vioque}, Miguel and {Zari}, Eleonora and {Pantaleoni Gonz{\'a}lez}, Michelangelo and {Testi}, Leonardo},
        title = "{A novel method for estimating the far-ultraviolet flux, and a catalogue for disc-hosting stars in nearby star-forming regions}",
      journal = {\aap},
         year = 2025,
        month = mar,
       volume = {695},
          eid = {A74},
        pages = {A74},
          doi = {10.1051/0004-6361/202453011},
archivePrefix = {arXiv},
       eprint = {2501.18752},
 primaryClass = {astro-ph.EP},
       adsurl = {https://ui.adsabs.harvard.edu/abs/2025A&A...695A..74A}
}

@ARTICLE{2025A&A...698A.172R,
       author = {{Rogers}, Ciar{\'a}n and {de Marchi}, Guido and {Brandl}, Bernhard},
        title = "{Externally irradiated young stars in NGC 3603: A JWST NIRSpec catalogue of pre-main-sequence stars in a massive star formation region}",
      journal = {\aap},
         year = 2025,
        month = jun,
       volume = {698},
          eid = {A172},
        pages = {A172},
          doi = {10.1051/0004-6361/202453358},
archivePrefix = {arXiv},
       eprint = {2412.05650},
 primaryClass = {astro-ph.SR},
       adsurl = {https://ui.adsabs.harvard.edu/abs/2025A&A...698A.172R}
}

@ARTICLE{2012ApJ...756..137B,
       author = {{Bally}, John and {Youngblood}, Allison and {Ginsburg}, Adam},
        title = "{The Spindle: An Irradiated Disk and Bent Protostellar Jet in Orion}",
      journal = {\apj},
         year = 2012,
        month = sep,
       volume = {756},
       number = {2},
          eid = {137},
        pages = {137},
          doi = {10.1088/0004-637X/756/2/137},
       adsurl = {https://ui.adsabs.harvard.edu/abs/2012ApJ...756..137B}
}

@ARTICLE{2026arXiv260507016N,
       author = {{Narang}, Mayank and {Pontoppidan}, Klaus M. and {Salyk}, Colette and {Arulanantham}, Nicole and {Blake}, Geoffrey A. and {Banzatti}, Andrea and {Najita}, Joan and {Pascucci}, Ilaria and {Huang}, Jane and {Krijt}, Sebastiaan and {Oberg}, Karin and {Rosotti}, Giovanni and {Kaeufer}, Till and {Dahl}, Emma and {Cleeves}, L. Ilsedore and {Zhang}, Ke and {Green}, Joel},
        title = "{Characterizing the Extended Molecular Hydrogen Winds in Protoplanetary Disks from the JWST Disk Infrared Spectroscopic Chemistry Survey}",
      journal = {arXiv e-prints},
         year = 2026,
        month = may,
          eid = {arXiv:2605.07016},
        pages = {arXiv:2605.07016},
archivePrefix = {arXiv},
       eprint = {2605.07016},
 primaryClass = {astro-ph.SR},
       adsurl = {https://ui.adsabs.harvard.edu/abs/2026arXiv260507016N}
}

@ARTICLE{2014ApJ...795....1P,
       author = {{Pascucci}, I. and {Ricci}, L. and {Gorti}, U. and {Hollenbach}, D. and {Hendler}, N.~P. and {Brooks}, K.~J. and {Contreras}, Y.},
        title = "{Low Extreme-ultraviolet Luminosities Impinging on Protoplanetary Disks}",
      journal = {\apj},
         year = 2014,
        month = nov,
       volume = {795},
       number = {1},
          eid = {1},
        pages = {1},
          doi = {10.1088/0004-637X/795/1/1},
archivePrefix = {arXiv},
       eprint = {1407.1574},
 primaryClass = {astro-ph.SR},
       adsurl = {https://ui.adsabs.harvard.edu/abs/2014ApJ...795....1P}
}

@ARTICLE{2012ApJ...759...47S,
       author = {{Szul{\'a}gyi}, Judit and {Pascucci}, Ilaria and {{\'A}brah{\'a}m}, P{\'e}ter and {Apai}, D{\'a}niel and {Bouwman}, Jeroen and {Mo{\'o}r}, Attila},
        title = "{Observational Constraints on the Stellar Radiation Field Impinging on Transitional Disk Atmospheres}",
      journal = {\apj},
         year = 2012,
        month = nov,
       volume = {759},
       number = {1},
          eid = {47},
        pages = {47},
          doi = {10.1088/0004-637X/759/1/47},
archivePrefix = {arXiv},
       eprint = {1209.3860},
 primaryClass = {astro-ph.EP},
       adsurl = {https://ui.adsabs.harvard.edu/abs/2012ApJ...759...47S}
}

@ARTICLE{2025ApJ...979..120H,
       author = {{Hallatt}, Tim and {Lee}, Eve J.},
        title = "{On the Formation of Planets in the Milky Way's Thick Disk}",
      journal = {\apj},
         year = 2025,
        month = feb,
       volume = {979},
       number = {2},
          eid = {120},
        pages = {120},
          doi = {10.3847/1538-4357/ad9aa1},
archivePrefix = {arXiv},
       eprint = {2408.09319},
 primaryClass = {astro-ph.EP},
       adsurl = {https://ui.adsabs.harvard.edu/abs/2025ApJ...979..120H}
}

@ARTICLE{2024A&A...689A.338H,
       author = {{Huang}, Shuo and {Portegies Zwart}, Simon and {Wilhelm}, Maite J.~C.},
        title = "{Suppression of giant planet formation around low-mass stars in clustered environments}",
      journal = {\aap},
         year = 2024,
        month = sep,
       volume = {689},
          eid = {A338},
        pages = {A338},
          doi = {10.1051/0004-6361/202451051},
archivePrefix = {arXiv},
       eprint = {2407.19018},
 primaryClass = {astro-ph.EP},
       adsurl = {https://ui.adsabs.harvard.edu/abs/2024A&A...689A.338H}
}

@ARTICLE{2022MNRAS.515.4287W,
       author = {{Winter}, Andrew J. and {Haworth}, Thomas J. and {Coleman}, Gavin A.~L. and {Nayakshin}, Sergei},
        title = "{The growth and migration of massive planets under the influence of external photoevaporation}",
      journal = {\mnras},
         year = 2022,
        month = sep,
       volume = {515},
       number = {3},
        pages = {4287-4301},
          doi = {10.1093/mnras/stac1564},
archivePrefix = {arXiv},
       eprint = {2206.02818},
 primaryClass = {astro-ph.EP},
       adsurl = {https://ui.adsabs.harvard.edu/abs/2022MNRAS.515.4287W}
}

@ARTICLE{2023MNRAS.522.1939Q,
       author = {{Qiao}, Lin and {Coleman}, Gavin A.~L. and {Haworth}, Thomas J.},
        title = "{Planet formation via pebble accretion in externally photoevaporating discs}",
      journal = {\mnras},
         year = 2023,
        month = jun,
       volume = {522},
       number = {2},
        pages = {1939-1950},
          doi = {10.1093/mnras/stad944},
archivePrefix = {arXiv},
       eprint = {2303.15177},
 primaryClass = {astro-ph.EP},
       adsurl = {https://ui.adsabs.harvard.edu/abs/2023MNRAS.522.1939Q}
}

@ARTICLE{2023A&A...673L...2V,
       author = {{van Terwisga}, S.~E. and {Hacar}, A.},
        title = "{Survey of Orion Disks with ALMA (SODA). II. UV-driven disk mass loss in L1641 and L1647}",
      journal = {\aap},
         year = 2023,
        month = may,
       volume = {673},
          eid = {L2},
        pages = {L2},
          doi = {10.1051/0004-6361/202346135},
archivePrefix = {arXiv},
       eprint = {2304.05777},
 primaryClass = {astro-ph.EP},
       adsurl = {https://ui.adsabs.harvard.edu/abs/2023A&A...673L...2V}
}

@ARTICLE{2025AJ....169..296B,
       author = {{Bajaj}, Naman S. and {Pascucci}, Ilaria and {Beck}, Tracy L. and {Edwards}, Suzan and {Cabrit}, Sylvie and {Najita}, Joan R. and {Schwarz}, Kamber and {Semenov}, Dmitry and {Salyk}, Colette and {Gorti}, Uma and {Brittain}, Sean D. and {Krijt}, Sebastiaan and {Ruaud}, Maxime and {Page}, James Muzerolle},
        title = "{Class I/II Jets with JWST: Mass-loss Rates, Asymmetries, and Binary-induced Wigglings}",
      journal = {\aj},
         year = 2025,
        month = jun,
       volume = {169},
       number = {6},
          eid = {296},
        pages = {296},
          doi = {10.3847/1538-3881/adc73c},
archivePrefix = {arXiv},
       eprint = {2503.23319},
 primaryClass = {astro-ph.EP},
       adsurl = {https://ui.adsabs.harvard.edu/abs/2025AJ....169..296B}
}

@ARTICLE{Hohle_42Ori_2010,
       author = {{Hohle}, M.~M. and {Neuh{\"a}user}, R. and {Schutz}, B.~F.},
        title = "{Masses and luminosities of O- and B-type stars and red supergiants}",
      journal = {Astronomische Nachrichten},
         year = 2010,
        month = apr,
       volume = {331},
       number = {4},
        pages = {349},
          doi = {10.1002/asna.200911355},
archivePrefix = {arXiv},
       eprint = {1003.2335},
 primaryClass = {astro-ph.SR},
       adsurl = {https://ui.adsabs.harvard.edu/abs/2010AN....331..349H}
}

@ARTICLE{Dotter_MIST_2016,
       author = {{Dotter}, Aaron},
        title = "{MESA Isochrones and Stellar Tracks (MIST) 0: Methods for the Construction of Stellar Isochrones}",
      journal = {\apjs},
         year = 2016,
        month = jan,
       volume = {222},
       number = {1},
          eid = {8},
        pages = {8},
          doi = {10.3847/0067-0049/222/1/8},
archivePrefix = {arXiv},
       eprint = {1601.05144},
 primaryClass = {astro-ph.SR},
       adsurl = {https://ui.adsabs.harvard.edu/abs/2016ApJS..222....8D}
}

@ARTICLE{Choi_MIST_2016,
       author = {{Choi}, Jieun and {Dotter}, Aaron and {Conroy}, Charlie and {Cantiello}, Matteo and {Paxton}, Bill and {Johnson}, Benjamin D.},
        title = "{Mesa Isochrones and Stellar Tracks (MIST). I. Solar-scaled Models}",
      journal = {\apj},
         year = 2016,
        month = jun,
       volume = {823},
       number = {2},
          eid = {102},
        pages = {102},
          doi = {10.3847/0004-637X/823/2/102},
archivePrefix = {arXiv},
       eprint = {1604.08592},
 primaryClass = {astro-ph.SR},
       adsurl = {https://ui.adsabs.harvard.edu/abs/2016ApJ...823..102C}
}

@ARTICLE{Castelli_ATLAS9_2004,
       author = {{Castelli}, F. and {Kurucz}, R.~L.},
        title = "{Is missing Fe I opacity in stellar atmospheres a significant problem?}",
      journal = {\aap},
         year = 2004,
        month = may,
       volume = {419},
        pages = {725-733},
          doi = {10.1051/0004-6361:20040079},
       adsurl = {https://ui.adsabs.harvard.edu/abs/2004A&A...419..725C}
}

@ARTICLE{2025A&A...699A.194A,
       author = {{Arabhavi}, A.~M. and {Kamp}, I. and {Henning}, Th. and {van Dishoeck}, E.~F. and {Jang}, H. and {Waters}, L.~B.~F.~M. and {Christiaens}, V. and {Gasman}, D. and {Pascucci}, I. and {Perotti}, G. and {Grant}, S.~L. and {G{\"u}del}, M. and {Lagage}, P.-O. and {Barrado}, D. and {Caratti o Garatti}, A. and {Lahuis}, F. and {Kaeufer}, T. and {Kanwar}, J. and {Morales-Calder{\'o}n}, M. and {Schwarz}, K. and {Sellek}, A.~D. and {Tabone}, B. and {Temmink}, M. and {Vlasblom}, M. and {Patapis}, P.},
        title = "{MINDS: The very low-mass star and brown dwarf sample: Detections and trends in the inner disk gas}",
      journal = {\aap},
         year = 2025,
        month = jul,
       volume = {699},
          eid = {A194},
        pages = {A194},
          doi = {10.1051/0004-6361/202554109},
archivePrefix = {arXiv},
       eprint = {2506.02748},
 primaryClass = {astro-ph.EP},
       adsurl = {https://ui.adsabs.harvard.edu/abs/2025A&A...699A.194A}
}

@ARTICLE{2024ApJ...962L..16N,
       author = {{Narang}, Mayank and {Manoj}, P. and {Tyagi}, Himanshu and {Watson}, Dan M. and {Megeath}, S. Thomas and {Federman}, Samuel and {Rubinstein}, Adam E. and {Gutermuth}, Robert and {Caratti o Garatti}, Alessio and {Beuther}, Henrik and {Bourke}, Tyler L. and {Van Dishoeck}, Ewine F. and {Evans}, Neal J. and {Anglada}, Guillem and {Osorio}, Mayra and {Stanke}, Thomas and {Muzerolle}, James and {Looney}, Leslie W. and {Yang}, Yao-Lun and {Klaassen}, Pamela and {Karnath}, Nicole and {Atnagulov}, Prabhani and {Brunken}, Nashanty and {Fischer}, William J. and {Furlan}, Elise and {Green}, Joel and {Habel}, Nolan and {Hartmann}, Lee and {Linz}, Hendrik and {Nazari}, Pooneh and {Pokhrel}, Riwaj and {Rahatgaonkar}, Rohan and {Rocha}, Will R.~M. and {Sheehan}, Patrick and {Slavicinska}, Katerina and {Stutz}, Amelia M. and {Tobin}, John J. and {Tychoniec}, Lukasz and {Wolk}, Scott},
        title = "{Discovery of a Collimated Jet from the Low-luminosity Protostar IRAS 16253{\ensuremath{-}}2429 in a Quiescent Accretion Phase with the JWST}",
      journal = {\apjl},
         year = 2024,
        month = feb,
       volume = {962},
       number = {1},
          eid = {L16},
        pages = {L16},
          doi = {10.3847/2041-8213/ad1de3},
archivePrefix = {arXiv},
       eprint = {2310.14061},
 primaryClass = {astro-ph.SR},
       adsurl = {https://ui.adsabs.harvard.edu/abs/2024ApJ...962L..16N}
}

@ARTICLE{2025ApJ...983..110T,
       author = {{Tyagi}, Himanshu and {Manoj}, P. and {Narang}, Mayank and {Megeath}, S. Thomas and {Rocha}, Will R.~M. and {Brunken}, Nashanty and {Rubinstein}, Adam E. and {Gutermuth}, Robert and {Evans}, Neal J. and {Van Dishoeck}, Ewine F. and {Federman}, Samuel and {Watson}, Dan M. and {Neufeld}, David A. and {Anglada}, Guillem and {Beuther}, Henrik and {Caratti o Garatti}, Alessio and {Looney}, Leslie W. and {Nazari}, Pooneh and {Osorio}, Mayra and {Stanke}, Thomas and {Yang}, Yao-Lun and {Bourke}, Tyler L. and {Fischer}, William J. and {Furlan}, Elise and {Green}, Joel and {Habel}, Nolan and {Klaassen}, Pamela and {Karnath}, Nicole and {Linz}, Hendrik and {Muzerolle}, James and {Tobin}, John J. and {Atnagulov}, Prabhani and {Rahatgaonkar}, Rohan and {Sheehan}, Patrick and {Slavicinska}, Katerina and {Stutz}, Amelia M. and {Tychoniec}, Lukasz and {Wolk}, Scott and {Zakri}, Wafa},
        title = "{JWST-IPA: Chemical Inventory and Spatial Mapping of Ices in the Protostar HOPS 370{\textemdash}Evidence for an Opacity Hole and Thermal Processing of Ices}",
      journal = {\apj},
         year = 2025,
        month = apr,
       volume = {983},
       number = {2},
          eid = {110},
        pages = {110},
          doi = {10.3847/1538-4357/adb71f},
archivePrefix = {arXiv},
       eprint = {2410.06697},
 primaryClass = {astro-ph.SR},
       adsurl = {https://ui.adsabs.harvard.edu/abs/2025ApJ...983..110T}
}

@misc{2022zndo...7041998B,
       author = {{Bushouse}, Howard and {Eisenhamer}, Jonathan and ...},
        title = "{JWST Calibration Pipeline}",
         year = 2022,
        month = aug,
          eid = {10.5281/zenodo.7041998},
          doi = {10.5281/zenodo.7041998},
       adsurl = {https://ui.adsabs.harvard.edu/abs/2022zndo...7041998B}
}

@ARTICLE{2022MNRAS.512.3788Q,
       author = {{Qiao}, Lin and {Haworth}, Thomas J. and {Sellek}, Andrew D. and {Ali}, Ahmad A.},
        title = "{The evolution of protoplanetary discs in star formation and feedback simulations}",
      journal = {\mnras},
         year = 2022,
        month = may,
       volume = {512},
       number = {3},
        pages = {3788-3805},
          doi = {10.1093/mnras/stac684},
archivePrefix = {arXiv},
       eprint = {2203.04230},
 primaryClass = {astro-ph.EP},
       adsurl = {https://ui.adsabs.harvard.edu/abs/2022MNRAS.512.3788Q}
}

@ARTICLE{2023MNRAS.520.5331W,
       author = {{Wilhelm}, Maite J.~C. and {Portegies Zwart}, Simon and {Cournoyer-Cloutier}, Claude and {Lewis}, Sean C. and {Polak}, Brooke and {Tran}, Aaron and {Mac Low}, Mordecai-Mark},
        title = "{Radiation shielding of protoplanetary discs in young star-forming regions}",
      journal = {\mnras},
         year = 2023,
        month = apr,
       volume = {520},
       number = {4},
        pages = {5331-5353},
          doi = {10.1093/mnras/stad445},
archivePrefix = {arXiv},
       eprint = {2302.03721},
 primaryClass = {astro-ph.EP},
       adsurl = {https://ui.adsabs.harvard.edu/abs/2023MNRAS.520.5331W}
}

@INPROCEEDINGS{2023ASPC..534..567P,
       author = {{Pascucci}, I. and {Cabrit}, S. and {Edwards}, S. and {Gorti}, U. and {Gressel}, O. and {Suzuki}, T.~K.},
        title = "{The Role of Disk Winds in the Evolution and Dispersal of Protoplanetary Disks}",
    booktitle = {Protostars and Planets VII},
         year = 2023,
       editor = {{Inutsuka}, S. and {Aikawa}, Y. and {Muto}, T. and {Tomida}, K. and {Tamura}, M.},
       series = {Astronomical Society of the Pacific Conference Series},
       volume = {534},
        month = jul,
        pages = {567},
          doi = {10.48550/arXiv.2203.10068},
archivePrefix = {arXiv},
       eprint = {2203.10068},
 primaryClass = {astro-ph.EP},
       adsurl = {https://ui.adsabs.harvard.edu/abs/2023ASPC..534..567P}
}

@ARTICLE{2024A&A...687A..93A,
       author = {{Aru}, M.-L. and {Mauc{\'o}}, K. and {Manara}, C.~F. and {Haworth}, T.~J. and {Facchini}, S. and {McLeod}, A.~F. and {Miotello}, A. and {Petr-Gotzens}, M.~G. and {Robberto}, M. and {Rosotti}, G.~P. and {Vicente}, S. and {Winter}, A. and {Ansdell}, M.},
        title = "{Kaleidoscope of irradiated disks: MUSE observations of proplyds in the Orion Nebula Cluster. I. Sample presentation and ionization front sizes<xref rid=``FN2'' ref-type=``fn''/>}",
      journal = {\aap},
         year = 2024,
        month = jul,
       volume = {687},
          eid = {A93},
        pages = {A93},
          doi = {10.1051/0004-6361/202349004},
archivePrefix = {arXiv},
       eprint = {2403.12604},
 primaryClass = {astro-ph.SR},
       adsurl = {https://ui.adsabs.harvard.edu/abs/2024A&A...687A..93A}
}

@ARTICLE{2024ApJ...965L..13A,
       author = {{Arulanantham}, Nicole and {McClure}, M.~K. and {Pontoppidan}, Klaus and {Beck}, Tracy L. and {Sturm}, J.~A. and {Harsono}, D. and {Boogert}, A.~C.~A. and {Cordiner}, M. and {Dartois}, E. and {Drozdovskaya}, M.~N. and {Espaillat}, C. and {Melnick}, G.~J. and {Noble}, J.~A. and {Palumbo}, M.~E. and {Pendleton}, Y.~J. and {Terada}, H. and {van Dishoeck}, E.~F.},
        title = "{JWST MIRI MRS Images of Disk Winds, Water, and CO in an Edge-on Protoplanetary Disk}",
      journal = {\apjl},
         year = 2024,
        month = apr,
       volume = {965},
       number = {1},
          eid = {L13},
        pages = {L13},
          doi = {10.3847/2041-8213/ad35c9},
archivePrefix = {arXiv},
       eprint = {2402.12256},
 primaryClass = {astro-ph.SR},
       adsurl = {https://ui.adsabs.harvard.edu/abs/2024ApJ...965L..13A}
}

@ARTICLE{2023PASP..135d8003W,
       author = {{Wright}, Gillian S. and {Rieke}, George H. and {Glasse}, Alistair and {Ressler}, Michael and {Garc{\'\i}a Mar{\'\i}n}, Macarena and {Aguilar}, Jonathan and {Alberts}, Stacey and {{\'A}lvarez-M{\'a}rquez}, Javier and {Argyriou}, Ioannis and {Banks}, Kimberly and {Baudoz}, Pierre and {Boccaletti}, Anthony and {Bouchet}, Patrice and {Bouwman}, Jeroen and {Brandl}, Bernard R. and {Breda}, David and {Bright}, Stacey and {Cale}, Steven and {Colina}, Luis and {Cossou}, Christophe and {Coulais}, Alain and {Cracraft}, Misty and {De Meester}, Wim and {Dicken}, Daniel and {Engesser}, Michael and {Etxaluze}, Mireya and {Fox}, Ori D. and {Friedman}, Scott and {Fu}, Henry and {Gasman}, Danny and {G{\'a}sp{\'a}r}, Andr{\'a}s and {Gastaud}, Ren{\'e} and {Geers}, Vincent and {Glauser}, Adrian Michael and {Gordon}, Karl D. and {Greene}, Thomas and {Greve}, Thomas R. and {Grundy}, Timothy and {G{\"u}del}, Manuel and {Guillard}, Pierre and {Haderlein}, Peter and {Hashimoto}, Ryan and {Henning}, Thomas and {Hines}, Dean and {Holler}, Bryan and {Detre}, {\"O}rs Hunor and {Jahromi}, Amir and {James}, Bryan and {Jones}, Olivia C. and {Justtanont}, Kay and {Kavanagh}, Patrick and {Kendrew}, Sarah and {Klaassen}, Pamela and {Krause}, Oliver and {Labiano}, Alvaro and {Lagage}, Pierre-Olivier and {Lambros}, Scott and {Larson}, Kirsten and {Law}, David and {Lee}, David and {Libralato}, Mattia and {Lorenzo Alverez}, Jose and {Meixner}, Margaret and {Morrison}, Jane and {Mueller}, Migo and {Murray}, Katherine and {Mycroft}, Matthew and {Myers}, Richard and {Nayak}, Omnarayani and {Naylor}, Bret and {Nickson}, Bryony and {Noriega-Crespo}, Alberto and {{\"O}stlin}, G{\"o}ran and {O'Sullivan}, Brian and {Ottens}, Richard and {Patapis}, Polychronis and {Penanen}, Konstantin and {Pietraszkiewicz}, Martin and {Ray}, Tom and {Regan}, Michael and {Roteliuk}, Anthony and {Royer}, Pierre and {Samara-Ratna}, Piyal and {Samuelson}, Bridget and {Sargent}, Beth A. and {Scheithauer}, Silvia and {Schneider}, Analyn and {Schreiber}, J{\"u}rgen and {Shaughnessy}, Bryan and {Sheehan}, Evan and {Shivaei}, Irene and {Sloan}, G.~C. and {Tamas}, Laszlo and {Teague}, Kelly and {Temim}, Tea and {Tikkanen}, Tuomo and {Tustain}, Samuel and {van Dishoeck}, Ewine F. and {Vandenbussche}, Bart and {Weilert}, Mark and {Whitehouse}, Paul and {Wolff}, Schuyler},
        title = "{The Mid-infrared Instrument for JWST and Its In-flight Performance}",
      journal = {\pasp},
         year = 2023,
        month = apr,
       volume = {135},
       number = {1046},
          eid = {048003},
        pages = {048003},
          doi = {10.1088/1538-3873/acbe66},
       adsurl = {https://ui.adsabs.harvard.edu/abs/2023PASP..135d8003W}
}

@ARTICLE{1998ApJ...502L..71S,
       author = {{St{\"o}rzer}, H. and {Hollenbach}, D.},
        title = "{On the [O I] {\ensuremath{\lambda}}6300 Line Emission from the Photoevaporating Circumstellar Disks in the Orion Nebula}",
      journal = {\apjl},
         year = 1998,
        month = jul,
       volume = {502},
       number = {1},
        pages = {L71-L74},
          doi = {10.1086/311487},
       adsurl = {https://ui.adsabs.harvard.edu/abs/1998ApJ...502L..71S}
}

@ARTICLE{2026AJ....171...39V,
       author = {{Volz}, M. and {Espaillat}, C.~C. and {Pittman}, C.~V. and {Grant}, S.~L. and {Thanathibodee}, T. and {McClure}, M. and {Tabone}, B. and {Calvet}, N. and {Walter}, F.~M.},
        title = "{JWST Reveals Carbon-rich Chemistry in a Transitional Disk}",
      journal = {\aj},
         year = 2026,
        month = jan,
       volume = {171},
       number = {1},
          eid = {39},
        pages = {39},
          doi = {10.3847/1538-3881/ae1f16},
archivePrefix = {arXiv},
       eprint = {2511.08816},
 primaryClass = {astro-ph.SR},
       adsurl = {https://ui.adsabs.harvard.edu/abs/2026AJ....171...39V}
}

@ARTICLE{2023ApJ...958L...4E,
       author = {{Espaillat}, C.~C. and {Thanathibodee}, T. and {Pittman}, C.~V. and {Sturm}, J.~A. and {McClure}, M.~K. and {Calvet}, N. and {Walter}, F.~M. and {Franco-Hern{\'a}ndez}, R. and {Muzerolle Page}, J.},
        title = "{JWST Detects Neon Line Variability in a Protoplanetary Disk}",
      journal = {\apjl},
         year = 2023,
        month = nov,
       volume = {958},
       number = {1},
          eid = {L4},
        pages = {L4},
          doi = {10.3847/2041-8213/ad023d},
archivePrefix = {arXiv},
       eprint = {2311.07739},
 primaryClass = {astro-ph.SR},
       adsurl = {https://ui.adsabs.harvard.edu/abs/2023ApJ...958L...4E}
}

@ARTICLE{2010ApJ...712..274N,
       author = {{Najita}, Joan R. and {Carr}, John S. and {Strom}, Stephen E. and {Watson}, Dan M. and {Pascucci}, Ilaria and {Hollenbach}, David and {Gorti}, Uma and {Keller}, Luke},
        title = "{Spitzer Spectroscopy of the Transition Object TW Hya}",
      journal = {\apj},
         year = 2010,
        month = mar,
       volume = {712},
       number = {1},
        pages = {274-286},
          doi = {10.1088/0004-637X/712/1/274},
archivePrefix = {arXiv},
       eprint = {1002.4623},
 primaryClass = {astro-ph.SR},
       adsurl = {https://ui.adsabs.harvard.edu/abs/2010ApJ...712..274N}
}

@ARTICLE{2007ApJ...665..492L,
       author = {{Lahuis}, Fred and {van Dishoeck}, Ewine F. and {Blake}, Geoffrey A. and {Evans}, II, Neal J. and {Kessler-Silacci}, Jacqueline E. and {Pontoppidan}, Klaus M.},
        title = "{c2d Spitzer IRS Spectra of Disks around T Tauri Stars. III. [Ne II], [Fe I], and H$_{2}$ Gas-Phase Lines}",
      journal = {\apj},
         year = 2007,
        month = aug,
       volume = {665},
       number = {1},
        pages = {492-511},
          doi = {10.1086/518931},
archivePrefix = {arXiv},
       eprint = {0704.2305},
 primaryClass = {astro-ph},
       adsurl = {https://ui.adsabs.harvard.edu/abs/2007ApJ...665..492L}
}

@ARTICLE{2023ApJ...947....7B,
       author = {{Boyden}, Ryan D. and {Eisner}, Josh A.},
        title = "{Chemical Modeling of Orion Nebula Cluster Disks: Evidence for Massive, Compact Gas Disks with Interstellar Gas-to-dust Ratios}",
      journal = {\apj},
         year = 2023,
        month = apr,
       volume = {947},
       number = {1},
          eid = {7},
        pages = {7},
          doi = {10.3847/1538-4357/acaf77},
archivePrefix = {arXiv},
       eprint = {2212.12325},
 primaryClass = {astro-ph.EP},
       adsurl = {https://ui.adsabs.harvard.edu/abs/2023ApJ...947....7B}
}

@ARTICLE{2009ApJ...703.1203H,
       author = {{Hollenbach}, David and {Gorti}, U.},
        title = "{Diagnostic Line Emission from Extreme Ultraviolet and X-ray-illuminated Disks and Shocks Around Low-mass Stars}",
      journal = {\apj},
         year = 2009,
        month = oct,
       volume = {703},
       number = {2},
        pages = {1203-1223},
          doi = {10.1088/0004-637X/703/2/1203},
archivePrefix = {arXiv},
       eprint = {0908.1975},
 primaryClass = {astro-ph.SR},
       adsurl = {https://ui.adsabs.harvard.edu/abs/2009ApJ...703.1203H}
}

@INPROCEEDINGS{2023ASPC..534..539M,
       author = {{Manara}, C.~F. and {Ansdell}, M. and {Rosotti}, G.~P. and {Hughes}, A.~M. and {Armitage}, P.~J. and {Lodato}, G. and {Williams}, J.~P.},
        title = "{Demographics of Young Stars and their Protoplanetary Disks: Lessons Learned on Disk Evolution and its Connection to Planet Formation}",
    booktitle = {Protostars and Planets VII},
         year = 2023,
       editor = {{Inutsuka}, S. and {Aikawa}, Y. and {Muto}, T. and {Tomida}, K. and {Tamura}, M.},
       series = {Astronomical Society of the Pacific Conference Series},
       volume = {534},
        month = jul,
        pages = {539},
          doi = {10.48550/arXiv.2203.09930},
archivePrefix = {arXiv},
       eprint = {2203.09930},
 primaryClass = {astro-ph.SR},
       adsurl = {https://ui.adsabs.harvard.edu/abs/2023ASPC..534..539M}
}

@INPROCEEDINGS{2012ASInC...4...35M,
       author = {{Mann}, Rita K. and {Williams}, Jonathan P.},
        title = "{Protoplanetary disk masses in the Orion nebula cluster}",
    booktitle = {Astronomical Society of India Conference Series},
         year = 2012,
       series = {Astronomical Society of India Conference Series},
       volume = {4},
        month = jan,
        pages = {35},
       adsurl = {https://ui.adsabs.harvard.edu/abs/2012ASInC...4...35M}
}

@ARTICLE{2014ApJ...784...82M,
       author = {{Mann}, Rita K. and {Di Francesco}, James and {Johnstone}, Doug and {Andrews}, Sean M. and {Williams}, Jonathan P. and {Bally}, John and {Ricci}, Luca and {Hughes}, A. Meredith and {Matthews}, Brenda C.},
        title = "{ALMA Observations of the Orion Proplyds}",
      journal = {\apj},
         year = 2014,
        month = mar,
       volume = {784},
       number = {1},
          eid = {82},
        pages = {82},
          doi = {10.1088/0004-637X/784/1/82},
archivePrefix = {arXiv},
       eprint = {1403.2026},
 primaryClass = {astro-ph.SR},
       adsurl = {https://ui.adsabs.harvard.edu/abs/2014ApJ...784...82M}
}

@ARTICLE{2010ApJ...725..430M,
       author = {{Mann}, Rita K. and {Williams}, Jonathan P.},
        title = "{A Submillimeter Array Survey of Protoplanetary Disks in the Orion Nebula Cluster}",
      journal = {\apj},
         year = 2010,
        month = dec,
       volume = {725},
       number = {1},
        pages = {430-442},
          doi = {10.1088/0004-637X/725/1/430},
archivePrefix = {arXiv},
       eprint = {1010.1962},
 primaryClass = {astro-ph.SR},
       adsurl = {https://ui.adsabs.harvard.edu/abs/2010ApJ...725..430M}
}

@ARTICLE{Habing1968,
       author = {{Habing}, H.~J.},
        title = "{The interstellar radiation density between 912 A and 2400 A}",
      journal = {\bain},
         year = 1968,
        month = jan,
       volume = {19},
        pages = {421},
       adsurl = {https://ui.adsabs.harvard.edu/abs/1968BAN....19..421H}
}

@ARTICLE{2024A&A...687A..96F,
       author = {{Franceschi}, Riccardo and {Henning}, Thomas and {Tabone}, Beno{\^\i}t and {Perotti}, Giulia and {Caratti o Garatti}, Alessio and {Bettoni}, Giulio and {van Dishoeck}, Ewine F. and {Kamp}, Inga and {Absil}, Olivier and {G{\"u}del}, Manuel and {Olofsson}, G{\"o}ran and {Waters}, L.~B.~F.~M. and {Arabhavi}, Aditya M. and {Christiaens}, Valentin and {Gasman}, Danny and {Grant}, Sierra L. and {Jang}, Hyerin and {Rodgers-Lee}, Donna and {Samland}, Matthias and {Schwarz}, Kamber and {Temmink}, Milou and {Barrado}, David and {Boccaletti}, Anthony and {Geers}, Vincent and {Lagage}, Pierre-Olivier and {Pantin}, Eric and {Ray}, Tom P. and {Scheithauer}, Silvia and {Vandenbussche}, Bart and {Wright}, Gillian},
        title = "{MINDS: Mid-infrared atomic and molecular hydrogen lines in the inner disk around a low-mass star}",
      journal = {\aap},
         year = 2024,
        month = jul,
       volume = {687},
          eid = {A96},
        pages = {A96},
          doi = {10.1051/0004-6361/202348034},
archivePrefix = {arXiv},
       eprint = {2404.11942},
 primaryClass = {astro-ph.SR},
       adsurl = {https://ui.adsabs.harvard.edu/abs/2024A&A...687A..96F}
}

@ARTICLE{2019MNRAS.485.3895H,
       author = {{Haworth}, Thomas J. and {Clarke}, Cathie J.},
        title = "{The first multidimensional view of mass loss from externally FUV irradiated protoplanetary discs}",
      journal = {\mnras},
         year = 2019,
        month = may,
       volume = {485},
       number = {3},
        pages = {3895-3908},
          doi = {10.1093/mnras/stz706},
archivePrefix = {arXiv},
       eprint = {1903.03644},
 primaryClass = {astro-ph.EP},
       adsurl = {https://ui.adsabs.harvard.edu/abs/2019MNRAS.485.3895H}
}

@ARTICLE{1998ApJ...499..758J,
       author = {{Johnstone}, Doug and {Hollenbach}, David and {Bally}, John},
        title = "{Photoevaporation of Disks and Clumps by Nearby Massive Stars: Application to Disk Destruction in the Orion Nebula}",
      journal = {\apj},
         year = 1998,
        month = may,
       volume = {499},
       number = {2},
        pages = {758-776},
          doi = {10.1086/305658},
       adsurl = {https://ui.adsabs.harvard.edu/abs/1998ApJ...499..758J}
}

@ARTICLE{2008ApJ...676..472B,
       author = {{Beck}, Tracy L. and {McGregor}, Peter J. and {Takami}, Michihiro and {Pyo}, Tae-Soo},
        title = "{Spatially Resolved Molecular Hydrogen Emission in the Inner 200 AU Environments of Classical T Tauri Stars}",
      journal = {\apj},
         year = 2008,
        month = mar,
       volume = {676},
       number = {1},
        pages = {472-489},
          doi = {10.1086/527528},
archivePrefix = {arXiv},
       eprint = {0711.3844},
 primaryClass = {astro-ph},
       adsurl = {https://ui.adsabs.harvard.edu/abs/2008ApJ...676..472B}
}

@ARTICLE{2024AJ....167..223S,
       author = {{Sellek}, Andrew D. and {Bajaj}, Naman S. and {Pascucci}, Ilaria and {Clarke}, Cathie J. and {Alexander}, Richard and {Xie}, Chengyan and {Ballabio}, Giulia and {Deng}, Dingshan and {Gorti}, Uma and {Gaspar}, Andras and {Morrison}, Jane},
        title = "{Modeling JWST MIRI-MRS Observations of T Cha: Mid-IR Noble Gas Emission Tracing a Dense Disk Wind}",
      journal = {\aj},
         year = 2024,
        month = may,
       volume = {167},
       number = {5},
          eid = {223},
        pages = {223},
          doi = {10.3847/1538-3881/ad34ae},
archivePrefix = {arXiv},
       eprint = {2403.09780},
 primaryClass = {astro-ph.EP},
       adsurl = {https://ui.adsabs.harvard.edu/abs/2024AJ....167..223S}
}

@ARTICLE{2008ApJ...675.1361F,
       author = {{Fatuzzo}, Marco and {Adams}, Fred C.},
        title = "{UV Radiation Fields Produced by Young Embedded Star Clusters}",
      journal = {\apj},
         year = 2008,
        month = mar,
       volume = {675},
       number = {2},
        pages = {1361-1374},
          doi = {10.1086/527469},
archivePrefix = {arXiv},
       eprint = {0712.3487},
 primaryClass = {astro-ph},
       adsurl = {https://ui.adsabs.harvard.edu/abs/2008ApJ...675.1361F}
}

@ARTICLE{Zhou2026,
       author = {{Zhou}, Q.},
        title = "{JWST/MIRI NGC 1977}",
      journal = {\apjl},
         year = 2026,
        month = feb,
       volume = {},
       number = {},
        pages = {},
          doi = {},
       adsurl = {}
}

@ARTICLE{2015ApJ...801...31R,
       author = {{Rigliaco}, Elisabetta and {Pascucci}, I. and {Duchene}, G. and {Edwards}, S. and {Ardila}, D.~R. and {Grady}, C. and {Mendigut{\'\i}a}, I. and {Montesinos}, B. and {Mulders}, G.~D. and {Najita}, J.~R. and {Carpenter}, J. and {Furlan}, E. and {Gorti}, U. and {Meijerink}, R. and {Meyer}, M.~R.},
        title = "{Probing Stellar Accretion with Mid-infrared Hydrogen Lines}",
      journal = {\apj},
         year = 2015,
        month = mar,
       volume = {801},
       number = {1},
          eid = {31},
        pages = {31},
          doi = {10.1088/0004-637X/801/1/31},
archivePrefix = {arXiv},
       eprint = {1501.06210},
 primaryClass = {astro-ph.SR},
       adsurl = {https://ui.adsabs.harvard.edu/abs/2015ApJ...801...31R}
}

@ARTICLE{1999ApJ...517..209G,
       author = {{Goldsmith}, Paul F. and {Langer}, William D.},
        title = "{Population Diagram Analysis of Molecular Line Emission}",
      journal = {\apj},
         year = 1999,
        month = may,
       volume = {517},
       number = {1},
        pages = {209-225},
          doi = {10.1086/307195},
       adsurl = {https://ui.adsabs.harvard.edu/abs/1999ApJ...517..209G}
}

@ARTICLE{2023A&A...675A.111A,
       author = {{Argyriou}, Ioannis and {Glasse}, Alistair and {Law}, David R. and {Labiano}, Alvaro and {{\'A}lvarez-M{\'a}rquez}, Javier and {Patapis}, Polychronis and {Kavanagh}, Patrick J. and {Gasman}, Danny and {Mueller}, Michael and {Larson}, Kirsten and {Vandenbussche}, Bart and {Glauser}, Adrian M. and {Royer}, Pierre and {Dicken}, Daniel and {Harkett}, Jake and {Sargent}, Beth A. and {Engesser}, Michael and {Jones}, Olivia C. and {Kendrew}, Sarah and {Noriega-Crespo}, Alberto and {Brandl}, Bernhard and {Rieke}, George H. and {Wright}, Gillian S. and {Lee}, David and {Wells}, Martyn},
        title = "{JWST MIRI flight performance: The Medium-Resolution Spectrometer}",
      journal = {\aap},
         year = 2023,
        month = jul,
       volume = {675},
          eid = {A111},
        pages = {A111},
          doi = {10.1051/0004-6361/202346489},
archivePrefix = {arXiv},
       eprint = {2303.13469},
 primaryClass = {astro-ph.IM},
       adsurl = {https://ui.adsabs.harvard.edu/abs/2023A&A...675A.111A}
}

@ARTICLE{2015PASP..127..595W,
       author = {{Wright}, G.~S. and {Wright}, David and {Goodson}, G.~B. and {Rieke}, G.~H. and {Aitink-Kroes}, Gabby and {Amiaux}, J. and {Aricha-Yanguas}, Ana and {Azzollini}, Ruym{\'a}n and {Banks}, Kimberly and {Barrado-Navascues}, D. and {Belenguer-Davila}, T. and {Blommaert}, J.~A.~D.~L. and {Bouchet}, Patrice and {Brandl}, B.~R. and {Colina}, L. and {Detre}, {\"O}rs and {Diaz-Catala}, Eva and {Eccleston}, Paul and {Friedman}, Scott D. and {Garc{\'\i}a-Mar{\'\i}n}, Macarena and {G{\"u}del}, Manuel and {Glasse}, Alistair and {Glauser}, Adrian M. and {Greene}, T.~P. and {Groezinger}, Uli and {Grundy}, Tim and {Hastings}, Peter and {Henning}, Th. and {Hofferbert}, Ralph and {Hunter}, Faye and {Jessen}, N.~C. and {Justtanont}, K. and {Karnik}, Avinash R. and {Khorrami}, Mori A. and {Krause}, Oliver and {Labiano}, Alvaro and {Lagage}, P.-O. and {Langer}, Ulrich and {Lemke}, Dietrich and {Lim}, Tanya and {Lorenzo-Alvarez}, Jose and {Mazy}, Emmanuel and {McGowan}, Norman and {Meixner}, M.~E. and {Morris}, Nigel and {Morrison}, Jane E. and {M{\"u}ller}, Friedrich and {rgaard-Nielson}, H.-U. N{\o} and {Olofsson}, G{\"o}ran and {O'Sullivan}, Brian and {Pel}, J.-W. and {Penanen}, Konstantin and {Petach}, M.~B. and {Pye}, J.~P. and {Ray}, T.~P. and {Renotte}, Etienne and {Renouf}, Ian and {Ressler}, M.~E. and {Samara-Ratna}, Piyal and {Scheithauer}, Silvia and {Schneider}, Analyn and {Shaughnessy}, Bryan and {Stevenson}, Tim and {Sukhatme}, Kalyani and {Swinyard}, Bruce and {Sykes}, Jon and {Thatcher}, John and {Tikkanen}, Tuomo and {van Dishoeck}, E.~F. and {Waelkens}, C. and {Walker}, Helen and {Wells}, Martyn and {Zhender}, Alex},
        title = "{The Mid-Infrared Instrument for the James Webb Space Telescope, II: Design and Build}",
      journal = {\pasp},
         year = 2015,
        month = jul,
       volume = {127},
       number = {953},
        pages = {595},
          doi = {10.1086/682253},
archivePrefix = {arXiv},
       eprint = {1508.02333},
 primaryClass = {astro-ph.IM},
       adsurl = {https://ui.adsabs.harvard.edu/abs/2015PASP..127..595W}
}

@ARTICLE{2015PASP..127..646W,
       author = {{Wells}, Martyn and {Pel}, J.-W. and {Glasse}, Alistair and {Wright}, G.~S. and {Aitink-Kroes}, Gabby and {Azzollini}, Ruym{\'a}n and {Beard}, Steven and {Brandl}, B.~R. and {Gallie}, Angus and {Geers}, V.~C. and {Glauser}, A.~M. and {Hastings}, Peter and {Henning}, Th. and {Jager}, Rieks and {Justtanont}, K. and {Kruizinga}, Bob and {Lahuis}, Fred and {Lee}, David and {Martinez-Delgado}, I. and {Mart{\'\i}nez-Galarza}, J.~R. and {Meijers}, M. and {Morrison}, Jane E. and {M{\"u}ller}, Friedrich and {Nakos}, Thodori and {O'Sullivan}, Brian and {Oudenhuysen}, Ad and {Parr-Burman}, P. and {Pauwels}, Evert and {Rohloff}, R.-R. and {Schmalzl}, Eva and {Sykes}, Jon and {Thelen}, M.~P. and {van Dishoeck}, E.~F. and {Vandenbussche}, Bart and {Venema}, Lars B. and {Visser}, Huib and {Waters}, L.~B.~F.~M. and {Wright}, David},
        title = "{The Mid-Infrared Instrument for the James Webb Space Telescope, VI: The Medium Resolution Spectrometer}",
      journal = {\pasp},
         year = 2015,
        month = jul,
       volume = {127},
       number = {953},
        pages = {646},
          doi = {10.1086/682281},
archivePrefix = {arXiv},
       eprint = {1508.03070},
 primaryClass = {astro-ph.IM},
       adsurl = {https://ui.adsabs.harvard.edu/abs/2015PASP..127..646W}
}

@ARTICLE{2025ApJ...985..224T,
       author = {{Tofflemire}, Benjamin M. and {Manara}, Carlo F. and {Banzatti}, Andrea and {Pontoppidan}, Klaus M. and {Najita}, Joan and {Nisini}, Brunella and {Whelan}, Emma T. and {Campbell-White}, Justyn and {Alqubelat}, Hala and {Kraus}, Adam L. and {Rab}, Christian and {Houge}, Adrien and {Krijt}, Sebastiaan and {Muzerolle}, James and {Fiorellino}, Eleonora and {Benisty}, Myriam and {Tychoniec}, Lukasz and {Salyk}, Colette and {Bourdarot}, Guillaume and {Hyden}, Jacob},
        title = "{Coordinated Space- and Ground-based Monitoring of Accretion Bursts in a Protoplanetary Disk: Establishing Mid-infrared Hydrogen Lines as Accretion Diagnostics for JWST/MIRI}",
      journal = {\apj},
         year = 2025,
        month = jun,
       volume = {985},
       number = {2},
          eid = {224},
        pages = {224},
          doi = {10.3847/1538-4357/adcc23},
archivePrefix = {arXiv},
       eprint = {2504.08029},
 primaryClass = {astro-ph.SR},
       adsurl = {https://ui.adsabs.harvard.edu/abs/2025ApJ...985..224T}
}

@ARTICLE{2015A&A...577A..42B,
       author = {{Baraffe}, Isabelle and {Homeier}, Derek and {Allard}, France and {Chabrier}, Gilles},
        title = "{New evolutionary models for pre-main sequence and main sequence low-mass stars down to the hydrogen-burning limit}",
      journal = {\aap},
         year = 2015,
        month = may,
       volume = {577},
          eid = {A42},
        pages = {A42},
          doi = {10.1051/0004-6361/201425481},
archivePrefix = {arXiv},
       eprint = {1503.04107},
 primaryClass = {astro-ph.SR},
       adsurl = {https://ui.adsabs.harvard.edu/abs/2015A&A...577A..42B}
}

@ARTICLE{2025AJ....169..165B,
       author = {{Banzatti}, Andrea and {Salyk}, Colette and {Pontoppidan}, Klaus M. and {Carr}, John S. and {Zhang}, Ke and {Arulanantham}, Nicole and {Krijt}, Sebastiaan and {{\"O}berg}, Karin I. and {Cleeves}, L. Ilsedore and {Najita}, Joan R. and {Pascucci}, Ilaria and {Blake}, Geoffrey A. and {Romero-Mirza}, Carlos E. and {Bergin}, Edwin A. and {Cieza}, Lucas A. and {Pinilla}, Paola and {Long}, Feng and {Mallaney}, Patrick and {Xie}, Chengyan and {Waggoner}, Abygail R. and {Kaeufer}, Till and {The Jdiscs Collaboration}},
        title = "{Water in Protoplanetary Disks with JWST-MIRI: Spectral Excitation Atlas and Radial Distribution from Temperature Diagnostic Diagrams and Doppler Mapping}",
      journal = {\aj},
         year = 2025,
        month = mar,
       volume = {169},
       number = {3},
          eid = {165},
        pages = {165},
          doi = {10.3847/1538-3881/ada962},
archivePrefix = {arXiv},
       eprint = {2409.16255},
 primaryClass = {astro-ph.EP},
       adsurl = {https://ui.adsabs.harvard.edu/abs/2025AJ....169..165B}
}

@ARTICLE{2025arXiv251203456S,
       author = {{Shridharan}, B and {Manoj}, P and {Pathak}, Vinod Chandra and {Garatti}, Alessio Caratti O and {Banerjee}, Bihan and {Henning}, Th. and {Kamp}, I. and {van Dischoeck}, E. and {Tyagi}, H. and {Arun}, R. and {Mathew}, B. and {G{\"u}del}, M. and {Lagage}, P.-O.},
        title = "{Improving Accretion Diagnostics for Young Stellar Objects with Mid-infrared Hydrogen lines from JWST/MIRI}",
      journal = {arXiv e-prints},
         year = 2025,
        month = dec,
          eid = {arXiv:2512.03456},
        pages = {arXiv:2512.03456},
          doi = {10.48550/arXiv.2512.03456},
archivePrefix = {arXiv},
       eprint = {2512.03456},
 primaryClass = {astro-ph.SR},
       adsurl = {https://ui.adsabs.harvard.edu/abs/2025arXiv251203456S}
}

@ARTICLE{2024ApJ...967..103B,
       author = {{Boyden}, Ryan D. and {Eisner}, Josh A.},
        title = "{Constraining Free{\textendash}Free Emission and Photoevaporative Mass-loss Rates for Known Proplyds and New VLA{\textendash}identified Candidate Proplyds in NGC 1977}",
      journal = {\apj},
         year = 2024,
        month = jun,
       volume = {967},
       number = {2},
          eid = {103},
        pages = {103},
          doi = {10.3847/1538-4357/ad3cd5},
archivePrefix = {arXiv},
       eprint = {2404.04437},
 primaryClass = {astro-ph.EP},
       adsurl = {https://ui.adsabs.harvard.edu/abs/2024ApJ...967..103B}
}

@ARTICLE{2024AJ....167..127B,
       author = {{Bajaj}, Naman S. and {Pascucci}, Ilaria and {Gorti}, Uma and {Alexander}, Richard and {Sellek}, Andrew and {Morrison}, Jane and {Gaspar}, Andras and {Clarke}, Cathie and {Xie}, Chengyan and {Ballabio}, Giulia and {Deng}, Dingshan},
        title = "{JWST MIRI MRS Observations of T Cha: Discovery of a Spatially Resolved Disk Wind}",
      journal = {\aj},
         year = 2024,
        month = mar,
       volume = {167},
       number = {3},
          eid = {127},
        pages = {127},
          doi = {10.3847/1538-3881/ad22e1},
archivePrefix = {arXiv},
       eprint = {2403.01060},
 primaryClass = {astro-ph.EP},
       adsurl = {https://ui.adsabs.harvard.edu/abs/2024AJ....167..127B}
}

@ARTICLE{2024ApJ...963..158P,
       author = {{Pontoppidan}, Klaus M. and {Salyk}, Colette and {Banzatti}, Andrea and {Zhang}, Ke and {Pascucci}, Ilaria and {{\"O}berg}, Karin I. and {Long}, Feng and {Romero-Mirza}, Carlos E. and {Carr}, John and {Najita}, Joan and {Blake}, Geoffrey A. and {Arulanantham}, Nicole and {Andrews}, Sean and {Ballering}, Nicholas P. and {Bergin}, Edwin and {Calahan}, Jenny and {Cobb}, Douglas and {Colmenares}, Maria Jose and {Dickson-Vandervelde}, Annie and {Dignan}, Anna and {Green}, Joel and {Heretz}, Phoebe and {Herczeg}, Gregory and {Kalyaan}, Anusha and {Krijt}, Sebastiaan and {Pauly}, Tyler and {Pinilla}, Paola and {Trapman}, Leon and {Xie}, Chengyan},
        title = "{High-contrast JWST-MIRI Spectroscopy of Planet-forming Disks for the JDISC Survey}",
      journal = {\apj},
         year = 2024,
        month = mar,
       volume = {963},
       number = {2},
          eid = {158},
        pages = {158},
          doi = {10.3847/1538-4357/ad20f0},
archivePrefix = {arXiv},
       eprint = {2311.17020},
 primaryClass = {astro-ph.EP},
       adsurl = {https://ui.adsabs.harvard.edu/abs/2024ApJ...963..158P}
}

@ARTICLE{2021MNRAS.503.4172H,
       author = {{Haworth}, Thomas J.},
        title = "{Warm millimetre dust in protoplanetary discs near massive stars}",
      journal = {\mnras},
         year = 2021,
        month = may,
       volume = {503},
       number = {3},
        pages = {4172-4182},
          doi = {10.1093/mnras/stab728},
archivePrefix = {arXiv},
       eprint = {2103.03950},
 primaryClass = {astro-ph.SR},
       adsurl = {https://ui.adsabs.harvard.edu/abs/2021MNRAS.503.4172H}
}

@ARTICLE{2025ApJ...991..128R,
       author = {{Romero-Mirza}, Carlos E. and {{\"O}berg}, Karin I. and {Banzatti}, Andrea and {Tabone}, Beno{\^\i}t and {Najita}, Joan and {Blake}, Geoffrey A. and {Bergin}, Edwin A. and {Krijt}, Sebastiaan and {Law}, Charles J. and {Long}, Feng and {Huang}, Jane and {Wilner}, David J. and {Andrews}, Sean M. and {Czekala}, Ian and {Teague}, Richard and {Aikawa}, Yuri and {Bae}, Jaehan and {Le Gal}, Romane and {Alarc{\'o}n}, Felipe and {The Jdiscs Collaboration}},
        title = "{JWST-MIRI Observations of the Irradiated Chemistry in the Inner Disk Cavity of GM Aur}",
      journal = {\apj},
         year = 2025,
        month = oct,
       volume = {991},
       number = {2},
          eid = {128},
        pages = {128},
          doi = {10.3847/1538-4357/adf8db},
       adsurl = {https://ui.adsabs.harvard.edu/abs/2025ApJ...991..128R}
}

@article{HITRAN2020,
author = {I.~E. {Gordon} AND L.~S. {Rothman} AND R.~J. {Hargreaves} AND R. {Hashemi} AND E.~V. {Karlovets} AND F.~M. {Skinner} AND E.~K. {Conway} AND C. {Hill} AND R.~V. {Kochanov} AND Y. {Tan} AND P. {Wcis{\l}o} AND A.~A. {Finenko} AND K. {Nelson} AND P.~F. {Bernath} AND M. {Birk} AND V. {Boudon} AND A. {Campargue} AND K.~V. {Chance} AND A. {Coustenis} AND B.~J. {Drouin} AND J.-M. {Flaud} AND R.~R. {Gamache} AND J.~T. {Hodges} AND D. {Jacquemart} AND E.~J. {Mlawer} AND A.~V. {Nikitin} AND V.~I. {Perevalov} AND M. {Rotger} AND J. {Tennyson} AND G.~C. {Toon} AND H. {Tran} AND V.~G. {Tyuterev} AND E.~M. {Adkins} AND A. {Baker} AND A. {Barbe} AND E. {Can{\`{e}}} AND A.~G. {Cs{'{a}}sz{'{a}}r} AND A. {Dudaryonok} AND O. {Egorov} AND A.~J. {Fleisher} AND H. {Fleurbaey} AND A. {Foltynowicz} AND T. {Furtenbacher} AND J.~J. {Harrison} AND J.-M. {Hartmann} AND V.-M. {Horneman} AND X. {Huang} AND T. {Karman} AND J. {Karns} AND S. {Kassi} AND I. {Kleiner} AND V. {Kofman} AND F. {Kwabia-Tchana} AND N.~N. {Lavrentieva} AND T.~J. {Lee} AND D.~A. {Long} AND A.~A. {Lukashevskaya} AND O.~M. {Lyulin} AND V.~Yu. {Makhnev} AND W. {Matt} AND S.~T. {Massie} AND M. {Melosso} AND S.~N. {Mikhailenko} AND D. {Mondelain} AND H.~S.~P. {M{"{u}}ller} AND O.~V. {Naumenko} AND A. {Perrin} AND O.~L. {Polyansky} AND E. {Raddaoui} AND P.~L. {Raston} AND Z.~D. {Reed} AND M. {Rey} AND C. {Richard} AND R. {T{'{o}}bi{'{a}}s} AND I. {Sadiek} AND D.~W. {Schwenke} AND E. {Starikova} AND K. {Sung} AND F. {Tamassia} AND S.~A. {Tashkun} AND J. {Vander Auwera} AND I.~A. {Vasilenko} AND A.~A. {Vigasin} AND G.~L. {Villanueva} AND B. {Vispoel} AND G. {Wagner} AND A. {Yachmenev} AND S.~N. {Yurchenko}},
title = {The {HITRAN2020} Molecular Spectroscopic Database},
journal = {Journal of Quantitative Spectroscopy and Radiative Transfer},
year = {2022},
volume = {277},
pages = {107949},
doi = {10.1016/j.jqsrt.2021.107949},
}

@ARTICLE{2022MNRAS.512.2594H,
       author = {{Haworth}, Thomas J. and {Kim}, Jinyoung S. and {Qiao}, Lin and {Winter}, Andrew J. and {Williams}, Jonathan P. and {Clarke}, Cathie J. and {Owen}, James E. and {Facchini}, Stefano and {Ansdell}, Megan and {Kama}, Mikhel and {Ballabio}, Giulia},
        title = "{An APEX search for carbon emission from NGC 1977 proplyds}",
      journal = {\mnras},
         year = 2022,
        month = may,
       volume = {512},
       number = {2},
        pages = {2594-2603},
          doi = {10.1093/mnras/stac656},
archivePrefix = {arXiv},
       eprint = {2203.03928},
 primaryClass = {astro-ph.EP},
       adsurl = {https://ui.adsabs.harvard.edu/abs/2022MNRAS.512.2594H}
}

@ARTICLE{2025MNRAS.537..598K,
       author = {{Keyte}, Luke and {Haworth}, Thomas J.},
        title = "{Impact of photoevaporative winds in chemical models of externally irradiated protoplanetary discs}",
      journal = {\mnras},
         year = 2025,
        month = feb,
       volume = {537},
       number = {1},
        pages = {598-616},
          doi = {10.1093/mnras/staf047},
archivePrefix = {arXiv},
       eprint = {2501.05172},
 primaryClass = {astro-ph.EP},
       adsurl = {https://ui.adsabs.harvard.edu/abs/2025MNRAS.537..598K}
}

@ARTICLE{2025ApJ...991...94C,
       author = {{Calahan}, Jenny K. and {{\"O}berg}, Karin and {Booth}, Alice},
        title = "{The Impact of External Radiation on the Inner Disk Chemistry of Planet Formation}",
      journal = {\apj},
         year = 2025,
        month = sep,
       volume = {991},
       number = {1},
          eid = {94},
        pages = {94},
          doi = {10.3847/1538-4357/adfa09},
archivePrefix = {arXiv},
       eprint = {2508.06613},
 primaryClass = {astro-ph.EP},
       adsurl = {https://ui.adsabs.harvard.edu/abs/2025ApJ...991...94C}
}

@ARTICLE{2020ApJ...894...74B,
       author = {{Boyden}, Ryan D. and {Eisner}, Josh A.},
        title = "{Protoplanetary Disks in the Orion Nebula Cluster: Gas-disk Morphologies and Kinematics as Seen with ALMA}",
      journal = {\apj},
         year = 2020,
        month = may,
       volume = {894},
       number = {1},
          eid = {74},
        pages = {74},
          doi = {10.3847/1538-4357/ab86b7},
archivePrefix = {arXiv},
       eprint = {2003.12580},
 primaryClass = {astro-ph.EP},
       adsurl = {https://ui.adsabs.harvard.edu/abs/2020ApJ...894...74B}
}

@ARTICLE{2018ApJ...860...77E,
       author = {{Eisner}, J.~A. and {Arce}, H.~G. and {Ballering}, N.~P. and {Bally}, J. and {Andrews}, S.~M. and {Boyden}, R.~D. and {Di Francesco}, J. and {Fang}, M. and {Johnstone}, D. and {Kim}, J.~S. and {Mann}, R.~K. and {Matthews}, B. and {Pascucci}, I. and {Ricci}, L. and {Sheehan}, P.~D. and {Williams}, J.~P.},
        title = "{Protoplanetary Disk Properties in the Orion Nebula Cluster: Initial Results from Deep, High-resolution ALMA Observations}",
      journal = {\apj},
         year = 2018,
        month = jun,
       volume = {860},
       number = {1},
          eid = {77},
        pages = {77},
          doi = {10.3847/1538-4357/aac3e2},
archivePrefix = {arXiv},
       eprint = {1805.03669},
 primaryClass = {astro-ph.SR},
       adsurl = {https://ui.adsabs.harvard.edu/abs/2018ApJ...860...77E}
}

@ARTICLE{2025MNRAS.541.2917P,
       author = {{Peake}, Tyger and {Haworth}, Thomas J. and {Aru}, Mari-Liis and {Henney}, William J.},
        title = "{Line ratio identification of external photoevaporation}",
      journal = {\mnras},
         year = 2025,
        month = aug,
       volume = {541},
       number = {4},
        pages = {2917-2933},
          doi = {10.1093/mnras/staf1090},
archivePrefix = {arXiv},
       eprint = {2506.19788},
 primaryClass = {astro-ph.EP},
       adsurl = {https://ui.adsabs.harvard.edu/abs/2025MNRAS.541.2917P}
}

@ARTICLE{2023ApJ...954..127B,
       author = {{Ballering}, Nicholas P. and {Cleeves}, L. Ilsedore and {Haworth}, Thomas J. and {Bally}, John and {Eisner}, Josh A. and {Ginsburg}, Adam and {Boyden}, Ryan D. and {Fang}, Min and {Kim}, Jinyoung Serena},
        title = "{Isolating Dust and Free-Free Emission in ONC Proplyds with ALMA Band 3 Observations}",
      journal = {\apj},
         year = 2023,
        month = sep,
       volume = {954},
       number = {2},
          eid = {127},
        pages = {127},
          doi = {10.3847/1538-4357/ace901},
archivePrefix = {arXiv},
       eprint = {2308.07369},
 primaryClass = {astro-ph.EP},
       adsurl = {https://ui.adsabs.harvard.edu/abs/2023ApJ...954..127B}
}

@ARTICLE{2023A&A...679A..82M,
       author = {{Mauc{\'o}}, K. and {Manara}, C.~F. and {Ansdell}, M. and {Bettoni}, G. and {Claes}, R. and {Alcala}, J. and {Miotello}, A. and {Facchini}, S. and {Haworth}, T.~J. and {Lodato}, G. and {Williams}, J.~P.},
        title = "{Testing external photoevaporation in the {\ensuremath{\sigma}}-Orionis cluster with spectroscopy and disk mass measurements}",
      journal = {\aap},
         year = 2023,
        month = nov,
       volume = {679},
          eid = {A82},
        pages = {A82},
          doi = {10.1051/0004-6361/202347627},
archivePrefix = {arXiv},
       eprint = {2309.05651},
 primaryClass = {astro-ph.SR},
       adsurl = {https://ui.adsabs.harvard.edu/abs/2023A&A...679A..82M}
}

@ARTICLE{2023MNRAS.525.4129H,
       author = {{Haworth}, Thomas J. and {Reiter}, Megan and {O'Dell}, C. Robert and {Zeidler}, Peter and {Berne}, Olivier and {Manara}, Carlo F. and {Ballabio}, Giulia and {Kim}, Jinyoung S. and {Bally}, John and {Goicoechea}, Javier R. and {Aru}, Mari-Liis and {Gupta}, Aashish and {Miotello}, Anna},
        title = "{The VLT MUSE NFM view of outflows and externally photoevaporating discs near the orion bar$^{★}$}",
      journal = {\mnras},
         year = 2023,
        month = nov,
       volume = {525},
       number = {3},
        pages = {4129-4142},
          doi = {10.1093/mnras/stad2581},
archivePrefix = {arXiv},
       eprint = {2308.12342},
 primaryClass = {astro-ph.GA},
       adsurl = {https://ui.adsabs.harvard.edu/abs/2023MNRAS.525.4129H}
}

@ARTICLE{2007MNRAS.376.1350C,
       author = {{Clarke}, C.~J.},
        title = "{The photoevaporation of discs around young stars in massive clusters}",
      journal = {\mnras},
         year = 2007,
        month = apr,
       volume = {376},
       number = {3},
        pages = {1350-1356},
          doi = {10.1111/j.1365-2966.2007.11547.x},
archivePrefix = {arXiv},
       eprint = {astro-ph/0702112},
 primaryClass = {astro-ph},
       adsurl = {https://ui.adsabs.harvard.edu/abs/2007MNRAS.376.1350C}
}

@ARTICLE{2025MNRAS.tmp.1907C,
       author = {{Coleman}, Gavin A.~L. and {Haworth}, Thomas J. and {Schroetter}, Ilane and {Bern{\'e}}, Olivier},
        title = "{Using simultaneous mass accretion and external photoevaporation rates for d203-504 to constrain disc evolution processes}",
      journal = {\mnras},
         year = 2025,
        month = nov,
          doi = {10.1093/mnras/staf2015},
archivePrefix = {arXiv},
       eprint = {2510.12535},
 primaryClass = {astro-ph.EP},
       adsurl = {https://ui.adsabs.harvard.edu/abs/2025MNRAS.tmp.1907C}
}

@ARTICLE{2024Sci...383..988B,
       author = {{Bern{\'e}}, Olivier and {Habart}, Emilie and {Peeters}, Els and {Schroetter}, Ilane and {Canin}, Am{\'e}lie and {Sidhu}, Ameek and {Chown}, Ryan and {Bron}, Emeric and {Haworth}, Thomas J. and {Klaassen}, Pamela and {Trahin}, Boris and {Van De Putte}, Dries and {Alarc{\'o}n}, Felipe and {Zannese}, Marion and {Abergel}, Alain and {Bergin}, Edwin A. and {Bernard-Salas}, Jeronimo and {Boersma}, Christiaan and {Cami}, Jan and {Cuadrado}, Sara and {Dartois}, Emmanuel and {Dicken}, Daniel and {Elyajouri}, Meriem and {Fuente}, Asunci{\'o}n and {Goicoechea}, Javier R. and {Gordon}, Karl D. and {Issa}, Lina and {Joblin}, Christine and {Kannavou}, Olga and {Khan}, Baria and {Lacinbala}, Ozan and {Languignon}, David and {Le Gal}, Romane and {Maragkoudakis}, Alexandros and {Meshaka}, Raphael and {Okada}, Yoko and {Onaka}, Takashi and {Pasquini}, Sofia and {Pound}, Marc W. and {Robberto}, Massimo and {R{\"o}llig}, Markus and {Schefter}, Bethany and {Schirmer}, Thi{\'e}baut and {Simmer}, Thomas and {Tabone}, Benoit and {Tielens}, Alexander G.~G.~M. and {Vicente}, S{\'\i}lvia and {Wolfire}, Mark G. and {PDRs4All Team} and {Aleman}, Isabel and {Allamandola}, Louis and {Auchettl}, Rebecca and {Baratta}, Giuseppe Antonio and {Baruteau}, Cl{\'e}ment and {Bejaoui}, Salma and {Bera}, Partha P. and {Black}, John H. and {Boulanger}, Francois and {Bouwman}, Jordy and {Brandl}, Bernhard and {Brechignac}, Philippe and {Br{\"u}nken}, Sandra and {Buragohain}, Mridusmita and {Burkhardt}, Andrew and {Candian}, Alessandra and {Cazaux}, St{\'e}phanie and {Cernicharo}, Jose and {Chabot}, Marin and {Chakraborty}, Shubhadip and {Champion}, Jason and {Colgan}, Sean W.~J. and {Cooke}, Ilsa R. and {Coutens}, Audrey and {Cox}, Nick L.~J. and {Demyk}, Karine and {Meyer}, Jennifer Donovan and {Engrand}, C{\'e}cile and {Foschino}, Sacha and {Garc{\'\i}a-Lario}, Pedro and {Gavilan}, Lisseth and {Gerin}, Maryvonne and {Godard}, Marie and {Gottlieb}, Carl A. and {Guillard}, Pierre and {Gusdorf}, Antoine and {Hartigan}, Patrick and {He}, Jinhua and {Herbst}, Eric and {Hornekaer}, Liv and {J{\"a}ger}, Cornelia and {Janot-Pacheco}, Eduardo and {Kaufman}, Michael and {Kemper}, Francisca and {Kendrew}, Sarah and {Kirsanova}, Maria S. and {Knight}, Collin and {Kwok}, Sun and {Labiano}, {\'A}lvaro and {Lai}, Thomas S.-Y. and {Lee}, Timothy J. and {Lefloch}, Bertrand and {Le Petit}, Franck and {Li}, Aigen and {Linz}, Hendrik and {Mackie}, Cameron J. and {Madden}, Suzanne C. and {Mascetti}, Jo{\"e}lle and {McGuire}, Brett A. and {Merino}, Pablo and {Micelotta}, Elisabetta R. and {Morse}, Jon A. and {Mulas}, Giacomo and {Neelamkodan}, Naslim and {Ohsawa}, Ryou and {Paladini}, Roberta and {Palumbo}, Maria Elisabetta and {Pathak}, Amit and {Pendleton}, Yvonne J. and {Petrignani}, Annemieke and {Pino}, Thomas and {Puga}, Elena and {Rangwala}, Naseem and {Rapacioli}, Mathias and {Ricca}, Alessandra and {Roman-Duval}, Julia and {Roueff}, Evelyne and {Rouill{\'e}}, Ga{\"e}l and {Salama}, Farid and {Sales}, Dinalva A. and {Sandstrom}, Karin and {Sarre}, Peter and {Sciamma-O'Brien}, Ella and {Sellgren}, Kris and {Shannon}, Matthew J. and {Simonnin}, Adrien and {Shenoy}, Sachindev S. and {Teyssier}, David and {Thomas}, Richard D. and {Togi}, Aditya and {Verstraete}, Laurent and {Witt}, Adolf N. and {Wootten}, Alwyn and {Ysard}, Nathalie and {Zettergren}, Henning and {Zhang}, Yong and {Zhang}, Ziwei E. and {Zhen}, Junfeng},
        title = "{A far-ultraviolet{\textendash}driven photoevaporation flow observed in a protoplanetary disk}",
      journal = {Science},
         year = 2024,
        month = mar,
       volume = {383},
       number = {6686},
        pages = {988-992},
          doi = {10.1126/science.adh2861},
archivePrefix = {arXiv},
       eprint = {2403.00160},
 primaryClass = {astro-ph.GA},
       adsurl = {https://ui.adsabs.harvard.edu/abs/2024Sci...383..988B}
}

@ARTICLE{2023Natur.621...56B,
       author = {{Bern{\'e}}, Olivier and {Martin-Drumel}, Marie-Aline and {Schroetter}, Ilane and {Goicoechea}, Javier R. and {Jacovella}, Ugo and {Gans}, B{\'e}renger and {Dartois}, Emmanuel and {Coudert}, Laurent H. and {Bergin}, Edwin and {Alarcon}, Felipe and {Cami}, Jan and {Roueff}, Evelyne and {Black}, John H. and {Asvany}, Oskar and {Habart}, Emilie and {Peeters}, Els and {Canin}, Amelie and {Trahin}, Boris and {Joblin}, Christine and {Schlemmer}, Stephan and {Thorwirth}, Sven and {Cernicharo}, Jose and {Gerin}, Maryvonne and {Tielens}, Alexander and {Zannese}, Marion and {Abergel}, Alain and {Bernard-Salas}, Jeronimo and {Boersma}, Christiaan and {Bron}, Emeric and {Chown}, Ryan and {Cuadrado}, Sara and {Dicken}, Daniel and {Elyajouri}, Meriem and {Fuente}, Asunci{\'o}n and {Gordon}, Karl D. and {Issa}, Lina and {Kannavou}, Olga and {Khan}, Baria and {Lacinbala}, Ozan and {Languignon}, David and {Le Gal}, Romane and {Maragkoudakis}, Alexandros and {Meshaka}, Raphael and {Okada}, Yoko and {Onaka}, Takashi and {Pasquini}, Sofia and {Pound}, Marc W. and {Robberto}, Massimo and {R{\"o}llig}, Markus and {Schefter}, Bethany and {Schirmer}, Thi{\'e}baut and {Sidhu}, Ameek and {Tabone}, Benoit and {Van De Putte}, Dries and {Vicente}, S{\'\i}lvia and {Wolfire}, Mark G.},
        title = "{Formation of the methyl cation by photochemistry in a protoplanetary disk}",
      journal = {\nat},
         year = 2023,
        month = sep,
       volume = {621},
       number = {7977},
        pages = {56-59},
          doi = {10.1038/s41586-023-06307-x},
archivePrefix = {arXiv},
       eprint = {2401.03296},
 primaryClass = {astro-ph.GA},
       adsurl = {https://ui.adsabs.harvard.edu/abs/2023Natur.621...56B}
}

@ARTICLE{2025NatAs...9.1326S,
       author = {{Schroetter}, Ilane and {Bern{\'e}}, Olivier and {Bron}, Emeric and {Alarcon}, Felipe and {Amiot}, Paul and {Bergin}, Edwin A. and {Boersma}, Christiaan and {Cami}, Jan and {Coleman}, Gavin A.~L. and {Dartois}, Emmanuel and {Fuente}, Asuncion and {Goicoechea}, Javier R. and {Habart}, Emilie and {Haworth}, Thomas J. and {Joblin}, Christine and {Franck}, Le Petit and {Onaka}, Takashi and {Peeters}, Els and {R{\"o}lling}, Markus and {Tielens}, Alexander G.~G.~M. and {Zannese}, Marion},
        title = "{A solar C/O ratio in planet-forming gas at 1 au in a highly irradiated disk}",
      journal = {Nature Astronomy},
         year = 2025,
        month = sep,
       volume = {9},
        pages = {1326-1336},
          doi = {10.1038/s41550-025-02596-6},
archivePrefix = {arXiv},
       eprint = {2505.22314},
 primaryClass = {astro-ph.GA},
       adsurl = {https://ui.adsabs.harvard.edu/abs/2025NatAs...9.1326S}
}

@ARTICLE{2025arXiv250802576K,
       author = {{Kurtovic}, Nicolas T. and {Grant}, Sierra L. and {Temmink}, Milou and {Sellek}, Andrew D. and {van Dishoeck}, Ewine F. and {Henning}, Thomas and {Kamp}, Inga and {Christiaens}, Valentin and {Banzatti}, Andrea and {Gasman}, Danny and {Kaeufer}, Till and {Stapper}, Lucas M. and {Franceschi}, Riccardo and {G{\"u}del}, Manuel and {Lagage}, Pierre-Olivier and {Vlasblom}, Marissa and {Perotti}, Giulia and {Schwarz}, Kamber and {Somigliana}, Alice},
        title = "{MINDS. Young binary systems with JWST/MIRI: Variable water-rich primaries and extended emission}",
      journal = {arXiv e-prints},
         year = 2025,
        month = aug,
          eid = {arXiv:2508.02576},
        pages = {arXiv:2508.02576},
          doi = {10.48550/arXiv.2508.02576},
archivePrefix = {arXiv},
       eprint = {2508.02576},
 primaryClass = {astro-ph.EP},
       adsurl = {https://ui.adsabs.harvard.edu/abs/2025arXiv250802576K}
}

@ARTICLE{1987ApJ...321..516C,
       author = {{Churchwell}, E. and {Felli}, M. and {Wood}, D.~O.~S. and {Massi}, M.},
        title = "{Solar System--sized Condensations in the Orion Nebula}",
      journal = {\apj},
         year = 1987,
        month = oct,
       volume = {321},
        pages = {516},
          doi = {10.1086/165648},
       adsurl = {https://ui.adsabs.harvard.edu/abs/1987ApJ...321..516C}
}

@ARTICLE{1993ApJ...410..696O,
       author = {{O'Dell}, C.~R. and {Wen}, Zheng and {Hu}, Xihai},
        title = "{Discovery of New Objects in the Orion Nebula on HST Images: Shocks, Compact Sources, and Protoplanetary Disks}",
      journal = {\apj},
         year = 1993,
        month = jun,
       volume = {410},
        pages = {696},
          doi = {10.1086/172786},
       adsurl = {https://ui.adsabs.harvard.edu/abs/1993ApJ...410..696O}
}

@ARTICLE{2018MNRAS.481..452H,
       author = {{Haworth}, Thomas J. and {Clarke}, Cathie J. and {Rahman}, Wahidur and {Winter}, Andrew J. and {Facchini}, Stefano},
        title = "{The FRIED grid of mass-loss rates for externally irradiated protoplanetary discs}",
      journal = {\mnras},
         year = 2018,
        month = nov,
       volume = {481},
       number = {1},
        pages = {452-466},
          doi = {10.1093/mnras/sty2323},
archivePrefix = {arXiv},
       eprint = {1808.07484},
 primaryClass = {astro-ph.SR},
       adsurl = {https://ui.adsabs.harvard.edu/abs/2018MNRAS.481..452H}
}

@ARTICLE{2023MNRAS.526.4315H,
       author = {{Haworth}, Thomas J. and {Coleman}, Gavin A.~L. and {Qiao}, Lin and {Sellek}, Andrew D. and {Askari}, Kanaar},
        title = "{FRIED v2: a new grid of mass-loss rates for externally irradiated protoplanetary discs}",
      journal = {\mnras},
         year = 2023,
        month = dec,
       volume = {526},
       number = {3},
        pages = {4315-4334},
          doi = {10.1093/mnras/stad3054},
archivePrefix = {arXiv},
       eprint = {2310.03097},
 primaryClass = {astro-ph.EP},
       adsurl = {https://ui.adsabs.harvard.edu/abs/2023MNRAS.526.4315H}
}

@ARTICLE{2025ApJ...980..148S,
       author = {{Schwarz}, Kamber R. and {Samland}, Matthias and {Olofsson}, G{\"o}ran and {Henning}, Thomas and {Sellek}, Andrew and {G{\"u}del}, Manuel and {Tabone}, Beno{\^\i}t and {Kamp}, Inga and {Lagage}, Pierre-Olivier and {van Dishoeck}, Ewine F. and {Caratti o Garatti}, Alessio and {Glauser}, Adrian M. and {Ray}, Tom P. and {Arabhavi}, Aditya M. and {Christiaens}, Valentin and {Franceschi}, R. and {Gasman}, Danny and {Grant}, Sierra L. and {Kanwar}, Jayatee and {Kaeufer}, Till and {Kurtovic}, Nicolas T. and {Perotti}, Giulia and {Temmink}, Milou and {Vlasblom}, Marissa},
        title = "{MINDS. JWST-MIRI Observations of a Spatially Resolved Atomic Jet and Polychromatic Molecular Wind toward SY Cha}",
      journal = {\apj},
         year = 2025,
        month = feb,
       volume = {980},
       number = {1},
          eid = {148},
        pages = {148},
          doi = {10.3847/1538-4357/adaa79},
archivePrefix = {arXiv},
       eprint = {2409.11176},
 primaryClass = {astro-ph.SR},
       adsurl = {https://ui.adsabs.harvard.edu/abs/2025ApJ...980..148S}
}

@ARTICLE{2025A&A...699A.361V,
       author = {{van Dishoeck}, E.~F. and {Tychoniec}, {\L}. and {Rocha}, W.~R.~M. and {Slavicinska}, K. and {Francis}, L. and {van Gelder}, M.~L. and {Ray}, T.~P. and {Beuther}, H. and {Caratti o Garatti}, A. and {Brunken}, N.~G.~C. and {Chen}, Y. and {Devaraj}, R. and {Geers}, V.~C. and {Gieser}, C. and {Greene}, T.~P. and {Justtanont}, K. and {Le Gouellec}, V.~J.~M. and {Kavanagh}, P.~J. and {Klaassen}, P.~D. and {Janssen}, A.~G.~M. and {Navarro}, M.~G. and {Nazari}, P. and {Notsu}, S. and {Perotti}, G. and {Ressler}, M.~E. and {Reyes}, S.~D. and {Sellek}, A.~D. and {Tabone}, B. and {Tap}, C. and {Theijssen}, N.~C.~M.~A. and {Colina}, L. and {G{\"u}del}, M. and {Henning}, Th. and {Lagage}, P.-O. and {{\"O}stlin}, G. and {Vandenbussche}, B. and {Wright}, G.~S.},
        title = "{JWST Observations of Young protoStars (JOYS): Overview of program and early results}",
      journal = {\aap},
         year = 2025,
        month = jul,
       volume = {699},
          eid = {A361},
        pages = {A361},
          doi = {10.1051/0004-6361/202554444},
archivePrefix = {arXiv},
       eprint = {2505.08002},
 primaryClass = {astro-ph.GA},
       adsurl = {https://ui.adsabs.harvard.edu/abs/2025A&A...699A.361V}
}

@ARTICLE{2023ARA&A..61..287O,
       author = {{{\"O}berg}, Karin I. and {Facchini}, Stefano and {Anderson}, Dana E.},
        title = "{Protoplanetary Disk Chemistry}",
      journal = {\araa},
         year = 2023,
        month = aug,
       volume = {61},
        pages = {287-328},
          doi = {10.1146/annurev-astro-022823-040820},
archivePrefix = {arXiv},
       eprint = {2309.05685},
 primaryClass = {astro-ph.EP},
       adsurl = {https://ui.adsabs.harvard.edu/abs/2023ARA&A..61..287O}
}

@ARTICLE{2025OJAp....8E..54A,
       author = {{Allen}, Megan and {Anania}, Rossella and {Andersen}, Morten and {Aru}, Mari-Liis and {Ballabio}, Giulia and {Ballering}, Nicholas P. and {Beccari}, Giacomo and {Bern{\'e}}, Olivier and {Bik}, Arjan and {Boyden}, Ryan and {Coleman}, Gavin and {D{\'\i}az-Berrios}, Javiera and {Eatson}, Joseph W. and {Frediani}, Jenny and {Forbrich}, Jan and {Gkimisi}, Katia and {Goicoechea}, Javier R. and {Gupta}, Saumya and {Guarcello}, Mario G. and {Haworth}, Thomas J. and {Henney}, William J. and {Isella}, Andrea and {Itrich}, Dominika and {Keyte}, Luke and {Kim}, Jinyoung Serena and {Kuhn}, Michael and {Le Petit}, Frank and {Luo}, Lilian and {Manara}, Carlo and {Mauco}, Karina and {Meshaka}, Rapha{\"e}l and {Millstone}, Samuel and {Owen}, James E. and {Paine}, S{\'e}bastien and {Parker}, Richard and {Peake}, Tyger and {Peatt}, Megan and {Pinilla}, Paola and {Qiao}, Lin and {Ram{\'\i}rez-Tannus}, Mar{\'\i}a Claudia and {Ramsay}, Suzanne and {Reiter}, Megan and {Rogers}, Ciar{\'a}n and {Rosotti}, Giovanni and {Schroetter}, Ilane and {Sellek}, Andrew and {Testi}, Leonardo and {van Terwisga}, Sierk and {Vicente}, Silvia and {Walsh}, Catherine and {Winter}, Andrew and {Wright}, Nicholas J. and {Zeidler}, Peter},
        title = "{The past, present and future of observations of externally irradiated disks}",
      journal = {The Open Journal of Astrophysics},
         year = 2025,
        month = may,
       volume = {8},
          eid = {54},
        pages = {54},
          doi = {10.33232/001c.137538},
archivePrefix = {arXiv},
       eprint = {2502.12255},
 primaryClass = {astro-ph.SR},
       adsurl = {https://ui.adsabs.harvard.edu/abs/2025OJAp....8E..54A}
}

@ARTICLE{2025NatAs.tmp..147S,
       author = {{Schroetter}, Ilane and {Bern{\'e}}, Olivier and {Bron}, Emeric and {Alarcon}, Felipe and {Amiot}, Paul and {Bergin}, Edwin A. and {Boersma}, Christiaan and {Cami}, Jan and {Coleman}, Gavin A.~L. and {Dartois}, Emmanuel and {Fuente}, Asuncion and {Goicoechea}, Javier R. and {Habart}, Emilie and {Haworth}, Thomas J. and {Joblin}, Christine and {Franck}, Le Petit and {Onaka}, Takashi and {Peeters}, Els and {R{\"o}lling}, Markus and {Tielens}, Alexander G.~G.~M. and {Zannese}, Marion},
        title = "{A solar C/O ratio in planet-forming gas at 1 au in a highly irradiated disk}",
      journal = {Nature Astronomy},
         year = 2025,
        month = jul,
          doi = {10.1038/s41550-025-02596-6},
archivePrefix = {arXiv},
       eprint = {2505.22314},
 primaryClass = {astro-ph.GA},
       adsurl = {https://ui.adsabs.harvard.edu/abs/2025NatAs.tmp..147S}
}

@ARTICLE{2022EPJP..137.1132W,
       author = {{Winter}, Andrew J. and {Haworth}, Thomas J.},
        title = "{The external photoevaporation of planet-forming discs}",
      journal = {European Physical Journal Plus},
         year = 2022,
        month = oct,
       volume = {137},
       number = {10},
          eid = {1132},
        pages = {1132},
          doi = {10.1140/epjp/s13360-022-03314-1},
archivePrefix = {arXiv},
       eprint = {2206.11910},
 primaryClass = {astro-ph.EP},
       adsurl = {https://ui.adsabs.harvard.edu/abs/2022EPJP..137.1132W}
}

@ARTICLE{2020MNRAS.492.1279S,
       author = {{Sellek}, Andrew D. and {Booth}, Richard A. and {Clarke}, Cathie J.},
        title = "{The evolution of dust in discs influenced by external photoevaporation}",
      journal = {\mnras},
         year = 2020,
        month = feb,
       volume = {492},
       number = {1},
        pages = {1279-1294},
          doi = {10.1093/mnras/stz3528},
archivePrefix = {arXiv},
       eprint = {1912.06154},
 primaryClass = {astro-ph.EP},
       adsurl = {https://ui.adsabs.harvard.edu/abs/2020MNRAS.492.1279S}
}

@ARTICLE{2013ApJ...766L..23W,
       author = {{Walsh}, Catherine and {Millar}, T.~J. and {Nomura}, Hideko},
        title = "{Molecular Line Emission from a Protoplanetary Disk Irradiated Externally by a Nearby Massive Star}",
      journal = {\apjl},
         year = 2013,
        month = apr,
       volume = {766},
       number = {2},
          eid = {L23},
        pages = {L23},
          doi = {10.1088/2041-8205/766/2/L23},
archivePrefix = {arXiv},
       eprint = {1303.4903},
 primaryClass = {astro-ph.GA},
       adsurl = {https://ui.adsabs.harvard.edu/abs/2013ApJ...766L..23W}
}

@ARTICLE{2016ApJ...826L..15K,
       author = {{Kim}, Jinyoung Serena and {Clarke}, Cathie J. and {Fang}, Min and {Facchini}, Stefano},
        title = "{Proplyds Around a B1 Star: 42 Orionis in NGC 1977}",
      journal = {\apjl},
         year = 2016,
        month = jul,
       volume = {826},
       number = {1},
          eid = {L15},
        pages = {L15},
          doi = {10.3847/2041-8205/826/1/L15},
archivePrefix = {arXiv},
       eprint = {1606.08271},
 primaryClass = {astro-ph.SR},
       adsurl = {https://ui.adsabs.harvard.edu/abs/2016ApJ...826L..15K}
}
\bibliographystyle{aasjournalv7}

\end{document}